\documentclass[manuscript]{acmart}

\usepackage{amsmath}
\usepackage{graphicx}
\usepackage{color}
\usepackage{xspace}
\usepackage{comment}
\usepackage{subcaption}
\usepackage{tabularx}
\usepackage{paralist}
\newcommand{\todo}[1]{{\color{red} TODO: {#1}}}
\usepackage{colortbl}
\usepackage{pgf} 
\usepackage{arydshln} 
\definecolor{ao}{rgb}{0.0, 0.5, 0.0}
\def\cca#1{\cellcolor{ao!#1}\ifdim #1pt>49pt\color{white}\fi{#1}}
\definecolor{mygr}{rgb}{0.0, 0.5, 0.0}
\definecolor{mybl}{rgb}{0.0, 0.0, 0.5}
\def\ccb#1{\cellcolor{mybl!#1}\ifdim #1pt>50pt\color{white}\fi{#1}}
\def\ccg#1{\cellcolor{mygr!#1}\ifdim #1pt>50pt\color{white}\fi{#1}}
\usepackage{multirow}
\usepackage{tcolorbox}
\newcommand{\precision}{\% Resolved\xspace}

\newcommand{\ccaTest}[2]{%
  \ifnum #1>40
    \cellcolor{green!#1}%
  \else
    \ifnum #1>20
      \cellcolor{yellow!#1}%
    \else
      \cellcolor{red!#1}%
    \fi
  \fi
  \ifnum #2>49\color{white}\fi #1%
}

\newcommand{\ClaudeText}[1]{\textcolor[HTML]{1f77b4}{\textbf{\textit{#1}}}}
\newcommand{\OpenAIText}[1]{\textcolor[HTML]{2ca02c}{\textbf{\textit{#1}}}}
\newcommand{\ClaudeOpenAIText}[1]{\textcolor[HTML]{9467bd}{\textbf{\textit{#1}}}}
\newcommand{\OpenSourceText}[1]{\textcolor[HTML]{ffd700}{\textbf{\textit{#1}}}}
\newcommand{\MultiVendorText}[1]{\textcolor[HTML]{ff7f0e}{\textbf{\textit{#1}}}}
\newcommand{\UnknownText}[1]{\textcolor[HTML]{000000}{\textbf{\textit{#1}}}}

\title{Dissecting the SWE-Bench Leaderboards: Profiling Submitters and Architectures of LLM- and Agent-Based Repair Systems
}

\author{Matias Martinez}
\affiliation{%
  \institution{Universitat Politècnica de Catalunya}
  \city{Barcelona}
  \country{Spain}}
\email{matias.martinez@upc.edu}

\author{Xavier Franch}
\affiliation{%
  \institution{Universitat Politècnica de Catalunya}
  \city{Barcelona}
  \country{Spain}}
\email{xavier.franch@upc.edu}

\newcommand{\rqleaderboarc}{{What are the key characteristics of the submissions to the SWE-Bench Lite and SWE-Bench Verified leaderboards across dimensions such as submitter origin, product type, availability, and LLMs used?}}

\newcommand{\rqcorecomponents}{What is the high-level architecture of the approaches behind the submissions to the SWE-Bench Lite and SWE-Bench Verified leaderboards?}

\newcommand{\rqpipeline}{How are the phases of the end-to-end software maintenance pipeline implemented in submissions to the SWE-Bench Lite and SWE-Bench Verified leaderboards?}

\begin{document}

\newcommand{\swelite}{\texttt{SWE-Bench Lite}\xspace}
\newcommand{\sweverif}{\texttt{SWE-Bench Verified}\xspace}
\newcommand{\swebench}{\texttt{SWE-Bench}\xspace}
\newcommand{\defectsJ}{\texttt{Defects4J}\xspace}
\newcommand{\approach}[1]{\texttt{#1}}
\newcommand{\agent}[1]{\emph{#1}}
\newcommand{\group}[1]{\emph{#1}}
\newcommand{\new}[1]{\textcolor{blue}{#1}}
\newcommand{\company}[1]{\texttt{#1}}
\newcommand{\llm}[1]{\texttt{#1}}
\newcommand{\resultsBoth}[2]{(#1\% on \texttt{Lite}, #2\% on \texttt{Verified})}
\newcommand{\resultsLite}[1]{({#1}\% on \texttt{Lite})}
\newcommand{\resultsVerified}[1]{({#1}\% on \texttt{Verified})}
\newcommand{\totalsubmitters}{71\xspace}
\newcommand{\totalapproaches}{80\xspace}
\newcommand{\totalentries}{178\xspace}

\begin{abstract}
The rapid progress in Automated Program Repair (APR) has been driven by advances in AI, particularly large language models (LLMs) and agent-based systems. SWE-Bench is a recent benchmark designed to evaluate LLM-based repair systems using real issues and pull requests mined from 12 popular open-source Python repositories. Its public leaderboards—SWE-Bench Lite and SWE-Bench Verified—have become central platforms for tracking progress and comparing solutions. However, because the submission process does not require detailed documentation, the architectural design and origin of many solutions remain unclear. In this paper, we present the first comprehensive study of all submissions to the SWE-Bench Lite (79 entries) and Verified (99 entries) leaderboards, analyzing \totalapproaches unique approaches across dimensions such as submitter type, product availability, LLM usage, and system architecture.
Our findings reveal the dominance of proprietary LLMs (especially Claude 3.5), the presence of both agentic and non-agentic designs, and a contributor base spanning from individual developers to large tech companies.

\end{abstract}
\maketitle

\section{Introduction}

The field of Automated Program Repair (APR) has progressed rapidly over the last decade. Early approaches relied on techniques such as search-based methods —for example, GenProg \cite{LeGoues2011genprog}— and constraint-based methods, including SemFix~\cite{nguyen2013Semfix} and Nopol \cite{Xuan2017nopol}.
More recently, advances in Artificial Intelligence (AI) —particularly in deep learning and transformer architectures— have significantly influenced the field. Many current state-of-the-art techniques \cite{ruan2024specrovercodeintentextraction, ma2025alibabalingmaagent, bouzenia2024RepairAgent} are based on agents powered by large language models (LLMs), which can autonomously handle the entire repair process, from understanding the issue to validating the generated patch.

This progress has been significantly facilitated by the creation of bug benchmarks, which have allowed researchers to focus more on developing repair techniques rather than on how to evaluate them. 
They also serve as a common ground for evaluating and comparing the performance of  APR techniques.
One of the most widely adopted benchmarks is Defects4J \cite{Just2014Defects4J}, which contains over 800 real bugs mined from open-source Java libraries. It also provides a framework for reproducing bugs, visualizing ground truth patches, and other essential functionalities.
Defects4J has been widely used to evaluate a wide range of techniques, from traditional search-based approaches \cite{martinez2017automatic, Liu2020:ontheefficiency} to emerging methods, including deep learning-based techniques \cite{deeprepair2019, Chen2021SequenceR}, transformer models \cite{Ye2022RewardRepair, Smytzek2024FixKit, Lutellier2020Coconut, Nan2021Cure}, early explorations with LLMs \cite{Jiang2023Impact, Xia2024ChatRepair}, and, more recently, agentic APR systems \cite{bouzenia2024RepairAgent, ye2025adversarial}.

Recently, new benchmarks have emerged to address evolving scenarios, most notably \swebench{}~\cite{jimenez2024SWEBenchLLMs}, which includes 2,294 instances mined from pull requests and issues across 12 widely used Python repositories.
\swebench{} differs from Defects4J and similar benchmarks (e.g. \cite{Madeiral2019bears,lin20217Quixbugs}) in several important ways.
First, each instance in \swebench{} does not come with test cases or scripts that expose the bug and validate candidate fixes. 
In contrast, Defects4J provides such tests —mined from the same commits that introduce the fix or from subsequent ones.
The \swebench{} setup is more aligned with real-world scenarios that developers typically face,
where test cases are not  available when a bug is reported.
Second, \swebench{} provides a framework for running experiments and computing metrics such as precision (percentage of resolved issues) and project-level repair statistics, similar to what projects like RepairThemAll~\cite{Durieux:2019:RepairThemAll} have built on top of Defects4J and other benchmarks.
Finally, the creators of \swebench{} actively maintain public leaderboards where evaluation results are published and compared. 
Researchers and developers can submit their results via pull requests, including output generated by the evaluation framework and accompanying execution logs.
Such leaderboards were uncommon in earlier APR research, which often lacked platforms for sharing and comparing results.

The initial results published on the \swebench{} leaderboard correspond to experiments conducted by its authors to assess the repair capabilities of large language models, specifically, a fine-tuned version of CodeLlama~\cite{roziere2024codellama}.
Subsequently, the same authors introduced and evaluated an approach based on an Agent backed by an LLM (SWE-Agent).
A key feature of \swebench{} is the inclusion of issue descriptions, extracted from the GitHub platform, which enables the evaluation of LLM-based and agentic systems in reasoning over both natural language and source code, and in generating patches.
Since its release, \swebench{} has been adopted by researchers to benchmark their repair approaches and the experimental results included in their articles (e.g., \cite{Zhang2024AutoCoderRover, ma2025alibabalingmaagent, xia2024agentlessdemystifyingllmbasedsoftware, ruan2024specrovercodeintentextraction, li2025patchpilot}). Moreover, many of them have submitted their results to the leaderboards.


The \swebench{} leaderboard enables the research community —and the public at large— to regularly track the progress of intelligent systems (e.g., those based on LLMs or agents) in addressing software issues.
While this openness to submissions plays a crucial role in fostering innovation, it introduces a limitation from a research perspective: submissions are not required to disclose how the reported results were obtained.
For example, although some entries are linked to academic publications such as AutoCoderRover \cite{Zhang2024AutoCoderRover} or Agentless~\cite{xia2024agentlessdemystifyingllmbasedsoftware}, many others (e.g., from IBM or Amazon) are not. Rather than viewing this as a weakness of the leaderboard, we argue that it enables broader participation by organizations beyond academia, such as startups and tech companies,  who may wish to showcase the capabilities of their technologies while protecting their intellectual property.
This lack of publicly available details has motivated our study.
{\bf
In this paper, we present the first in-depth study of the \swebench{} leaderboards. We aim to analyze and categorize the solutions submitted to them, in order to learn who are the contributors to the benchmark and better understand the architectural choices and features that are driving progress in automated program repair.
}
We focus on two leaderboards hosted on the \swebench{} website:
\begin{inparaenum}
\item \swelite{}, which consists of 300 instances selected from the full benchmark; and
\item \sweverif{}, which includes 500 instances curated and verified by OpenAI~\cite{chowdhury2024swebenchverified}.
\end{inparaenum}

To carry out our study, we examine each entry in the \swebench{} leaderboards and collect the available metadata, including the submission pull request, submitter link, and precision score (\precision).
We then searched both academic publications and grey literature (e.g., blog posts, LinkedIn posts) to gather additional information about each submission. 
This includes the nature of the submitter (e.g., company or academic institution), any software artifact related to the submission, the LLMs used, the openness of the solution, and relevant architectural details of the solution.
To describe the high-level architecture of each submitted approach, we focus on three dimensions:
\begin{inparaenum}
 \item the repair workflow authoring,
    \item  the autonomy over the execution path, and
    \item  the number of LLM-based Agents, if the solution is agent-based.
\end{inparaenum}

Finally, we adopt the end-to-end software maintenance pipeline defined by Liu et al.~\cite{liu2024SurveyAgentsLLM4SE}, originally proposed to categorize LLM-based agents.
This pipeline consists of the following phases:
\begin{inparaenum}[\it a)]
\item Preprocessing,
\item Issue Reproduction,
\item Issue Localization,
\item Task Decomposition,
\item Patch Generation,
\item Patch Verification, and
\item Ranking.
\end{inparaenum}
For each phase, we analyze and discuss the different strategies employed by the submissions to the \swebench{} leaderboards.


In total, we analyze 79 entries from \swelite{}, 99 entries from \sweverif{}, comprising \totalapproaches{} unique approaches.\footnote{An approach can be evaluated on both \swelite{} and \swelite{} leaderboards, and some approaches have multiple entries, e.g., each using different LLMs.}
Our results show that, while there is a diversity of submitter types (e.g., academia, industry, and collaborations), the majority of submissions come from industry, particularly from small companies and large publicly traded corporations such as Amazon, IBM, and Google.
We also identified submissions made by individual developers, highlighting how access to powerful language and code models —whether commercial (e.g. GTP 4 from OpenAI, Claude 4 from Anthropic) or open-source (e.g., Llama from Meta, Qwen from Alibaba)— is enabling the creation of sophisticated repair systems.

We also found that the majority of submissions achieving state-of-the-art results rely on proprietary LLMs —most notably, Claude 4 Sonnet.
In terms of system architecture, submissions are highly diverse: they range from single-LLM solutions without agentic capabilities, to complex multi-agent systems with emergent workflows. 
Notably, no single architecture consistently achieves state-of-the-art performance. Rather, high-performing submissions follow a variety of architectural strategies, suggesting that multiple design paradigms can be effective.
We also present a discussion of \swebench{} that covers patch quality concerns, debates over agentic architectures, the openness of solutions, as well as limitations and potential saturation of the benchmark.
{
To the best of our knowledge, this paper presents the first comprehensive characterization of all submitters and solutions published on the \swebench{} leaderboards.} 
In fact, the type of analysis of architectures and pipelines that we have carried out could be eventually applied to similar benchmarks. 

\label{sec:rq}
The research questions that guide our research are:

\begin{enumerate}

\item RQ 1:  \emph{\rqleaderboarc}

\item RQ: 2 \emph{\rqcorecomponents}

\item RQ 3: \emph{\rqpipeline}

\end{enumerate}

The paper continues as follows.
Section \ref{sec:methodology} presents the methodology of our study.
Section \ref{sec:results} presents the results.
Section \ref{sec:discussion} presents a discussion of the results.
Section \ref{sec:ttv} presents the threats to validity.
Section \ref{sec:relatedwork} presents the related work.
Section \ref{sec:conclusion} concludes the paper.

\section{Methodology}
\label{sec:methodology}

In this section, we describe the methods used to answer the research questions.
In Section \ref{sec:method:inspection} we present the method to inspect the \swebench{} dashboards and to respond to RQ 1.
Sections~\ref{sec:methodology:architecture} and \ref{sec:methodology:stages} outline the methods used to analyze the submitted approaches in relation to RQ2 and RQ3.
Finally, Section~\ref{sec:method:collectedData} details the data collected to support our analysis and answer the research questions.

\subsection{Inspection of \swelite{}  and \sweverif{} Leaderboards}
\label{sec:method:inspection}

We present the method for collecting information from two \swebench{} leaderboards: \swelite{} and \sweverif{}.
We decided not to analyse the other two \swebench{}  leaderboards, namely \texttt{Full} and \texttt{Multimodal}, for the following reasons: \begin{inparaenum}[\it a)] \item \texttt{Full}: 
all solutions submitted to \texttt{Full} are also included in, at least, one of the two considered leaderboards, either \texttt{Lite} or \texttt{Verify}; \item \texttt{Multimodal}: we restrict our research to language- (either natural or programming) based agents. \end{inparaenum}

\subsubsection{Data Collection Procedure}
\label{sec:method:inspection:datacollection}
To collect data, we inspect the \swelite{} and \sweverif{} leaderboards as follows.
First, we visit each leaderboard page and retrieve from each entry (that is, each submission): 
Submission name, Resolved rate (\%),and Site (URL).
Then, for each entry, we conduct the following steps.
\begin{inparaenum}[\it a)] \item We visit the listed site to gather valuable information about the approach (such as its architecture and design decisions) and to find any references related to experiments conducted on SWE-Bench.
\item While browsing a page, we manually navigate the entire site and follow any external links that may be useful for our goal (e.g., link to a blog or a paper).
\item We supplement these data with a dedicated Google search to capture additional pages that are not accessible through the website linked to the leaderboard.
The Google query we create is \texttt{"<Name\_Entry> + SWE-Bench"}; in  \texttt{"<Name\_Entry>"}, we remove the date and the model's details.
For example, given the entry \texttt{"Gru(2024-12-08)"}, the name used in the Google search is \texttt{"GRU"}.
\item We inspect the pages obtained with the web search, and store those that describe the approach or the experiment conducted on \swebench{}.
\item We also inspect the data submitted for each entry, notably the \texttt{README.md}\footnote{Example of \texttt{README.md} file included in a commit to \swebench{}: \url{evaluation/lite/20240509_amazon-q-developer-agent-20240430-dev/README.md}} and \texttt{metadata.yaml} files, available in the \swebench's repository\footnote{\url{https://github.com/SWE-bench/experiments/tree/main/evaluation/lite}} dedicated to collect the row information from the submissions.
\item We also visit the LinkedIn of the submitter, and eventually some of its members, in order to detect posts that discuss the experiment.

In addition to collecting the submitter information for each entry, we also identify the underlying approach associated with it.
For example, the entries \texttt{"SWE-agent + GPT-4o (2024-05-13)"} and \texttt{"SWE-agent + GPT-4 (1106)"} originate from the same research group (Princeton University) and are based on the same approach, \texttt{SWE-agent}, but differ in the LLM used (GPT-4o and GPT-4, respectively).
\end{inparaenum}
We collected information about the LLMs used in each submission from multiple sources:
\begin{inparaenum}[\it a)]
\item the submission name (e.g., \llm{GPT-4o} and \llm{GPT-4}, as in the entries mentioned above);
\item the \texttt{metadata.yaml} file—if available—specifically the \texttt{model} field;
\item the \texttt{README.md} file;
\item associated scientific articles and grey literature (blog posts).
\end{inparaenum}

Availability and configuration of data has sometimes restricted the type of analysis. For instance, we initially planned an analysis of contributions per country, but at the end we discarded this option since it was not entirely clear in many leaderboard entries.

\subsubsection{Description of data to collect}
We apply content analysis~\cite{kovrigin2024importancereasoningcontextretrieval} to the collected data in order to elicit coding schemas related to several attributes. We apply a combination of deductive and inductive coding: we predefine the list of themes (corresponding to the different attributes we wanted to analyse), and for each of them, we start with an empty set of codes that we gradually refine as we process the leaderboards' content. We present below the result.

\paragraph{Submitter Category}
\label{sec:method:inspection:submitterCategory}

We collect the type of organization that submitted a solution to \swebench{}.
Our inductive coding approach results in the following codes grouped into categories:

\begin{enumerate}
\item \textbf{Academia}: The submission is made by members affiliated with academic institutions and is accompanied by a research article whose authors are all from academia. We distinguish among \textbf{Single Academy} institutions and \textbf{Collaboration Academia}, when different academic institutions collaborate in a submissions.

\item \textbf{Industry}: The submission is made by members affiliated with a company. We further classify these companies by size, using the Microsoft-LinkedIn classification system\footnote{Microsoft Company Size Codes: \url{https://learn.microsoft.com/en-us/linkedin/shared/references/reference-tables/company-size-codes}}, which estimates the company size based on LinkedIn user data. For instance, Google’s LinkedIn page reports 10,001+ employees \footnote{\url{https://www.linkedin.com/company/google/about/}}. 

\begin{enumerate}
\item \textbf{Small}: Companies with fewer than 50 employees (Codes A-B-C).
\item \textbf{Medium}: Companies with fewer than 500 employees (Codes D-E).
\item \textbf{Large}: Private companies (not publicly traded) with more than 500 employees (Codes F-G-H-I).
\item \textbf{Large-Publicly Traded}: Publicly traded companies with more than 500 employees (e.g., Google, NASDAQ:GOOG).
\item \textbf{Unknown}: Companies for which we were unable to subclassify due to lack of information.
\end{enumerate}
In addition, similar to the academia case, we record \textbf{Collaboration Industry} when the submission comes from a joint effort between industry-affiliated contributors.

\item \textbf{Academia-Industry:} The submitter has a blend of academia and industry. We have found two different types:
\begin{enumerate}
\item \textbf{Collaboration Academia-Industry}: collaboration among academic and industry-affiliated contributors, as evidenced by an article co-authored by individuals from both domains.
\item \textbf{Academic Spin-off}: A company that is a spin-off from an academic institution, as explicitly indicated on its website, including reference to the originating university.
\end{enumerate}

\item \textbf{Open-Source Community}: The submission is made by a non-profit or community-driven organization.

\item \textbf{Single Developer}: The submission is made by an individual acting in a personal capacity.

\item \textbf{Unknown}: We have not found enough information as to classify the submitter.
\end{enumerate}

When an entry has more than a single source (e.g., an Arxiv paper and a LinkedIn post), we check them all in order to capture as much detail as possible.
Nevertheless, when summarizing the results to address the research question, we report only one source per entry, specifically, the one that provides the most detailed information about the approach.

\paragraph{Product and Accessibility} 

We analyze how solutions submitted to \swebench{} are made accessible to users or stakeholders, focusing on two key dimensions.
First, we assess the \textbf{type of product}, if any, associated with each submission. 
In particular, we focus on two dimensions.
\begin{itemize}
    \item {\bf Product Purpose}: describes \emph{what} the product does.
We identify ten different purposes:
\begin{inparaenum}[\it a)]
\item Agent Framework,
\item Coding Assistant,
\item Development Assistant,
\item Development Framework,
\item Development Platform,
\item Issue Resolution,
\item None,
\item Other-Inference Service,
\item Platform Code Optimization/Improvement,
\item Platform Code Representation,
\item Platform Integration, and 
\item Problem Solving.
\end{inparaenum} An additional code is introduced for products for which we cannot identify the form.

\item {\bf Product Form}: Describes \emph{how} the product is delivered or used.
We identify five different forms:
\begin{inparaenum}[\it a)]
    \item Cloud platform,
 \item Command-line tool,
 \item Github plugin,
 \item IDE plugin, and
 \item Library.
\end{inparaenum} An additional code is introduced for entries without form.
\end{itemize}

For example, the entries corresponding to the \texttt{Amazon Q Developer Agent} approach -submitted to both \swelite{} and \sweverif{}- are linked to a Coding Assistant, distributed as an IDE plugin for the IntelliJ platform.\footnote{Amazon Q plugin: \url{https://plugins.jetbrains.com/plugin/24267-amazon-q/?b=jb&p=build&s=hero}}

Second, we record the \textbf{availability of the product}. Our deductive coding on both leaderboards results in four codes representing different levels of availability: 
\begin{enumerate}
\item \textbf{Publicly Available Product (PAP)}:
A product based on the submitted solution is publicly accessible, either commercially (paid) or for free. 
This includes  plugins for desktop IDEs (e.g., IntelliJ) and cloud-based applications (e.g., cloud-based IDEs).  
The product must be publicly accessible, for example, as an IDE plug-in available in plug-in/extensions marketplaces, such as those for Visual Studio\footnote{\url{https://marketplace.visualstudio.com/vscode}} or IntelliJ\footnote{\url{https://plugins.jetbrains.com/}}.
Importantly, this category does not refer to the availability of source code; for instance, an IntelliJ plugin may be either open- or closed-source.
\item \textbf{Upon Request (UR)}:
The product is not openly accessible but can be obtained upon request or through a waiting list. This includes early-access solutions and invite-only services.
\item \textbf{Business-to-Business (B2B)}:
The solution is only available under business agreements (B2B) and is not accessible to the general public. 
\item \textbf{Non-Commercial Solution (NCS)}: 
The solution corresponds to a non-commercial tool. 
In this case, no associated product is distributed through plugin marketplaces nor is there a deployed solution such as an online IDE. 
This category typically includes tools shared as research artifacts or open-source code (e.g., on GitHub).
\item \textbf{Unavailable (UN)}: 
The solution is not available under any of the previously defined levels.
\end{enumerate}
Note that in this analysis, we focus on products associated to each entry -if any- rather than on the code used in the \swebench{} experiment (which could be eventually the same).

\paragraph{Open-source solution}

We identify entries whose source code is open and available on platforms such as GitHub from those whose code is closed. 
Again, this assessment does not take into account the openness of the underlying language models.

\paragraph{LLMs Employed in Solutions}

We identify the underlying LLMs used by each solution, noting that multiple models may be involved.
This information is usually presented in the submission name (shown in the leaderboard) but also eventually in the \texttt{metadata.yaml} file required for the \swebench{} submission.
If it is not available there, we search for this information in other sources such as preprint, scientific articles or blog posts.\footnote{If an article presents evaluations of multiple models, we select the one whose reported precision (i.e., \% Resolved) matches that of the corresponding leaderboard entry.}
Note that these articles may also report additional results (e.g., using other models) that are not included in the leaderboard. However, for this research question, we exclusively consider the result corresponding to the leaderboard submission, disregarding the others.

\subsection{Architecture of approaches in \swebench{}}
\label{sec:methodology:architecture}

In the previous step, we collected artifacts related to each leaderboard entry, which include -if available- published papers, ArXiv documents, Blog post, LinkedIn post, documentation on Github repository (Readme.me) or information contained in the commit done for the submission.
We now analyze these artifacts in order to extract information that describes the approaches associated with the submissions.

In RQ2, the information that we aim at capturing is the architecture of the approaches.
For that, we have elicited three different dimensions.

\subsubsection{Workflow Authoring}
\label{sec:metodology:systemarchitecture:workflow}

The first dimension is the presence or absence of a predefined workflow, that is, a set of steps conducted to repair issues. 

\begin{inparaenum}
    \item  {\bf Workflow authored by humans}: 
    The overall flow is predefined by humans. 
    Traditional APR systems such as GenProg ~\cite{legoues2012genprogtse} usually implement a workflow consisting of two or three main steps, which are executed sequentially: first, a Fault Localization stage, followed by Patch Generation stage, and finally Patch validation stage.
    That workflow is usually hard-coded in the approach implementation.
    Note that inside one of the workflow steps, autonomy (see next dimension) may be present, nevertheless the control structure of the principal workflow is fixed. 
    
    \item  {\bf Emergent workflow}:  There is no predefined workflow crafted by humans. It corresponds to fully agentic systems which are driven entirely by one or more autonomous agents.
    The execution path is dynamic and reactive, based on goals, environment or feedback.
\end{inparaenum}

\subsubsection{Autonomy over execution path: Control flow autonomy}
\label{sec:metodology:systemarchitecture:controlflow}

This dimension captures the degree of freedom the system has in determining its execution trajectory when solving a task:

\begin{enumerate}
    \item \textbf{Emergent Autonomy}: No predefined workflow is provided. The system (typically an agent) autonomously determines the flow of execution (including which steps to take, their ordering, and whether to repeat or skip them) based entirely on environmental feedback and internal reasoning. Examples include tool-using agents equipped only with a goal and a set of available actions (e.g., view and edit a source code file).

    \item \textbf{Scaffolded Execution}: A human provides a structured workflow, including stages (e.g., issue localization, patch generation, validation) and allowed transitions between them. 
    The system —typically leveraging an LLM— is free to make decisions within and across these stages.
    This structure  -known as a \emph{Scaffold}- allows local autonomy within a globally defined flow. 
  
    \item {\bf Fixed Execution}: The control flow is fully predefined by a human and executed deterministically. The system follows a strict sequence of steps with no dynamic decision-making or autonomy. This pattern characterizes traditional APR pipelines, such as Nopol~\cite{Xuan2017nopol}, and early LLM-based systems~\cite{Jiang2023Impact}. 
    
\end{enumerate}

\subsubsection{Number of Agents}
\label{sec:metodology:systemarchitecture:numberofagents}
A quick observation of the leaderboard entries reveals, at least by name, that most of the solutions are based on Agents (e.g., \texttt{SWE-Agent}).
Nevertheless, we define the following categories:
\begin{inparaenum}[\it a)]
    \item No Agent. This includes repair systems that just prompt LLMs, e.g., using Chain of Thoughts;
    \item Single Agent;
    \item Multiple Agents.
\end{inparaenum}

Please note that some combinations are not allowed, e.g. an approach cannot have an emergent workflow and be non-agentic. 

\subsection{LLM-based Agents for End-to-end Software Maintenance}
\label{sec:methodology:stages}

For responding RQ3, we follow the taxonomy of LLM-based
agents for software engineering task by Liu et al. \cite{liu2024SurveyAgentsLLM4SE}.
In particular, we focus on the pipeline of LLM-based Agents for End-to-end Software Maintenance that the authors present, which is composed of the following phases:
\begin{inparaenum}[\it a)]
    \item Preprocessing,
    \item Issue Reproduction,
    \item Issue Localization,
    \item Task Decomposition,
    \item Patch Generation,
    \item Patch Verification,
    \item Ranking.
\end{inparaenum}
For each of these phases, we discuss how the approaches face them.
Even those not all the solutions are based on agents, the phases are also present in ``traditional'' APR approaches such as the \emph{Generate-and-Validate} \cite{LeGoues2011genprog}, which conduct three of these phases: 
\begin{inparaenum}[\it a)]
    \item Bug localization (related to the mentioned Issue Localization),
    \item Patch Generation,
    \item Patch Verification.
\end{inparaenum}

\subsection{Collected Data}
\label{sec:method:collectedData}

\begin{table}[t!]
\centering
\scriptsize
\begin{tabular}{|l|rr|rr|rr|rr|rr|rr|rr|}
\hline
Type & \multicolumn{2}{c|}{Scientific Articles} & \multicolumn{2}{c|}{Blog} & \multicolumn{2}{c|}{SWECommit}& \multicolumn{2}{c|}{Github} & \multicolumn{2}{c}{Linkedin}  & \multicolumn{2}{c|}{Whitepaper}  & \multicolumn{2}{c|}{Nothing}\\
\hline
Leaderboard & Lite & Verified & Lite & Verified & Lite & Verified & Lite & Verified & Lite & Verified & Lite & Verified & Lite & Verified \\
\hline
Total & 39 & 40 & 18 & 38 & 10 & 6 & 6 & 0 & 1 & 1 &  1 & 0 & 3 & 3 \\
\hline
\end{tabular}
\caption{Distribution of approach documentation types across the \swelite{} and \sweverif{} leaderboards. 
Each column shows the number of entries associated with a specific documentation source.}
\label{tab:results:descriptionPivot}
\end{table}

We collected the entries published in the leaderboards until July 17th.\footnote{Version of  \swebench{} studied: \url{https://github.com/SWE-bench/swe-bench.github.io/commit/c53e6f9b8ea0b17060ef3a7718d3020e529f8a22}}
Table \ref{tab:results:descriptionPivot} presents the distribution of artifact types used to describe each leaderboard entry.
In this study, we analyze these artifacts, together with the information available in the \swebench{} leaderboards -available online, to answer the research questions.
Let us discuss the distribution of the type artifacts that we analyze.
In both the \swelite{} and \sweverif{} leaderboards, the most common artifact type is scientific articles (including preprints on arXiv), accounting for 39 and 40 entries, respectively.
The scientific venues and journals where these articles have been published include both Software Engineering and AI/Machine Learning conferences.
In the Software Engineering field, we identified publications at ICSE 2025~\cite{ruan2024specrovercodeintentextraction}, FSE 2025~\cite{xia2024agentlessdemystifyingllmbasedsoftware,ma2025alibabalingmaagent} and ISSTA 2024~\cite{Zhang2024AutoCoderRover} and 2025~\cite{ma2024lingmaswegpt}.
In the AI and Machine Learning domain, relevant articles were published at ICLR 2024~\cite{wang2025solvedSWEBench}, ICLR 2025~\cite{ouyang2024repographenhancingaisoftware, wang2024openhandsopenplatformai}, ICML 2025~\cite{zainullina2025guidedsearchstrategies, li2025patchpilot, sohrabizadeh2025nemotroncortexa} and NeurIPS 2024~\cite{yang2024sweagentagent}.

Blog posts represent the second most frequent type of artifact used to describe the experiments conducted on \swebench{}, to announce results, and to explain the proposed solutions.
We associated 18 and 38 entries to blog posts in \swelite{} and \sweverif{}, respectively.
There is also one entry, CodeFixer from Globant, where the description was provided in a white paper available on the company's website, and was not published on arXiv or any other pre-print repository at the time of writing this article.
Moreover, some entries provide descriptions of the experiments and/or approaches in the \texttt{README.me} files included in the commits to SWE-Bench that introduce the experimental results. This is the case for 10 and 6 submissions in \swelite{} and \sweverif{}, respectively.
We also detect entries (3 on each leaderboard) not related to any artifact describing the experiment or the approach.

\section{Results}
\label{sec:results}

\subsection{RQ1: \rqleaderboarc}
\label{sec:leadboard}

We begin with an overview of the number of entries and distinct approaches submitted to the leaderboards.
Then, we analyze the approaches along four dimensions:
\begin{inparaenum}
\item submitter profile,
\item product type,
\item availability, and
\item underlying LLM(s) used.
\end{inparaenum}

\subsubsection{Overview and Temporal Analysis of \swebench{} Leaderboard Submissions}

By July 17th 2025, we count, in total, 79 and 99 entries on \swelite{} and \sweverif{}, respectively. 
We identified 52 and 50 distinct approaches, respectively, including Agentless \cite{xia2024agentlessdemystifyingllmbasedsoftware} and AutoCodeRover \cite{Zhang2024AutoCoderRover}.
Among these, 30 approaches appear only in \swelite{}, 28 only in \sweverif{}, and 22 in both.
The complete list of approaches and submissions is provided in the appendix~\cite{Apprendix}.

Fig. \ref{fig:date_submission} shows the cumulative number of entries on each leaderboard, taking the dates from the leaderboards themselves.
The vertical dotted red line shows the release date of \sweverif{} (August 13th 2024), considered to be the date in which OpenAI released the blog article that describes the leaderboard\footnote{\url{https://openai.com/index/introducing-swe-bench-verified/}}. We observe that there are entries whose date precedes that official release, being the reason that the mentioned article also included, in addition to well defined construction criteria of \sweverif{}, the results of an evaluation of open-source approaches (\emph{open-source scaffolds} according to that article) such as AutoCoderRover, Agentless and SWE-Agent on \sweverif{}, which were already present in \swelite{}.
The \sweverif{} leaderboard includes these entries with the same date as \swelite{}.\footnote{Added in the leaderboard by this commit: \url{https://github.com/SWE-bench/swe-bench.github.io/commit/8d66692d5f55101c3c0f53daac3f8a479d29406e}}
For example, the entry \texttt{SWE-agent + GPT 4 (1106)} is present in both leaderboards with the date \texttt{2024-04-02} (a date earlier than \sweverif{}'s release).

The figure shows how the leaderboards have evolved along five stages:
\begin{enumerate}
    \item Nonexistence of \sweverif{}. From Oct. 2023 until May 2024, all submission were in \swelite{} only.
    \item Prevalence of \swelite{}. Until mid-September 2024, both repositories co-existed but still the growth rate of \swelite{} was greater than that of \sweverif{}'s.
    \item Prevalence of \sweverif{}. In only two months, until mid-Nov. 2024, \sweverif{} experienced a sudden growth until reaching the same size as \swelite{} (41 entries each).
    \item Similarity of both leaderboards. Until mid-Feb. 2025, the number of entries remained approximately the same for both \swelite{} and \sweverif{}.
    \item Reduced activity on \swelite{}. While submissions to \sweverif{} continue to grow steadily, activity on \swelite{} has significantly slowed, with only a few recent entries—such as those involving Claude 4 LLM— being added.
\end{enumerate}

Figure~\ref{fig:resolved_evolve} shows the evolution of the \precision{} metric, which is the primary measure reported in both leaderboards. 
Despite a period among Oct. 2024 and Nov. 2024 in which some solutions uploaded to \sweverif{} were not optimal, the precision is notably higher in \sweverif{} than in \swelite{}.  Specifically, the median and maximum precision are 46.9\% and 75.2\% for \sweverif{}, compared to 31.5\% and 60.0\% for \swelite{}.
This difference may be attributed to the construction of \sweverif{}, which was derived from the original \swebench{} by manually filtering out instances considered too hard or unsolvable, with the aim of increasing the benchmark's robustness and reliability~\cite{chowdhury2024swebenchverified}.

\begin{figure}[t]
\begin{subfigure}[h]{0.49\textwidth}
  \centering
\includegraphics[width=\textwidth]{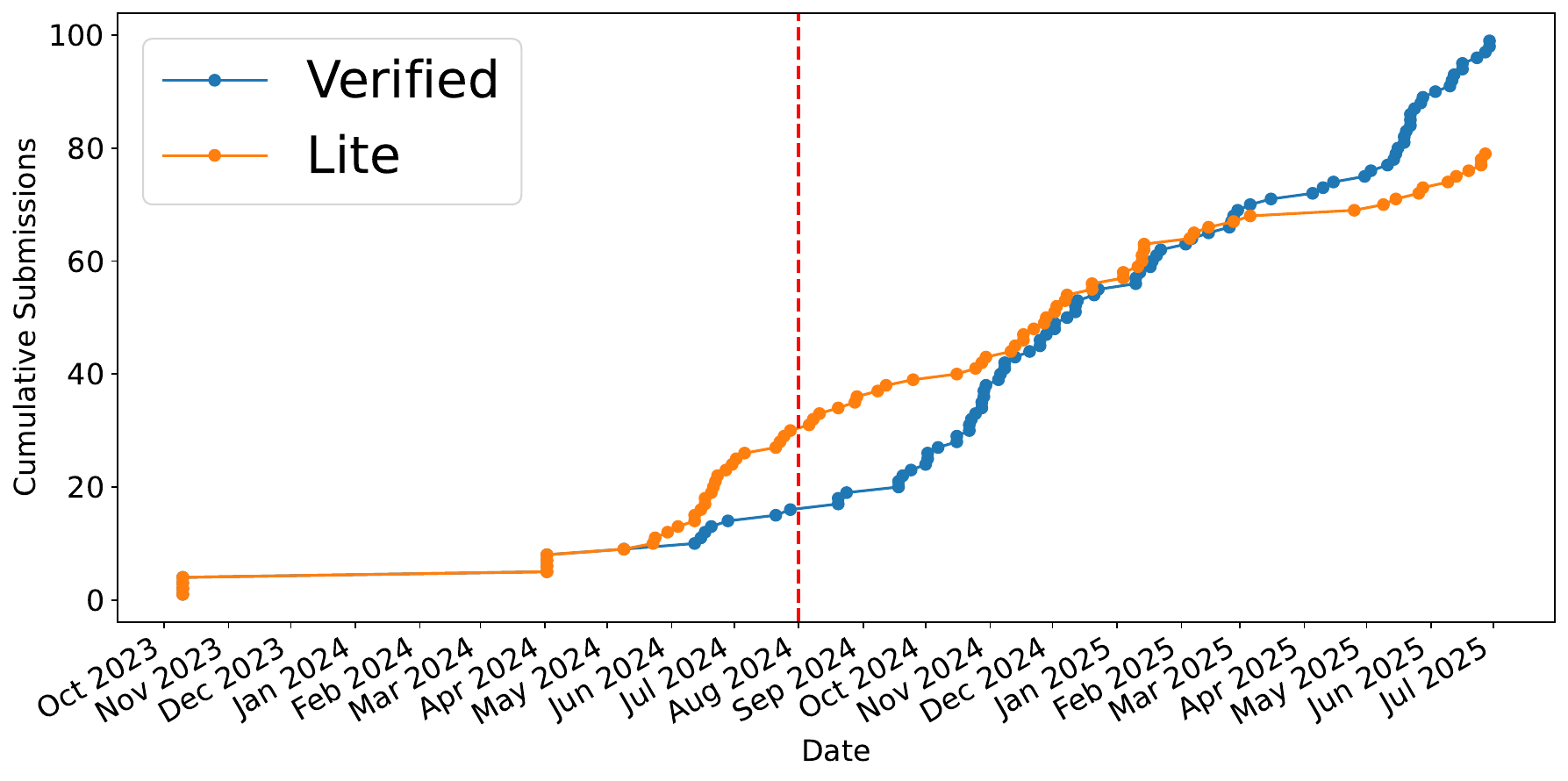}
\caption{Entries by date. The vertical dotted line correspond to the official release of \sweverif{} by OpenAI.}
\label{fig:date_submission}
\end{subfigure}
~
\begin{subfigure}[h]{0.49\textwidth}
  \centering
\includegraphics[width=\textwidth]{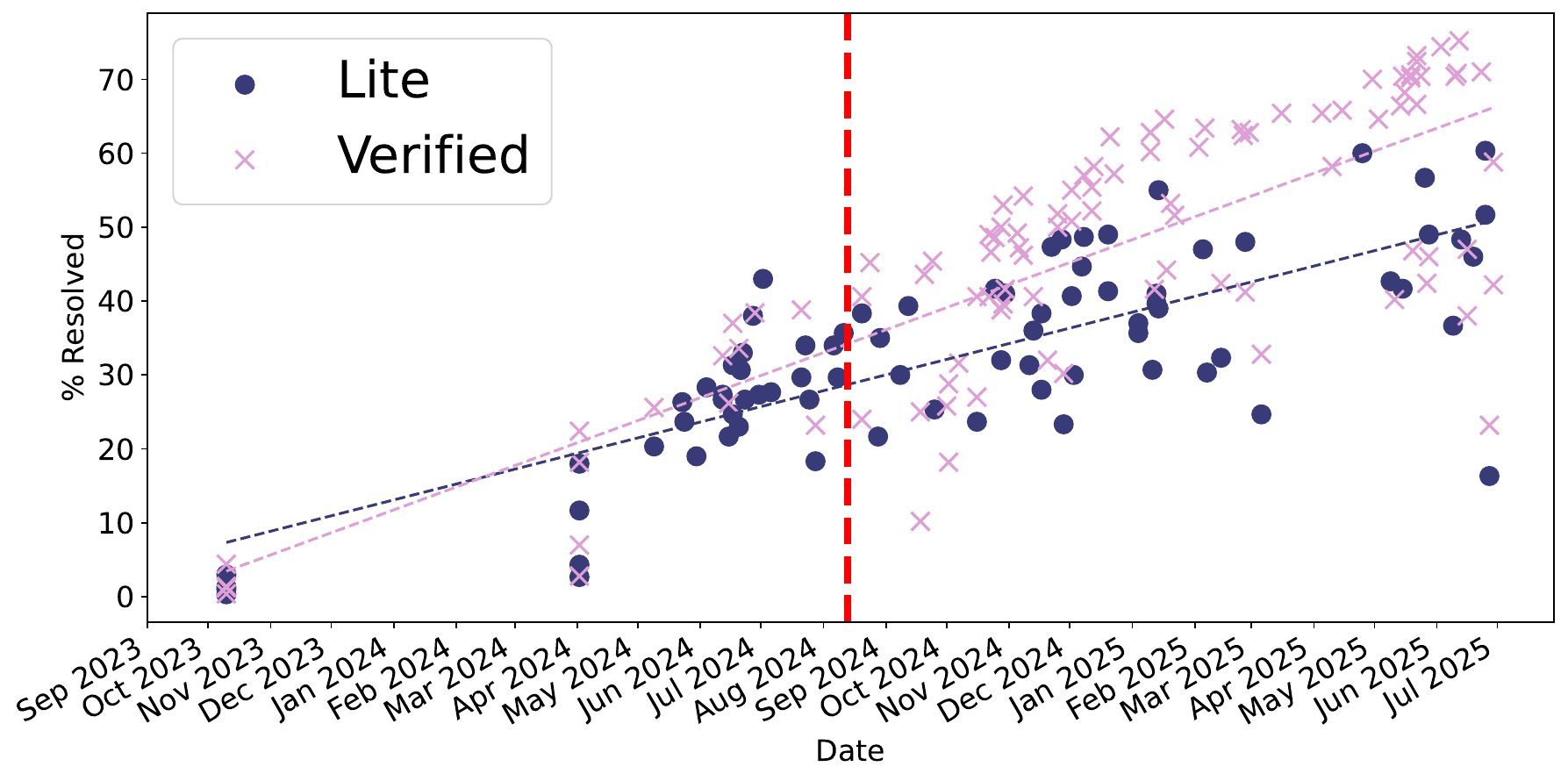}
\caption{Evolution of \precision{}. The vertical line correspond to the official release of \sweverif{} by OpenAI.}
\label{fig:resolved_evolve}
\end{subfigure}
\caption{}
\end{figure}

\begin{table}[t!]
\centering

\begin{tabular}{|l|llll|llll|rr|}
\hline
&\multicolumn{4}{|c|}{\swelite{}} & \multicolumn{4}{|c|}{\sweverif{}} & \multicolumn{2}{|c|}{Total} \\ 
\cline{2-9}
Submitter Type&\multicolumn{2}{|c|}{} & \multicolumn{2}{|c|}{\precision{}}
&\multicolumn{2}{|c|}{} & \multicolumn{2}{|c|}{\precision{}} & \multicolumn{2}{|c|}{} \\
\cline{2-9}
&\#E & \#S & Median & Max
&\#E & \#S & Median & Max &\#E & \#S \\

\hline

Single Academy (SA) & 19 & 4 & \ccb{23} & \ccg{56.67} & 17 & 3 & \ccb{23.2} & \ccg{66.6} & 36 & 4 \\

Collaboration Academia (CA) & 6 & 3 & \ccb{41.34} & \ccg{60.33} & 3 & 2 & \ccb{32.8} & \ccg{46} & 9 & 4 \\

 \hdashline
Total Academia & 25 & 7 & \ccb{27.33} & \ccg{60.33} & 20 & 5 & \ccb{31.5} & \ccg{66.6} & 45 & 8 \\

\hline
Academia Spinoff (AS) & 2 & 1 & \ccb{24.84} & \ccg{30.67} & 4 & 2 & \ccb{42.3} & \ccg{51.6} & 6 & 2 \\

Collaboration Ac-Ind (CAI) & 10 & 9 & \ccb{29.84} & \ccg{47} & 7 & 6 & \ccb{53} & \ccg{64.6} & 17 & 12 \\
 \hdashline
Total Academia-Industry  & 12 & 10 & \ccb{29.835} & \ccg{47} & 11 & 8 & \ccb{46.2} & \ccg{64.6} & 23 & 14 \\
\hline

Company-Large (CL) & 4 & 2 & \ccb{34.84} & \ccg{39.33} & 4 & 1 & \ccb{68.5} & \ccg{75.2} & 8 & 2 \\
Company-Large-Publicly Traded (CLP) & 7 & 5 & \ccb{27.33} & \ccg{48.33} & 21 & 8 & \ccb{39.6} & \ccg{68.2} & 28 & 10 \\
Company-Medium (CM) & 1 & 1 & \ccb{49} & \ccg{49} & 8 & 4 & \ccb{63.9} & \ccg{73.2} & 9 & 4 \\
Company-Small (CS) & 19 & 15 & \ccb{38} & \ccg{60} & 31 & 16 & \ccb{51.8} & \ccg{74.4} & 50 & 23 \\
Company-Unknown (CU) & 1 & 1 & \ccb{44.67} & \ccg{44.67} & - & - & - & - & 1 & 1 \\

Collaboration Industry (CI) & - & - & - & - & 1 & 1 & \ccb{46.8} & \ccg{46.8} & 1 & 1 \\
 \hdashline
Total Industry  & 32 & 24 & \ccb{35.8} & \ccg{60} & 65 & 30 & \ccb{50} & \ccg{75.2} & 97 & 41 \\ 
\hline

OpenSource Community (OSC) & 6 & 2 & \ccb{28.68} & \ccg{39} & 2 & 2 & \ccb{67.1} & \ccg{70.8} & 8 & 3 \\

\hline
Single Developer (SD) & 2 & 2 & \ccb{29} & \ccg{30.33} & - & - & - & - & 2 & 2 \\
\hline
Unknown (UNK) & 2 & 2 & \ccb{42.17} & \ccg{42.67} & 1 & 1 & \ccb{41.6} & \ccg{41.6} & 3 & 3 \\

\hline
\end{tabular}
\caption{Descriptive statistics of submitter organizations. \#E shows the total number of entries, \#S is the number of distinct submitters.}
\label{tab:summarySubmitter}
\end{table}

\begin{figure}[t]
\begin{subfigure}[h]{0.33\textwidth}
  \centering
\includegraphics[width=0.95\textwidth]{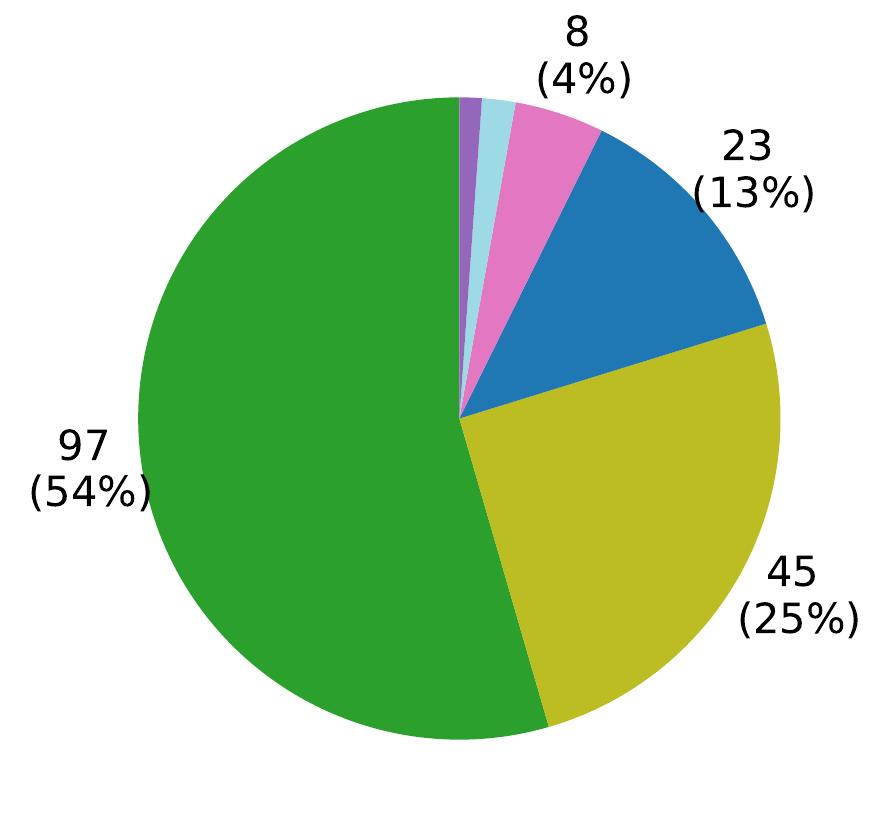}
\caption{Accumulated}
\label{fig:pieEntries_acum}
\end{subfigure}
\begin{subfigure}[h]{0.33\textwidth}
  \centering
\includegraphics[width=\textwidth]{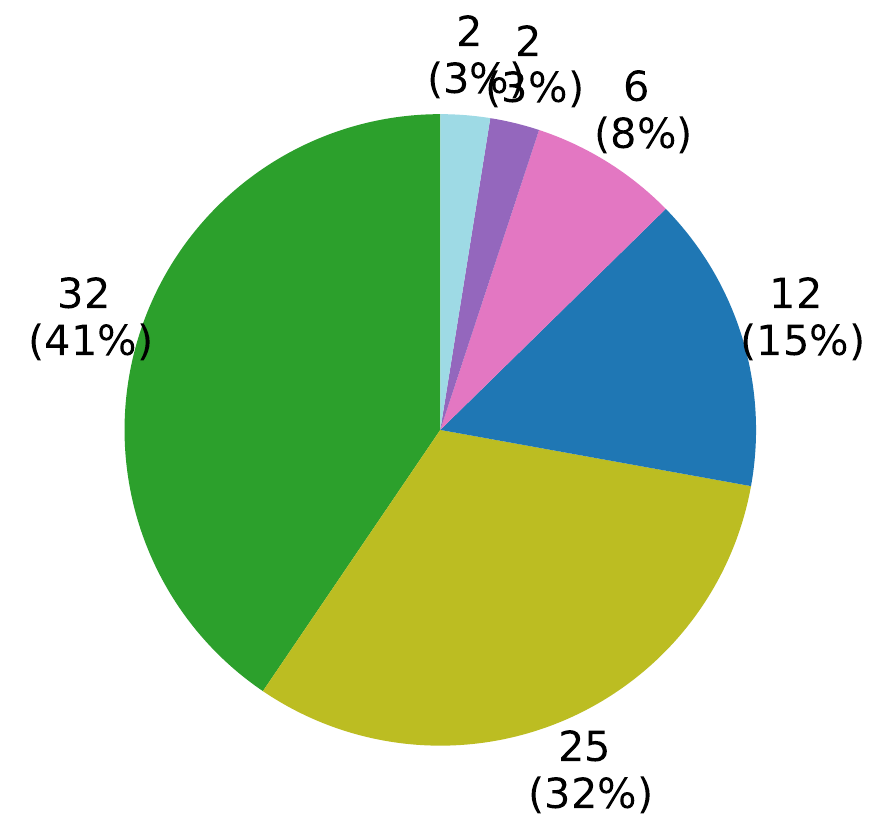}
\caption{\swelite{}}
\label{fig:pieEntries_lite}
\end{subfigure}
\begin{subfigure}[h]{0.33\textwidth}
  \centering
\includegraphics[width=0.90\textwidth]{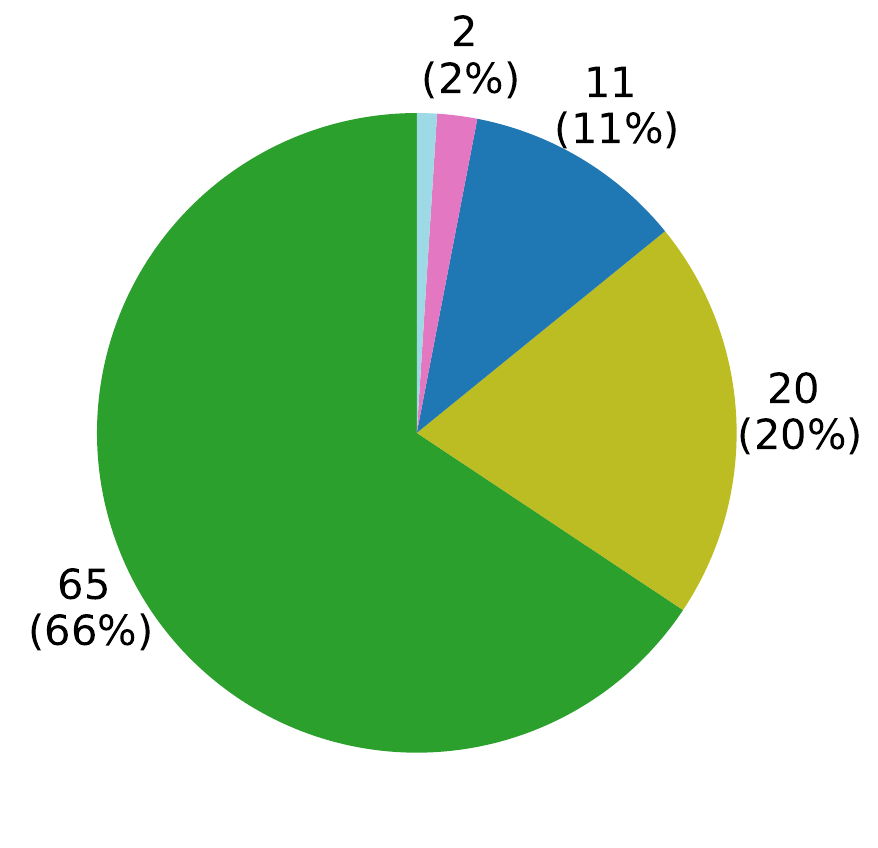}
\caption{\sweverif{}}
\label{fig:pieEntries_verif}
\end{subfigure}

\begin{subfigure}[h]{\textwidth}
  \centering
\includegraphics[width=\textwidth]{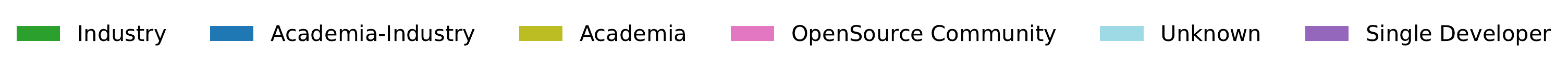}
\end{subfigure}

\caption{Total Entries by category of submitter: accumulated (left),  (center) and \sweverif{} (right).}
\label{fig:pie:entriesbysubmitterType}
\end{figure}


\subsubsection{Analysis of Submitter Organization}
\label{res:rq1:submitter}

\begin{figure}[t]
  \centering
\includegraphics[width=1.0\textwidth]
{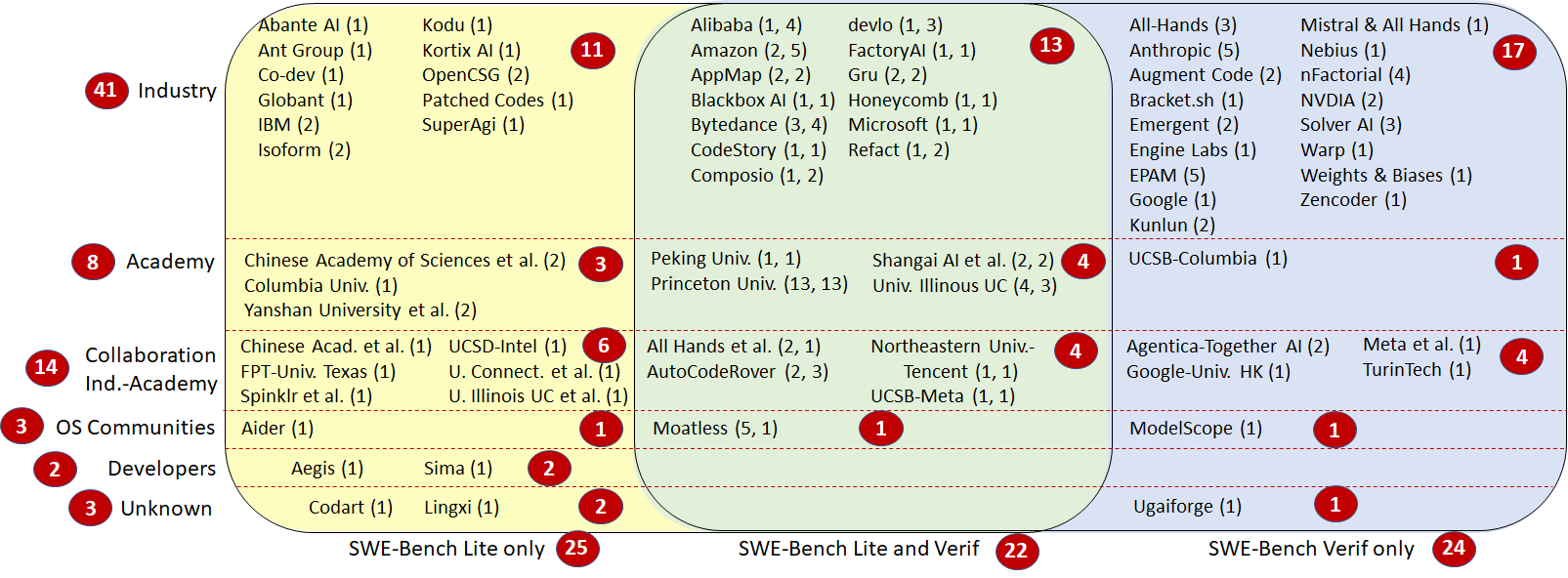}
\caption{Organizations per leaderboard. In parenthesis the number of submissions: for those organization with submissions to both leaderboard we append two numbers $(l,v)$: $l$ is the submissions to \swelite{} and $v$ to \sweverif{}. 
}
\label{fig:vennsubmissions}
\end{figure}

\paragraph{Descriptive statistics.}
\label{sec:results:submitter}
Figure~\ref{fig:vennsubmissions} presents the organizations that submitted results to the leaderboards (referred to as \emph{submitters}), grouped by submitter type (e.g., Industry, Academia).
The figure displays the name of each submitter along with the number of submissions (i.e., entries) made to each leaderboard.
Some submitter names have been abbreviated, particularly in the case of collaborations involving multiple organizations.
The full names are provided in the Appendix~\cite{Apprendix}.
There are 71 distinct submitters in total, with similar number of submitters to \swelite{} and to \sweverif{} (47 submitters vs. 46). 
Therefore, the ratio of submissions per submitter to \sweverif{} is greater than that of \swelite{} (2.15 vs. 1.68).

\begin{figure}[t]
\begin{subfigure}[h]{0.33\textwidth}
  \centering
\includegraphics[width=0.95\textwidth]{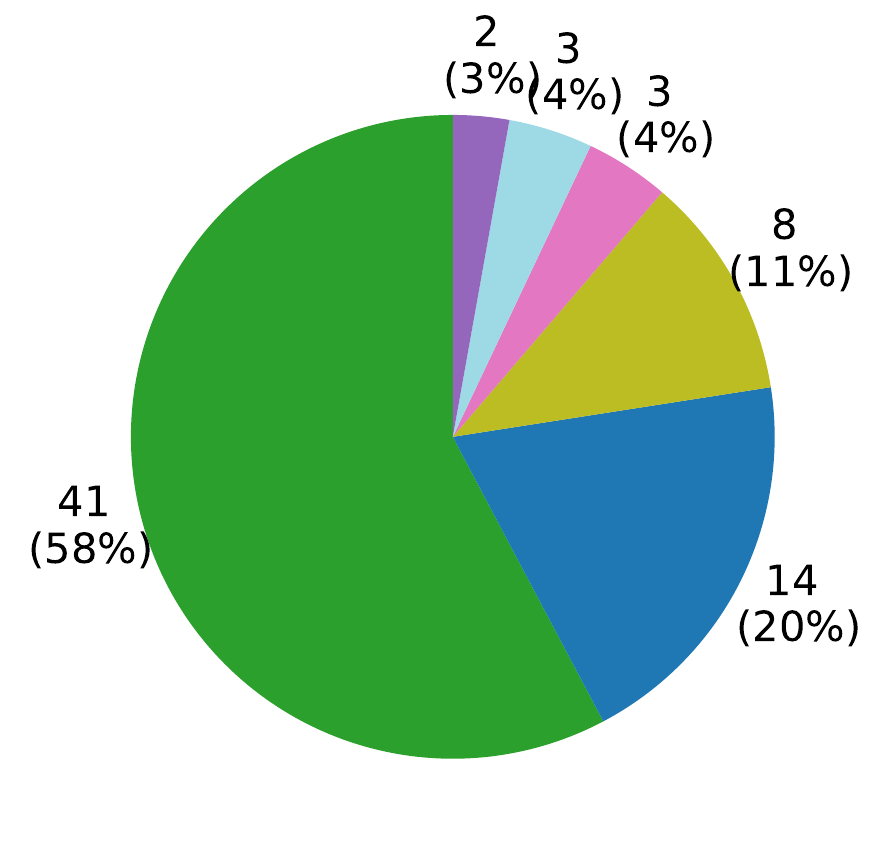}
\caption{Accumulated}
\label{fig:pieSubmittersAccum}
\end{subfigure}
\begin{subfigure}[h]{0.33\textwidth}
  \centering
\includegraphics[width=\textwidth]{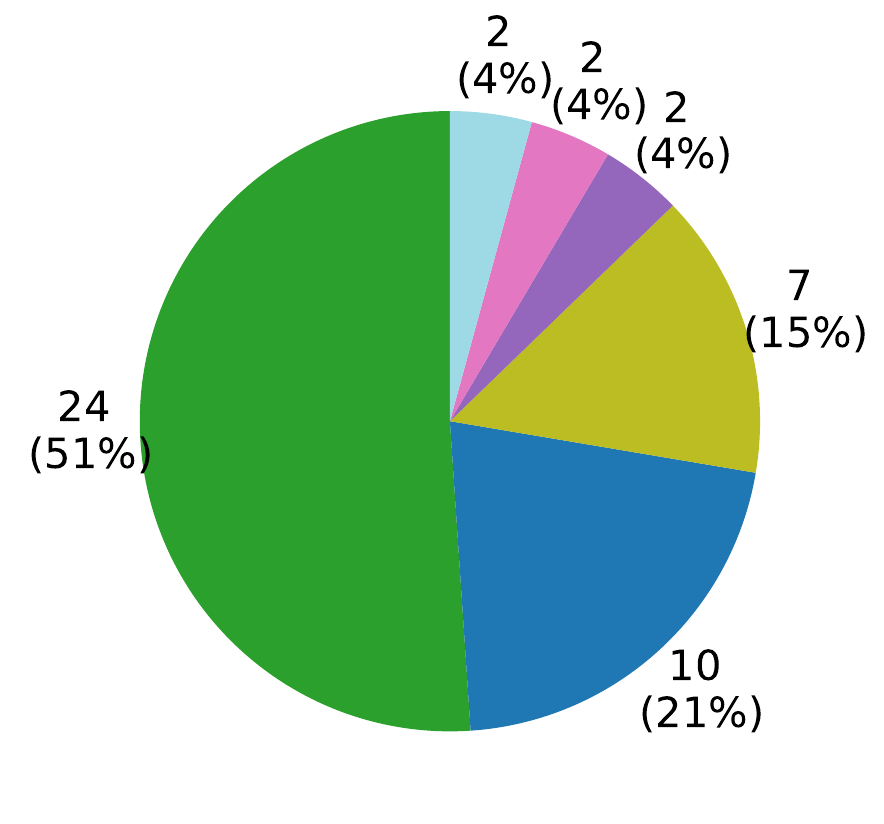}
\caption{\swelite{}}
\label{fig:pieSubmittersLite}
\end{subfigure}
\begin{subfigure}[h]{0.33\textwidth}
  \centering
\includegraphics[width=0.90\textwidth]{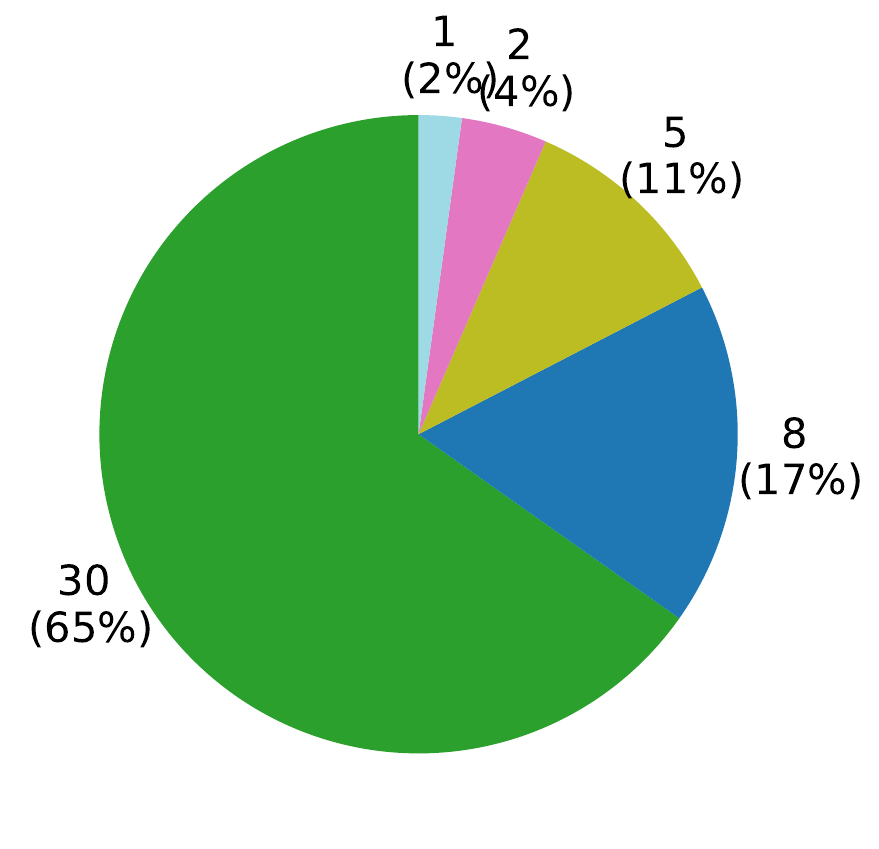}
\caption{\sweverif{}}
\label{fig:pieSubmittersVerif}
\end{subfigure}

\begin{subfigure}[h]{\textwidth}
  \centering
\includegraphics[width=\textwidth]{fig/legend_pie.pdf}
\end{subfigure}

\caption{Total Distinct Submitters  by category of submitter: accumulated (left),  (center) and \sweverif{} (right).}
\label{fig:pie:uniqueSubmittersbysubmitterType}
\end{figure}

Table~\ref{tab:summarySubmitter} summarizes, for each submitter type, the number of entries (\#E), the number of distinct submitters (\#S), and the median and maximum \precision{} scores.
The categories Industry and Academia are further broken down into sub-types, as described in Section \ref{sec:method:inspection:submitterCategory}.
As also illustrated in Figure~\ref{fig:pie:uniqueSubmittersbysubmitterType}, the majority of submitters are companies, accounting for 58\% of the total. This proportion rises to 78\% when including industry–academia collaborations.
We now examine the characteristics of each leaderboard individually.

The difference in the number of entries between \swelite{} (79) and \sweverif{} (99)
comes mainly from a larger number of industry submissions, as Figure \ref{fig:pie:entriesbysubmitterType} shows, increasing from 41\% (32 entries) to 66\% (65 entries). 
Especially notable is the increase in submissions from large publicly traded companies, which account for 21.2\% (21 entries) of all entries in \sweverif{}, compared to  8.8\% (7 entries) in \swelite{}.
The absolute number of academic submissions remains on a similar level (25 vs 20), and open source community solutions virtually disappear in \sweverif{}, the same as single developer solutions. 
Solutions from small companies prevail in both leaderboards, and together with solutions from academia, they represent half of submissions not only in total numbers, but also individually in each leaderboard.

\textbf{The case of \swelite{}}. The dominance of companies between the submitters is not as high as in \sweverif{} but still more than half (51\%), and from them, small companies prevail (up to 15 submitters, representing 32.6\% over the total). 
We also found two entries submitted by an Academia spin-off, AutoCodeRover, which has been acquired by an established company, Sonar, in Feb. 2025.\footnote{\url{https://www.sonarsource.com/company/press-releases/sonar-acquires-autocoderover-to-supercharge-developers-with-ai-agents/} (Last access Feb. 25, 2025)}

Large companies—including publicly traded and medium-sized firms—have also submitted their solutions, typically independently and without external collaboration (e.g., IBM, Amazon).
In contrast, some companies submitted exclusively through collaborative efforts, primarily with academic partners: Google \cite{su2025learnbyinteract}, Intel \cite{yu2025orcaLocallmAgent} and Meta \cite{wei2025SWE-RL}. Other collaborations have involved small companies founded in recent years such as All-Hands \cite{wang2024openhandsopenplatformai}.

This shows that the interest in \swebench{} is not limited to academia—where it originally emerged—or academic spin-offs, but also extends to established players in the software industry.
This growing diversity of contributors suggests that \swebench{} is becoming a widely accepted benchmark for evaluating AI-based software engineering tools across sectors.

Seven submitters to \swelite{} are from academia, one of them being Princeton University, the founder of \swebench{}. Another two universities from the USA (University of Illinois Urbana-Champaign and Columbia University) and four Chinese academic submitters (Peking University and  collaboration among Chinese universities and research centers) complete the academic landscape. 
We also found entries submitted by two open source communities (Moatless and Aider).
Last, there are two submissions done by single developers: (i)
\texttt{SIMA}, an experiment aimed at testing the multi-agent architectures proposed by Li et al. \cite{li2024MoreAgentsneed}; (ii) Aegis, a \emph{``side project that I'm using as a learning opportunity to get hands-on with applications of agents for software engineering''}.
It is important to note that these two entries from single developers were built on top of the open-source Moatless framework,
highlighting the importance of providing reusable artifacts and evaluation infrastructure, as Moatless does.

\textbf{The case of \sweverif{}}. As commented, the prevalence of industry in \sweverif{} increases significantly: 
the 65\% of the submitters and the 66\% of the entries are from industry.
Per type of companies, fluctuations are minor; we may remark, though, the involvement of NVIDIA, Google, Anthropic and EPAM. Concerning academia, a new collaboration between two U.S. universities (UCSB and Columbia) contributed to \sweverif{}.

\textbf{Submission to one or more leaderboards.}
The Venn diagram in Figure~\ref{fig:vennsubmissions} shows which organizations have submitted to \swelite{}, \sweverif{}, or both. Twenty-two contributors (31\%) submitted solutions to both leaderboards, most of them (14) with duplicated submissions (conversely, eight contributors submitted different solutions to both leaderboards). For example, Amazon submitted \texttt{Amazon Q Developer Agent (v20240430-dev)} to both leaderboards. 
On the other hand, 25 contributors have submitted only to \swelite{}, and 24 exclusively to \sweverif{}. 
For instance, IBM has submitted only to \swelite{} (with two entries), while EPAM has submitted only to \sweverif{} (with five entries).

\paragraph{Analysis of \% Resolver rate}

Table~\ref{tab:summarySubmitter} shows the median and maximum \precision{} achieved by each type of submitter across the two leaderboards. 

In \swelite{}, the highest precision (60.3\%) is achieved by \approach{ExpeRepair}~\cite{mu2025experepairdualmemoryenhancedllmbased}, a submission resulting from a collaboration between academic institutions. A close second (60\%) is the submission by a small company, Refact.ai.
Notably, recent entries from individual academic groups also surpass the 50\% mark, including \approach{SWE-Agent} (56.67\%) and \approach{SemAgent}-\cite{pabba2025semagentsemanticsawareprogram} (51\%).
Surprisingly, no medium or large company has outperformed these results so far.
Across all types of submitters, the median precision remains relatively consistent, ranging from 27\% to 29.8\%, with the exception of industry submissions, which exhibit a notably higher median (35.8\%).
We assess whether there are statistically significant differences in precision across submitter types using the Kruskal–Wallis test. 
The analysis reveals a statistically significant difference between Single Academia and Small Companies (H = 19.89, p = 0.0469).
This difference may be influenced by the early submissions to the leaderboard, which were primarily made by academic groups and tended to have lower precision.

In \sweverif{}, the results show a different landscape. 
Various types of companies —small, medium, and large— submitted results with the highest \precision{}, all exceeding 73\%.
However, other submitter categories also achieved competitive results. Notably, academia (e.g., Princeton University with \approach{SWE-Agent}) and the open-source community (e.g., \approach{Moatless}) reached precision scores above 66\%.
Excluding the open-source submissions —only two entries, considered outliers—the highest median precision is observed among medium and large companies, driven by submissions from Anthropic, which evaluated its state-of-the-art LLMs featuring hybrid reasoning capabilities.
In contrast, the median precision from academia is notably lower than that of the other categories.
The Kruskal–Wallis test returned a similar result as \swelite{} (H = 35.0745, p = 0.0001). 
Dunn’s post-hoc test further reveals that the precision scores from small, medium and large companies differ significantly from those of academia ($p\-values$ 0.00588, 0.00313 and 0.01786, resp.). 
This finding corroborates the trend discussed above.

\begin{figure*}[t!]
    \centering
    \begin{subfigure}[t]{0.5\textwidth}
        \centering
        \includegraphics[width=\textwidth]{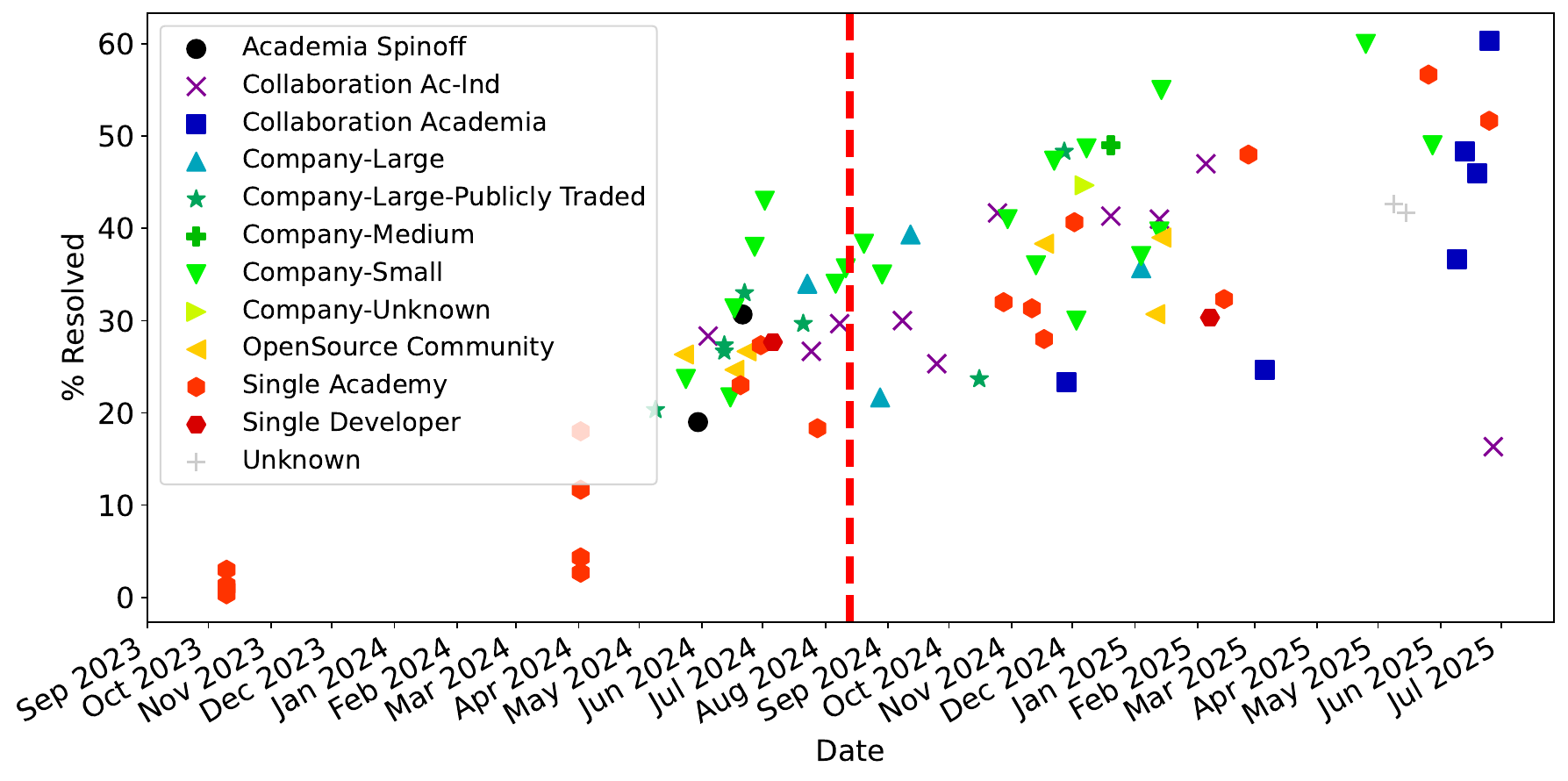}
        \caption{\swelite{}}
        \label{fig:evolLiteByOrigin}
    \end{subfigure}%
     ~ 
      \begin{subfigure}[t]{0.5\textwidth}
        \centering
        \includegraphics[width=\textwidth]{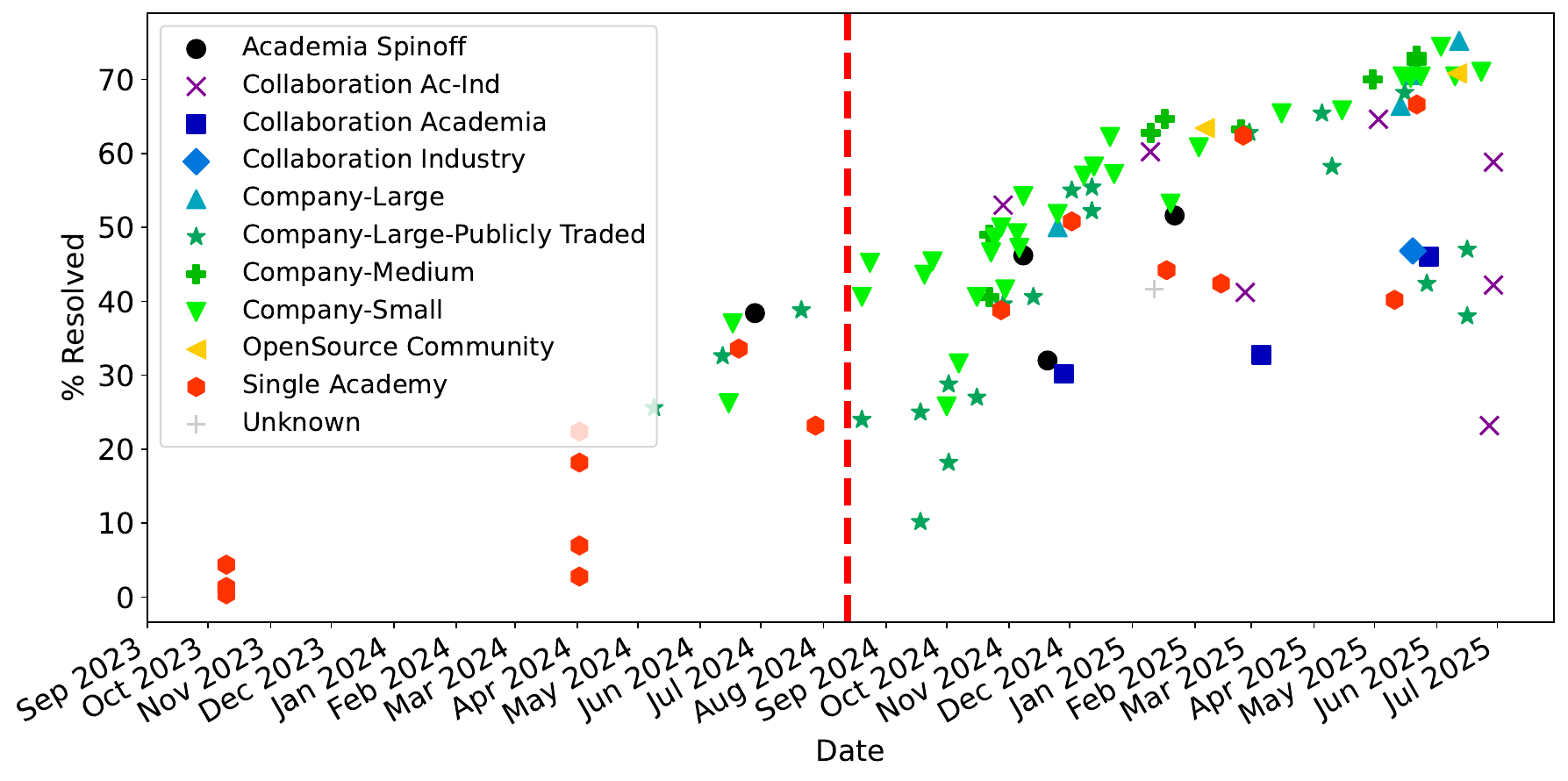}
        \caption{\sweverif{}}
        \label{fig:evolVerifByOrigin}
    \end{subfigure}%
    \caption{Evolution of \texttt{\% Resolved} by type of submitter.}
    \label{fig:evolPrecisionByOrigin}
\end{figure*}

\paragraph{Temporal Evolution of \precision{}}

Fig~\ref{fig:evolPrecisionByOrigin} shows the leaderboards' entries arranged according to the date and precision displayed in the leaderboards themselves.
Firstly, we observe that the initial entries from both leaderboards come from academia, in particular from the authors of the benchmark.
The benchmark was initially used to evaluate the ability of large language models (LLMs) to repair software issues~\cite{jimenez2024SWEBenchLLMs}, and was later applied to assess LLM-based agents~\cite{yang2024sweagentagent}. 
These early approaches achieved a maximum precision of around 20\%. 
Then, starting in May–June 2024, both leaderboards began receiving submissions from other organizations beyond academia.

In \swelite{}, as shown in  Figure~\ref{fig:evolLiteByOrigin} we observe that small companies have been constantly pushing the results. 
Notably, \approach{Isoform} achieved a state-of-the-art precision of 55\% on Jan. 2025, a months later Refact.ai achieved a 60\%.
Nevertheless, 
more recently —nearly a year after the leaderboard's launch— academic submissions have achieved competitive results, led by a collaboration between institutions~\cite{mu2025experepairdualmemoryenhancedllmbased}, and \approach{SWE-Agent}, from a single academic team~\cite{yang2024sweagentagent}.

In \sweverif{}, as shown in Fig.~\ref{fig:evolVerifByOrigin}, small companies have also consistently driven improvements in precision. For instance, \texttt{Refact.ai} achieved top precision (74.40\%) on June 2025.
Nevertheless, in 2025, other types of companies reached state-of-the-art performance as well —for example, the medium-sized AI leader \texttt{Anthropic} achieved 73.20\%, and the large company \texttt{Bytedance} currently holds the top entry with 75.2\%.
Submissions from academia have achieved competitive, though not top-ranking, results -driven by \texttt{Agentless-1.5} \cite{xia2024agentlessdemystifyingllmbasedsoftware} (50.8\%) and \texttt{SWE-Agent} \cite{yang2024sweagentagent} (66\%), both developed by single academic teams, as well as by a collaboration between the University of California (UCSB) and Meta, which produced \texttt{PatchPilot}~\cite{li2025patchpilot} with near state-of-the-art precision of 64.6\% on May 2025.

\begin{tcolorbox}
\underline{\bf{Answer to RQ 1 (Submitters):}}
Entries in the \swebench{} leaderboard originate from a wide variety of submitters, including (combinations or particularization of) academy, industry and individual developers (who may belong to open source communities). 
Small companies are dominant, but single academy has almost an equal share of entries in \swelite{}, although the number of distinct submitters is significantly higher in single academy (with an average of 4 entries per submitter) than in the rest of categories (less than 2 entries per submitter in average).  
This diversity underscores the benchmark’s accessibility and minimal barriers to participation, demonstrating that it is readily approachable for experimentation by a broad spectrum of contributors.
Since the introduction of \sweverif{}, companies (especially small ones) have been driving progress toward state-of-the-art performance.
\end{tcolorbox}


\subsubsection{Product and Accessibility}
\label{sec:results:product}

\newcolumntype{b}{X}
\newcolumntype{s}{>{\hsize=.05\hsize}X}
\newcolumntype{m}{>{\hsize=.5\hsize}X}

\setlength{\dashlinedash}{0.2pt}  
\setlength{\dashlinegap}{1pt}     
\setlength{\arrayrulewidth}{0.3pt} 

\begin{table}[t!]
\centering
\scriptsize
\begin{tabularx}{\textwidth}{|s|mm|sss|sss|ss|X|}
\hline

&\multicolumn{2}{c}{Product}&\multicolumn{3}{|c|}{Lite}&\multicolumn{3}{|c|}{Verified} &\multicolumn{2}{c|}{Total}& Approaches submitted to\\
\cline{2-11}
\rotatebox{0}{Av.} &Purpose   & Form   & \#A  & \#E & \% & \#A & \#E  & \% & \#A  & \#E &  Lite and Validated leaderboards \\
\hline

\multirow[t]{12}{*}{PAP} & Coding Assistant & IDE plugin & 8 & 12 & \ccb{33.5} & 7 & 15 & \ccb{44.2} & 10 & 27 & Alibaba Lingma Agent, Amazon Q Developer Agent, AppMap Navie, Augment Agent, Blackbox AI Agent, Bytedance MarsCode Agent, CodeShellAgent, CodeShellTester, Kodu-v1, OpenCSG Starship \\
\cdashline{2-12}
 & \multirow[t]{3}{*}{Development Assistant} & Cloud platform & 1 & 1 & \ccb{37} & 0 & 0 & \ccb{0} & 1 & 1 & Patched.Codes Patchwork \\
 &  & Command-line tool & 0 & 0 & \ccb{0} & 1 & 1 & \ccb{64.6} & 1 & 1 & W\&B Programmer O1 \\
 &  & Github plugin - Bot & 2 & 2 & \ccb{40.5} & 1 & 1 & \ccb{62.2} & 3 & 3 & AbanteAI MentatBot, CodeStory Aide, CodeStory Midwit Agent \\
\cdashline{2-12}
 & \multirow[t]{5}{*}{Development Platform} & Cloud platform & 5 & 6 & \ccb{36.2} & 7 & 17 & \ccb{51.8} & 8 & 23 & Composio SWE-Kit, Emergent E1, Engine Labs, Factory Code Droid, Kortix AI, OpenHands, Solver, devlo \\
 &  & IDE & 0 & 0 & \ccb{0} & 1 & 1 & \ccb{71} & 1 & 1 & Warp \\
 &  & Multiple (Command-line, Plug-in, Cloud) & 1 & 1 & \ccb{60} & 1 & 2 & \ccb{72.4} & 1 & 3 & Refact.ai Agent \\
 &  & Multiple (Plug-in,  Cloud) & 0 & 0 & \ccb{0} & 1 & 1 & \ccb{70} & 1 & 1 & Zencoder \\
 &  & Multiple (Plug-in, IDE, Terminal) & 0 & 0 & \ccb{0} & 1 & 2 & \ccb{72.9} & 1 & 2 & TRAE \\
\cdashline{2-12}
\cdashline{2-12}
 & Other-Inference Service & Cloud platform & 0 & 0 & \ccb{0} & 1 & 1 & \ccb{40.6} & 1 & 1 & Nebius AI \\
\cdashline{2-12}
 & Problem Solving & Foundation Model & 0 & 0 & \ccb{0} & 2 & 6 & \ccb{56.1} & 2 & 6 & Amazon Nova Premier, Tools + Claude \\
\cdashline{2-12}
 &\multicolumn{2}{c|}{Total}  
 & 17 & 22 & \ccb{36.5} & 23 & 47 & \ccb{53} & 30 & 69 & \\
\cline{1-12}  \cdashline{2-12}

\multirow[t]{5}{*}{NCS} & Agent Framework & Command-line tool & 1 & 5 & \ccb{30.7} & 1 & 1 & \ccb{70.8} & 1 & 6 & Moatless Tools \\
\cdashline{2-12}
 & Coding Assistant & Command-line tool & 1 & 1 & \ccb{26.3} & 0 & 0 & \ccb{0} & 1 & 1 & Aider \\
\cdashline{2-12}
 & Development Assistant & Command-line tool & 2 & 2 & \ccb{38.3} & 0 & 0 & \ccb{0} & 2 & 2 & Lingxi, SuperCoder2 \\
\cdashline{2-12}
 & Development Framework & Library & 0 & 0 & \ccb{0} & 1 & 1 & \ccb{63.4} & 1 & 1 & AgentScope \\
\cdashline{2-12}
 & Issue Resolution & Command-line tool & 14 & 29 & \ccb{27.3} & 9 & 23 & \ccb{32.8} & 17 & 52 & Aegis, Agentless, Agentless + RepoGraph, Agentless Lite, Co-PatcheR, CodeR, DARS Agent, DeepSWE, ExpeRepair, HyperAgent, KGCompass, MCTS-Refine, OrcaLoca + Agentless-1.5, RAG, SWE-Fixer, SWE-RL, SWE-agent \\
\cdashline{2-12}
 \cdashline{2-12}  
 &\multicolumn{2}{c|}{Total} 
& 18 & 37 & \ccb{28.3} & 11 & 25 & \ccb{33.6} & 22 & 62 & \\
 
\cline{1-12}  \cdashline{2-12}

\multirow[t]{3}{*}{B2B} & \multirow[t]{2}{*}{Development Platform} & Command-line tool & 0 & 0 & \ccb{0} & 1 & 5 & \ccb{39.6} & 1 & 5 & EPAM AI/Run Developer Agent \\
 &  & Multiple (Command-line, Plug-in, Cloud) & 1 & 1 & \ccb{48.3} & 0 & 0 & \ccb{0} & 1 & 1 & Globant Code Fixer Agent \\
 \cdashline{2-12}
 & Platform Code Optimization/Improvement & Unknown & 1 & 2 & \ccb{24.8} & 2 & 3 & \ccb{46.2} & 2 & 5 & AutoCodeRover, AutoCodeRover-v2 \\
  \cdashline{2-12}  
 &\multicolumn{2}{c|}{Total} & 2 & 3 & \ccb{30.7} & 3 & 8 & \ccb{42.9} & 4 & 11 & \\
\cline{1-12}  \cdashline{2-12}

\multirow[t]{4}{*}{UR} & Development Platform & Cloud platform & 1 & 1 & \ccb{30} & 1 & 1 & \ccb{52.2} & 2 & 2 & AIGCode Infant-Coder, Google Jules \\
 \cdashline{2-12}
 & Platform Code Optimization/Improvement & Cloud platform & 0 & 0 & \ccb{0} & 1 & 1 & \ccb{32} & 1 & 1 & Artemis Agent \\
 \cdashline{2-12}
 & Platform Code Representation & None & 0 & 0 & \ccb{0} & 1 & 1 & \ccb{53.2} & 1 & 1 & Bracket.sh \\
 \cdashline{2-12}
 & Platform Integration & None & 1 & 2 & \ccb{45} & 0 & 0 & \ccb{0} & 1 & 2 & Isoform \\
 \cdashline{2-12}  
 &\multicolumn{2}{c|}{Total} & 2 & 3 & \ccb{35} & 3 & 3 & \ccb{52.2} & 5 & 6 & \\
\cline{1-12}  
\cdashline{2-12}

& None & None & 13 & 14 & \ccb{35.7} & 10 & 16 & \ccb{46.1} & 19 & 30 & Aime-coder v1, Bytedance AutoSE, CORTEXA, Codart AI, CodeFuse-AAIS, Codev, Gru, Honeycomb, IBM AI Agent SWE-1, IBM Research Agent-101, Learn-by-interact, MASAI, PatchKitty\_PatchPilot, SIMA, SemAgent, Skywork-SWE, nFactorial, reproducedRG, ugaiforge \\

\cdashline{2-12}  
 &\multicolumn{2}{c|}{Total} &
13 & 14 & \ccb{35.7} & 10 & 16 & \ccb{46.1} & 19 & 30 & \\
 
\cline{1-12}

\end{tabularx}
\caption{Product Availability (column Av.), Product Purpose and Type. \#A is the number of distinct approaches, \#E is the number of leaderboard's entries, \% is the median \precision{}. The availabilities are PAP: Publicly Available Product, UR: Upon Request, B2B:  Business-to-Business, NCS: Non-Commercial Solution, UN: Unavailable.}
\label{tab:resultsProductTypeAndAvailability}
\end{table}


Table~\ref{tab:resultsProductTypeAndAvailability} presents the number of approaches (\#A) and entries (\#E) per product availability category, along with their median precision (\precision{}). We analyse below the contents of each category.



{\bf Publicly Available Product (PAP).} It includes publicly accessible commercial tools and accounts for 69 entries across both leaderboards, representing 30 distinct approaches.
This category includes products with different purposes, namely Coding Assistants, Development Platforms and Development Assistants.
The most common product type is \emph{Coding Assistant}, comprising 10 products available as IDE plug-ins and linked 27 entries. 
These include, for example, \texttt{Amazon Q Developer}, \texttt{Alibaba Lingma}, and \texttt{AppMap Navie}, which are evaluated on \swebench{} and available as extensions for popular development environments such as \texttt{IntelliJ JetBrains IDE} and \texttt{Visual Studio Code}.
The second major product type is \emph{Development Platform}, typically delivered as cloud-based services. 
These platforms generally allow developers to specify new features using natural language and integrate directly with code repositories (e.g., GitHub). There are 23 entries linked to these platforms across both leaderboards and include solutions such as \texttt{OpenHands}, \texttt{Solver}, and \texttt{Factory Code Droid}.
Other solutions such as Refact.ai, TRAE or Zencoder offer different product forms in addition to the cloud Services, such as IDE plugins.
Notably, the submissions related  \emph{Development Platform} products achieve the highest median \precision{} across both leaderboards.
A third type includes \emph{Development Assistant}, which helps developers in particular development task. Three of the entries, such as \texttt{CodeStory Aide} (later rebranded as \texttt{AgentFarm}), materialize this assistant as automated bots integrated into GitHub workflows, interacting with developers via pull requests, while the two others as a command-line tool and a cloud platform. 
The submissions related PAP products achieve the highest median \precision{} across both leaderboards (36.5\% and 53\%), closely followed by those from UR (Under Request) category.


The {\bf Non-Commercial Solutions (NCS)} category accounts for 62 entries across both leaderboards, corresponding to 22 distinct approaches.
The dominant product type is \emph{Issue Resolution}, comprising 17 approaches such as Agentless, SWE-agent, HyperAgent, and AgentScope.
This category also includes extensions of these tools, for instance, OrcaLoca~\cite{yu2025orcaLocallmAgent}, built on top of Agentless-1.5, improves the issue localization component.
We also identify a Code Assistant, \texttt{Aider}, which differs from the previously mentioned assistant in two key aspects: first, it is not implemented as a plug-in but as a command-line tool; second, it is an open-source project developed independently, without backing from a company aiming to commercialize the product or related technologies.


The {\bf Business-to-Business (B2B)} category includes 4 distinct approaches and 11 entries.
It covers two main product types: \emph{Development Platform} and \emph{Platform Code Optimization}.
These tools are typically integrated into enterprise workflows and tailored for internal use.


The {\bf Under Request (UR)} category is associated with 6 entries from 5 approaches across both leaderboards.
It comprises a variety of product types, including \emph{Platform Integration}, \emph{Code Representation}, \emph{Code Optimization}, and even a \emph{Development Plugin}.
These products are typically available only upon request and are represented by a small number of entries (at most two each).
Moreover, submissions in this category are generally limited to a single leaderboard, rather than appearing in both.

Finally, we grouped under {\bf UN} (unknown) those entries for which no associated product could be identified.
This group comprises 16 approaches and 28 entries, with a median \precision{} of 28\% on \swelite{} and 49\% on \sweverif{}.

Note that the products presented above are not necessarily the exact artifacts used to conduct the \swebench{} experiments.
In some cases, such as \texttt{Agentless} or \texttt{SWE-Agent}, the systems were specifically designed for that purpose, and the artifacts available are used for replicating the study. In contrast, some tools, such as general-purpose Code Assistants, were not originally designed for repair tasks, and their use in \swebench{} likely reflects the initiative of submitters to adapt existing capabilities to meet the benchmark requirements.

\paragraph{Temporal Evolution of \precision{}}

\begin{figure*}[t!]
    \centering
    \begin{subfigure}[t]{0.5\textwidth}
        \centering
        \includegraphics[width=\textwidth]{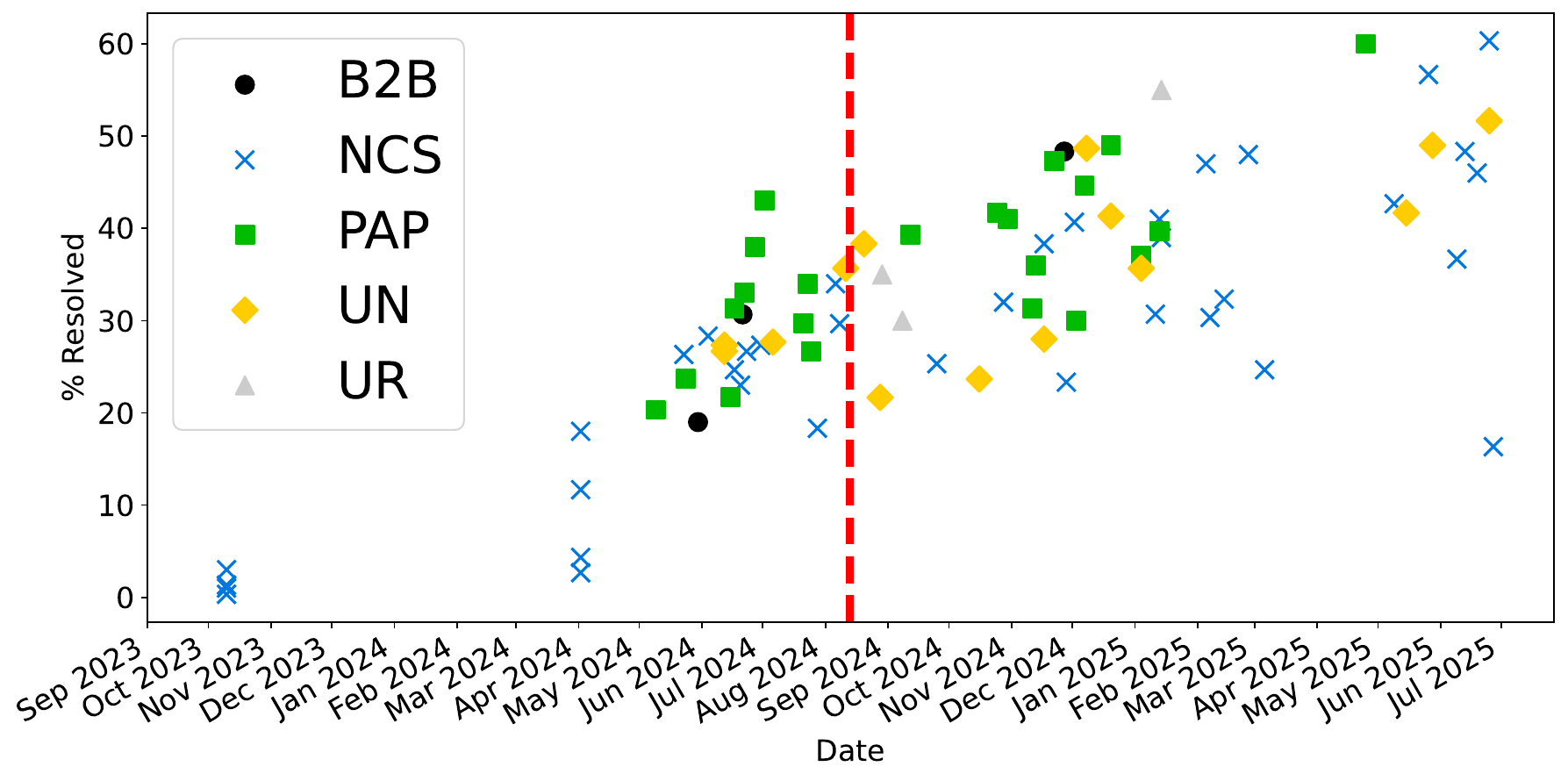}
        \caption{\swelite{}}
        \label{fig:evolProductAvailLite}
    \end{subfigure}%
     ~ 
      \begin{subfigure}[t]{0.5\textwidth}
        \centering
        \includegraphics[width=\textwidth]{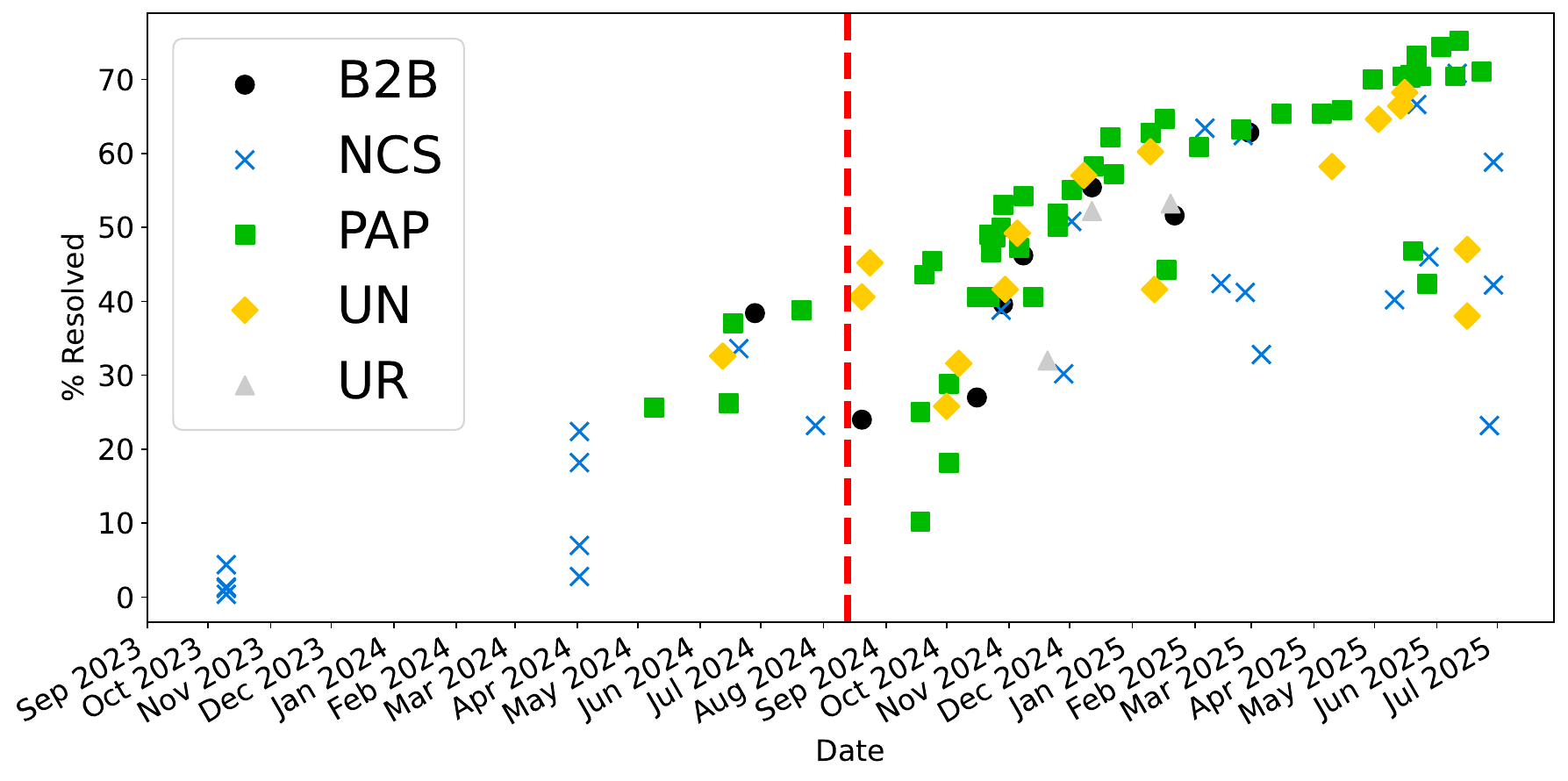}
        \caption{\sweverif{}}
        \label{fig:evolProductAvailVerified}
    \end{subfigure}
    \caption{Evolution of \texttt{\% Resolved} per product availability. The availabilities are PAP: Publicly Available Product, UR: Upon Request, B2B:  Business-to-Business, NCS: Non-Commercial Solution, UN: Unavailable.}
    \label{fig:evolProductAvailability}
\end{figure*}

Fig.~\ref{fig:evolProductAvailability} presents the leaderboard entries over time, showing their submission dates and the corresponding precision scores.
In \sweverif{}, entries associated with publicly available products (PAP) have consistently achieved state-of-the-art results up to the present date.
In contrast, the trend in \swelite{} differs: since early 2025, only a few PAP-related entries have been published, with the majority originating from non-commercial systems (NCS).
This may be attributed to the observation that industrial actors —who are more likely to release commercial products— tend to submit more to \sweverif{} than to \swelite{}, as previously discussed.

\begin{tcolorbox}
\underline{\bf{Answer to RQ 1 (Product):}}
Entries in the \swebench{} leaderboard are made available to the community in different ways: publicly accessible commercial tools, non-commercial solutions, available under request or as B2B solutions. Entries associated with Publicly Available Products and Under-Request solutions stand out for their precision, particularly in \sweverif{}.
A non-negligible number of entries (19\% over total) couldn't be classified.
\end{tcolorbox}

\subsubsection{Open-Source Availability of Solutions}
\label{sec:results:opensource}

\begin{table}[t!]
\centering
\begin{tabular}{|l|rr|ll|rr|ll|}
\hline
&\multicolumn{4}{c|}{\swelite{}} & \multicolumn{4}{c|}{\sweverif{}} \\ 
\cline{2-9}
&\multicolumn{2}{c|}{} & \multicolumn{2}{|c|}{\precision{}}
&\multicolumn{2}{c|}{} & \multicolumn{2}{c|}{\precision{}}\\
\cline{2-9}
&\#E & \#S  & Median & Max
&\#E & \#S & Median & Max\\
\hline
Open-source  &53 & 27 & \ccb{30.33} & \ccg{60.33} & 52 & 23 & \ccb{44.2} & \ccg{75.2} \\
Closed-source & 26 & 20 & \ccb{35.67} & \ccg{55} & 47 & 24 & \ccb{50} & \ccg{73.2} \\
\hline
\end{tabular}
\caption{Open-source vs Closed-source solutions on \swelite{} and \sweverif{}.  \#E shows the total number of entries, \#S is the number of distinct submitters.}
\label{tab:results:opensource}

\end{table}

\begin{figure*}[t!]
    \centering
    \begin{subfigure}[t]{0.5\textwidth}
        \centering
        \includegraphics[width=\textwidth]{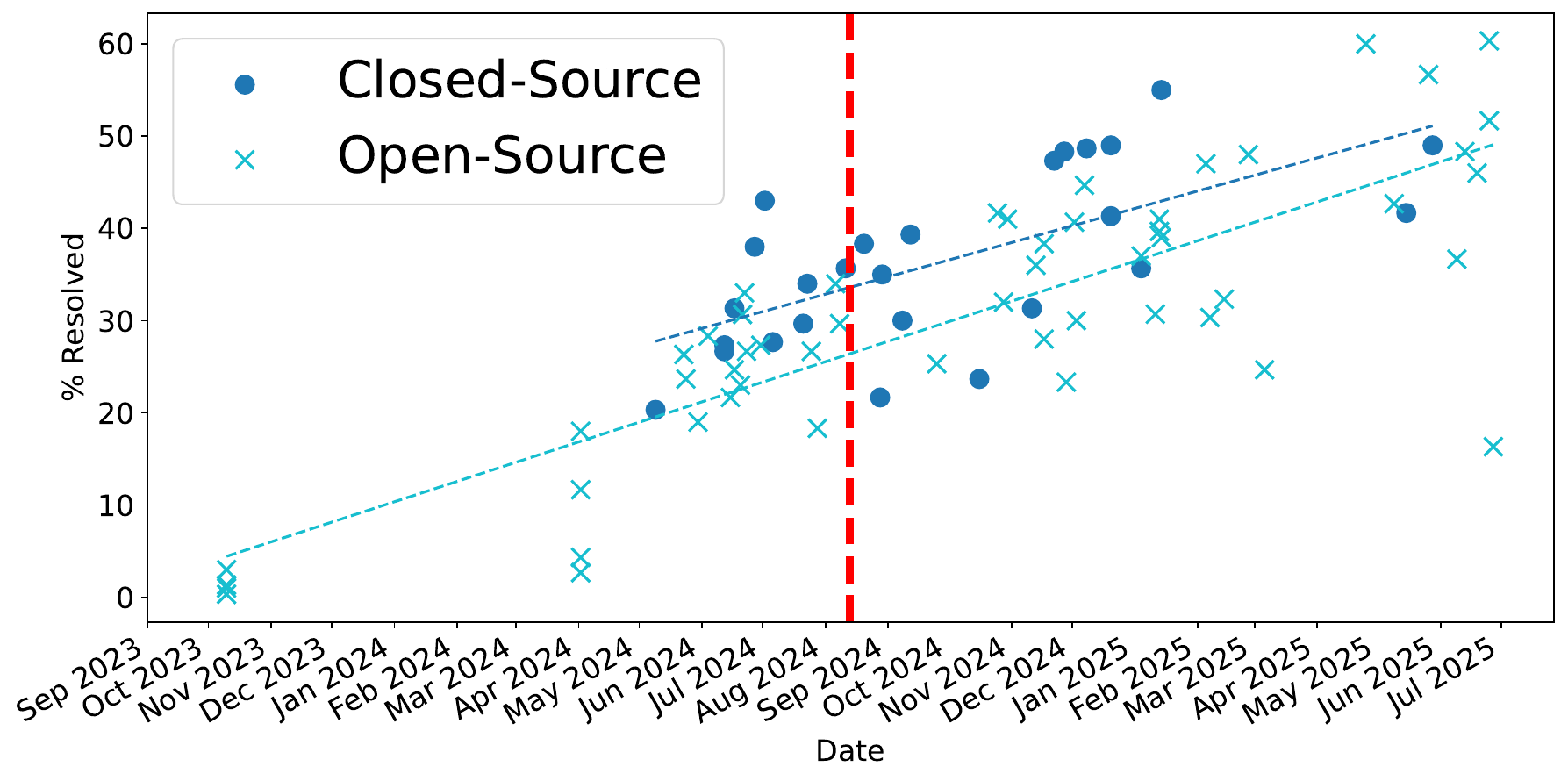}
        \caption{\swelite{}}
        \label{fig:evolisOpenLite}
    \end{subfigure}%
     ~ 
      \begin{subfigure}[t]{0.5\textwidth}
        \centering
        \includegraphics[width=\textwidth]{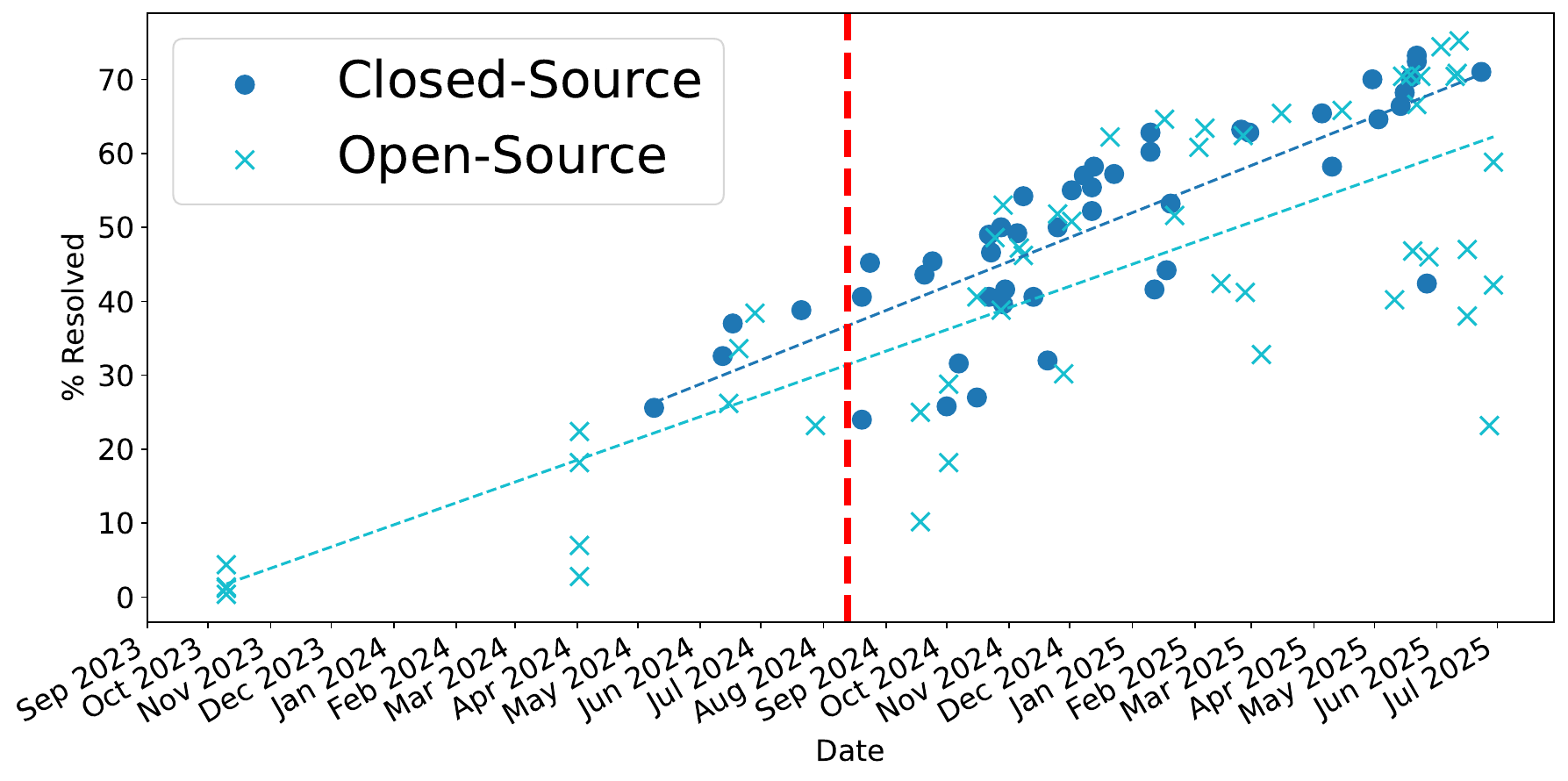}
        \caption{\sweverif{}}
        \label{fig:evolisOpenVerif}
    \end{subfigure}%
    \caption{Evolution of \texttt{\% Resolved} for Open and Closed source submissions.}
    \label{fig:evolOpenSource}
\end{figure*}

Table~\ref{tab:results:opensource} presents the number of entries associated with open-source and closed-source solutions.
In \swelite{}, the 67\% of entries (53 in total, representing 27 unique solutions) are open-source.
In contrast, \sweverif{} that percentage  is lower: 52\%.
As previously discussed, this leaderboard has attracted proportionally more submissions from industry actors, including startups and established companies, whose solutions are typically not released as open-source.
In both leaderboards, entries associated with closed-source systems exhibit a higher median \precision{}.
However, the top-ranked entries in both \swelite{} and \sweverif{} are open-source.

%
\paragraph{Temporal Evolution of \precision{}}

Fig.\ref{fig:evolOpenSource} shows the evolution of open- and closed- source entries  over time.
In \swelite{}, until early 2025, state-of-the-art results were achieved exclusively by closed-source solutions.
However, since then, just few close-solution were submitted and open-source solution recently achieve state-of-the-art results.
In \sweverif{} the trend is different: As illustrated in Fig.\ref{fig:evolisOpenVerif}, throughout 2025, open-source and closed-source submissions have alternated in holding the top rank.

\begin{tcolorbox}
\underline{\bf{Answer to RQ 1 (Open \& Closed source):}}
While closed-source solutions exhibit a higher median \precision{}, emerging open-source tools are showing competitive performance, with several of them achieving state-of-the-art results on both \swelite{} and \sweverif{} in 2025.
\end{tcolorbox}

\subsubsection{Large language Model Employed in Submissions}
\label{sec:results:llms}

\begin{table}[t!]
\scriptsize
\centering
\begin{tabular}{|l|rll|l|rll|}

\hline
\multicolumn{4}{|c|}{\swelite{}} & \multicolumn{4}{|c|}{\sweverif{}} \\ 
\hline
LLMs& \#E &  \multicolumn{2}{|c|}{\precision{}}   & LLMs& \#E & \multicolumn{2}{|c|}{\precision{}}   \\
\cline{3-4}
\cline{7-8}
&  &  Med & Max & &  & Med &  Max \\
\hline
Claude 4 Sonnet & 2 & \ccb{58.5} & \ccg{60.33} & Claude 3.7 Sonnet+Claude 4 Sonnet+Claude 4 Opus+Gemini 2.5 Pro & 1 & \ccb{75.2} & \ccg{75.2} \\
Claude 3.7 Sonnet+o4-mini & 1 & \ccb{60} & \ccg{60} & Claude 4 Sonnet+o4-mini & 1 & \ccb{74.4} & \ccg{74.4} \\
No Info & 9 & \ccb{35.67} & \ccg{55} & Claude 4 Opus & 1 & \ccb{73.2} & \ccg{73.2} \\
Claude 3.7 Sonnet+Gemini 2.5 Pro & 1 & \ccb{51.67} & \ccg{51.67} & Claude 4 Sonnet & 5 & \ccb{70.4} & \ccg{72.4} \\
Claude 3.5 Haiku+Claude 3.5 Sonnet+Gemini 2.5 Pro & 1 & \ccb{49} & \ccg{49} & Claude 3.7 Sonnet+Claude 4 Sonnet+GPT-4+Gemini 2.5 Pro & 1 & \ccb{71} & \ccg{71} \\
GPT-4 & 8 & \ccb{25.17} & \ccg{48.67} & Claude 3.7 Sonnet+o1+o4-mini+Gemini 2.5 Pro & 1 & \ccb{70.6} & \ccg{70.6} \\
Claude 3.5 Sonnet & 17 & \ccb{41} & \ccg{48.33} & Claude 3.7 Sonnet+o3+o4-mini & 1 & \ccb{70.4} & \ccg{70.4} \\
Claude 3.5 Sonnet+o3-mini+o4-mini & 1 & \ccb{48.33} & \ccg{48.33} & Claude 3.7 Sonnet+o3+Gemini 2.5 Pro & 1 & \ccb{70.2} & \ccg{70.2} \\
Claude 3.7 Sonnet & 1 & \ccb{48} & \ccg{48} & Claude 3.7 Sonnet+o4-mini & 1 & \ccb{70} & \ccg{70} \\
Claude 3.5 Sonnet+DeepSeek R1 & 1 & \ccb{47} & \ccg{47} & Claude 3.7 Sonnet+o1+o3+o3-mini & 1 & \ccb{68.2} & \ccg{68.2} \\
Claude 3.5 Sonnet+GPT-4o & 4 & \ccb{35} & \ccg{43} & Claude 3.7 Sonnet & 3 & \ccb{63.2} & \ccg{66.4} \\
Claude 3.5 Haiku+Claude 3.5 Sonnet & 1 & \ccb{41.67} & \ccg{41.67} & No Info & 18 & \ccb{45.3} & \ccg{65.8} \\
GPT-4o & 14 & \ccb{28.835} & \ccg{39.33} & Claude 3.7 Sonnet+o1 & 1 & \ccb{65.4} & \ccg{65.4} \\
Deepseek V3 (*) & 2 & \ccb{33.685} & \ccg{36.67} & o1 & 1 & \ccb{64.6} & \ccg{64.6} \\
Claude 3.5 Sonnet+GPT-4 & 1 & \ccb{33} & \ccg{33} & gpt-4o-mini & 2 & \ccb{58.9} & \ccg{64.6} \\
o3-mini & 2 & \ccb{31.33} & \ccg{32.33} & Claude 3.5 Sonnet+Qwen2.5 & 1 & \ccb{63.4} & \ccg{63.4} \\
Claude 3+GPT-4o & 2 & \ccb{25.83} & \ccg{26.33} & Claude 3.5 Sonnet & 15 & \ccb{51.6} & \ccg{62.8} \\
Qwen2.5 (*) & 3 & \ccb{23.33} & \ccg{24.67} & Claude 3.5 Sonnet+o3-mini & 1 & \ccb{60.8} & \ccg{60.8} \\
LLama 3+Mistral-Large+Qwen2.5+Granite (*)& 1 & \ccb{23.67} & \ccg{23.67} & Qwen3 (*)& 2 & \ccb{50.5} & \ccg{58.8} \\
GPT-4+GPT-4o & 1 & \ccb{21.67} & \ccg{21.67} & Claude 3.5 Sonnet+Deepseek & 1 & \ccb{58.2} & \ccg{58.2} \\
Claude 3 & 2 & \ccb{8} & \ccg{11.67} & Claude 3.5 Sonnet+GPT-4o & 5 & \ccb{47.2} & \ccg{57.2} \\
Claude 2 & 1 & \ccb{3} & \ccg{3} & Gemini 2 Flash & 2 & \ccb{48.2} & \ccg{52.2} \\
LLama 3 (*) & 2 & \ccb{1.165} & \ccg{1.33} & Claude 3.5 Sonnet+gpt-4o-mini & 1 & \ccb{48.6} & \ccg{48.6} \\
GPT3/3.5 & 1 & \ccb{0.33} & \ccg{0.33} & Qwen2.5 (*)& 11 & \ccb{30.2} & \ccg{47} \\
- & - & - & - & DevStral Small (*)& 1 & \ccb{46.8} & \ccg{46.8} \\
- & - & - & - & o3-mini & 1 & \ccb{42.4} & \ccg{42.4} \\
- & - & - & - & Amazon Nova Premier & 1 & \ccb{42.4} & \ccg{42.4} \\
- & - & - & - & LLama 3 (*)& 1 & \ccb{41.2} & \ccg{41.2} \\
- & - & - & - & Claude 3.5 Haiku & 1 & \ccb{40.6} & \ccg{40.6} \\
- & - & - & - & LLama 3+Qwen2.5 (*) & 1 & \ccb{40.6} & \ccg{40.6} \\
- & - & - & - & GPT-4o & 7 & \ccb{27} & \ccg{38.8} \\
- & - & - & - & GPT-4 & 2 & \ccb{12.6} & \ccg{22.4} \\
- & - & - & - & Claude 3 & 2 & \ccb{12.6} & \ccg{18.2} \\
- & - & - & - & Claude 2 & 1 & \ccb{4.4} & \ccg{4.4} \\
- & - & - & - & CodeLlama (*)& 2 & \ccb{1.3} & \ccg{1.4} \\
- & - & - & - & GPT3/3.5 & 1 & \ccb{0.4} & \ccg{0.4} \\

\hline
\end{tabular}
\caption{Combinations of base LLMs used per solution. \#E is the number of entries that use the combination. \precision{} shows the median and maximum values. An asterisk (*) indicates that all models are open-source.}
\label{tab:results:llmCombination}
\end{table}

\begin{figure}[t]
        \centering
        \includegraphics[width=0.5\textwidth]{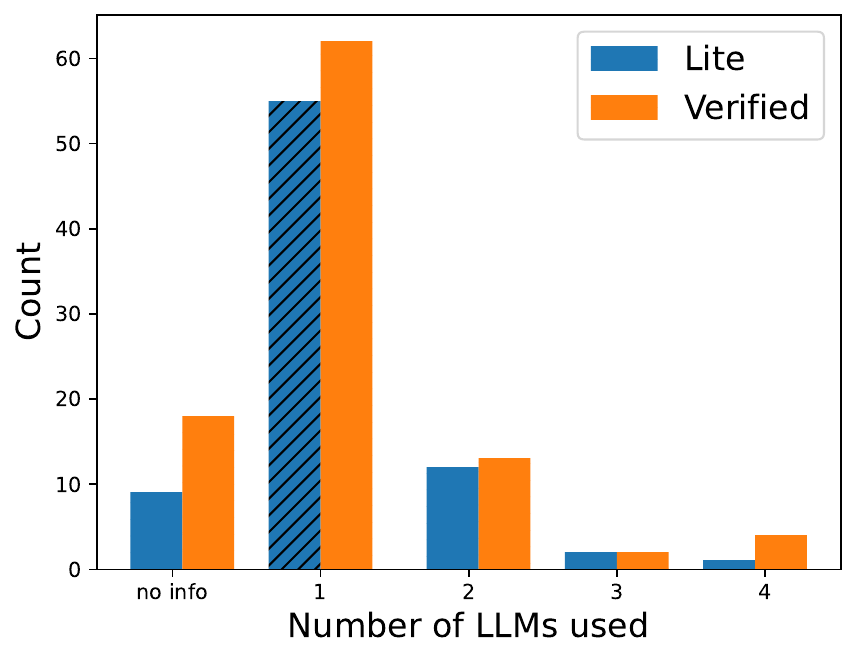}
        \caption{Distribution of LLMs used per leaderboard entry.}
        \label{fig:numberLLMsUsed}
    \end{figure}

Table~\ref{tab:results:llmCombination}  presents the base LLMs used in the submissions to both \swelite{} and \sweverif{}, and how frequent these LLMs are used together in the submissions.
The table is sorted by \precision{}.
As usual, some entries did not include enough information (see separate bar ``No Info'' in the figure. 

In \swelite{}, \llm{Claude 4 Sonnet} employed by the approach  \approach{ExpeRepair} \cite{mu2025experepairdualmemoryenhancedllmbased} achieved the highest performance with a preposition of 60.3\%.
This model has been used in two entries in total (column \#E): it was also used by \approach{SWE-agent}. 
\llm{Claude 3.5 Sonnet} is the most frequently used LLM, appearing in 16 entries as a standalone model, and in several others in combination with additional LLMs.
For instance, four entries combine both \llm{Claude 3.5 Sonnet} and \llm{GPT-4o}.

In \sweverif{}, a similar trend is observed: the state-of-the-art result (75.2\%), by \approach{TRAE} (June 2025) was achieved using a combination of multiples state-of-the-art models including \llm{Claude 4 Sonnet and Opus} \llm{Claude 3.7 Sonnet}.
Nevertheless, using \llm{Claude 4 Sonnet} or \llm{Claude 4 Opus} individually also led to competitive performance.
\llm{Claude 3.5 Sonnet} remains the most frequently used model, appearing in 17 entries as a standalone LLM and in five entries combined with others, such as \llm{GPT-4o}.
The second most frequently used LLM in isolation is \llm{Qwen2.5}, with 11 entries. Notably, it is an open-source model; however, the solutions employing it achieved lower performance, with precision scores up to 47\%.

\paragraph{Use of Multiple LLMs}
The majority of submissions report using a single LLM, as shown in Figure~\ref{fig:numberLLMsUsed}.
Excluding the most recent state-of-the-art models from the Claude 4 family, Table~\ref{tab:results:llmCombination} shows that top-performing results are often achieved through combinations of multiple LLMs.
Notably, most LLM combinations appear in only one submission per leaderboard, with a few exceptions: \texttt{Claude 3.5 Sonnet + GPT-4o} is used by four submissions in \swelite{} and five in \sweverif{}.

Solutions that use more than one LLM often assign different roles to each model within the repair process.
For example, the second-best solution in \swelite{}, \approach{Replit.ai Agent}, employs \llm{Claude 3.7 Sonnet} for orchestration and decision-making, and \llm{o4-mini} for deeper analysis and code generation. 
Similarly, \approach{InfantAgent}~\cite{lei2024InfantAgent} leverages closed-source models exclusively as the ``brain'' of its agent-based architecture, while smaller open-source models are employed for simpler, repetitive tasks.

Another example is \approach{Devlo}, which achieved a competitive 70.2\% on \sweverif{} by using three LLMs —\llm{Claude 3.7 Sonnet}, \llm{o3}, and \llm{Gemini 2.5 Pro}— during the patch generation phase to promote diversity in candidate patches.
In some cases, while one or more LLMs are responsible for generating patches, a different LLM is used for analyzing and selecting among them —a strategy referred to as ``LLM-as-a-judge'' or ``LLM-as-a-selector'', respectively.
For instance, a submission of \approach{TRAE} (which achieved 70.4\% in May 2025) uses \llm{o1} to select among patches generated by three different models: \llm{Claude 3.7 Sonnet}, \llm{Gemini 2.5 Pro} and \llm{o4-mini}.
Similarly,  \texttt{AgentScope}, with 63.4\% precision—  uses the open-source \llm{Qwen2.5} to select the best result among multiple trials made with \llm{Claude 3.5 Sonnet}.

\paragraph{Openness of LLMs}
\label{sec:results:openessmodels}

The majority of submissions rely on proprietary, closed-source LLMs accessed via API, most notably from the Claude 3 and 4 model families. 
However, a range of open-source models has also been employed, typically after fine-tuning for specific tasks such as patch evaluation.
Combinations of fully open-source models are indicated in Table \ref{tab:results:llmCombination} with an asterisk (*).
The most frequently used open-source model is \llm{Qwen2.5}, serving as the sole model in 3 entries from \swelite{} and 11 from \sweverif{}. 
Some entries rely exclusively on a set of open-source models, for example, one submission from IBM uses \emph{LLama}, \emph{Mistral-Large}, \llm{Qwen2.5} and \emph{Granite}.
In general, the precision of approach using only open-source LLMs is lower than those using closed LLMs, in both leaderboards.
Nevertheless, some solutions combine both closed- and open-source models. One such example is \texttt{AgentScope}, which achieved near state-of-the-art performance on \sweverif{} (63.4\%), at the time the result was published (February 2025), using \llm{Claude 3.5 Sonnet} in combination with \llm{Qwen2.5}.
Notably, no other submission using \llm{Claude 3.5 Sonnet} outperformed \texttt{AgentScope}, suggesting that properly fine-tuned open-source models can enhance the effectiveness of solutions built around closed-source LLMs.


\paragraph{Fine-Tuning Open-Source Models}
\label{sec:results:finetuning}

One of the key advantages of using open-source LLMs is the ability to fine-tune them for specific tasks.
Several submissions leveraged this flexibility to tailor models for different components of the repair pipeline.
For example, \approach{SWE-Fixer} \cite{xie2025SWEFixerTrainingopensourcellms} fine-tuned \llm{Qwen 2.5} twice: once for the bug retriever and once for the patch editor.
\approach{Nebius} fine-tuned two separate models: \llm{Qwen 2.5} for patch generation and LLaMA 3.1 for the patch critic.
DARS~\cite{aggarwal2025darsdynamicactionresampling} fine-tuned \llm{DeepSeek R1} to serve as a patch reviewer, while AgentScope, as mentioned before, fine-tuned \llm{Qwen-2.5} specifically for patch assessment.
SWE-agent-LM-32B applied fine-tuning to \llm{Qwen 2.5} using the SWE-smith dataset~\cite{yang2025swesmith}, which was generated from trajectories of SWE-agent~\cite{yang2024sweagentagent} running \llm{Claude 3.7 Sonnet}.
\llm{DevStral} is a fine-tuned version of \llm{Mistral-Small-3.1}, trained to address real-world GitHub issues.
Lingma SWE-GPT~\cite{ma2024lingmaswegpt} fine-tuned \llm{Qwen 2.5} for its specialized software engineering tasks.

Meta presented the \approach{SWE-RL} approach \cite{wei2025SWE-RL} for enabling LLMs to autonomously recover a developer’s reasoning processes and solutions.
The resulting reasoning model, \llm{Llama3-SWE-RL-70B}, fine-tuned \llm{Llama 3} using Reinforcement learning (RL). During the training, the policy LLM is tasked with solving a given issue through reasoning and producing the code changes, then converted into a consistent patch format for reward calculation.
\approach{SWE-RL} helps the LLM develop reasoning skills like self-reflection, exploring alternatives, and divide-and-conquer strategies.
\approach{Co-PatcheR} \cite{tang2025copatchercollaborativesoftwarepatching}  fine-tuned three distinct instances of \llm{Qwen2.5}, each tailored for a specific task —localization and generation, and test generation with and without assertions— using distillation data generated by \llm{Claude 3.7 Sonnet} with explicit reasoning chains.

Some of these fine-tuned models, such as \llm{DevStra}l, \llm{Lingma SWE-GPT}, \llm{SWE-agent-LM-32B}, there are publicly available on model repositories such as HuggingFace or ModelScope.
Others, such the case of the approaches \approach{DARS} and \approach{SWE-Fixer}, also uploaded the training data on this platform\footnote{https://huggingface.co/AGENTDARS, https://huggingface.co/datasets/SWE-bench/SWE-smith-trajectories}.

\paragraph{End-to-End Open Solutions}
\label{sec:results:end2endopeness}

Beyond the openness of the LLMs and the public availability of fine-tuned models, some submitters have also made their full solution code publicly available (e.g., agent scaffolds).
The open-source approaches, with at least one submission exclusively using open-source models, are 13:
\begin{inparaenum}[\it 1)]
\item \approach{AgentScope},
\item \approach{Alibaba Lingma Agent},
\item \approach{Co-PatcheR},
\item \approach{DARS Agent},
\item \approach{DeepSWE},
\item \approach{KGCompass},
\item \approach{MCTS-Refine},
\item \approach{Moatless Tools},
\item \approach{OpenHands},
\item \approach{RAG},
\item \approach{SWE-Fixer},
\item \approach{SWE-agent}, and 
\item \approach{Skywork-SWE}.
\end{inparaenum}
This level of openness means that, given appropriate infrastructure to deploy a LLM, others can fully replicate and build upon these systems, promoting transparency, reproducibility, and further research.

\paragraph{Temporal Evolution of \precision{} per LLM employed}

\begin{figure*}[t!]
    \centering
    \begin{subfigure}[t]{0.5\textwidth}
        \centering
        \includegraphics[width=\textwidth]{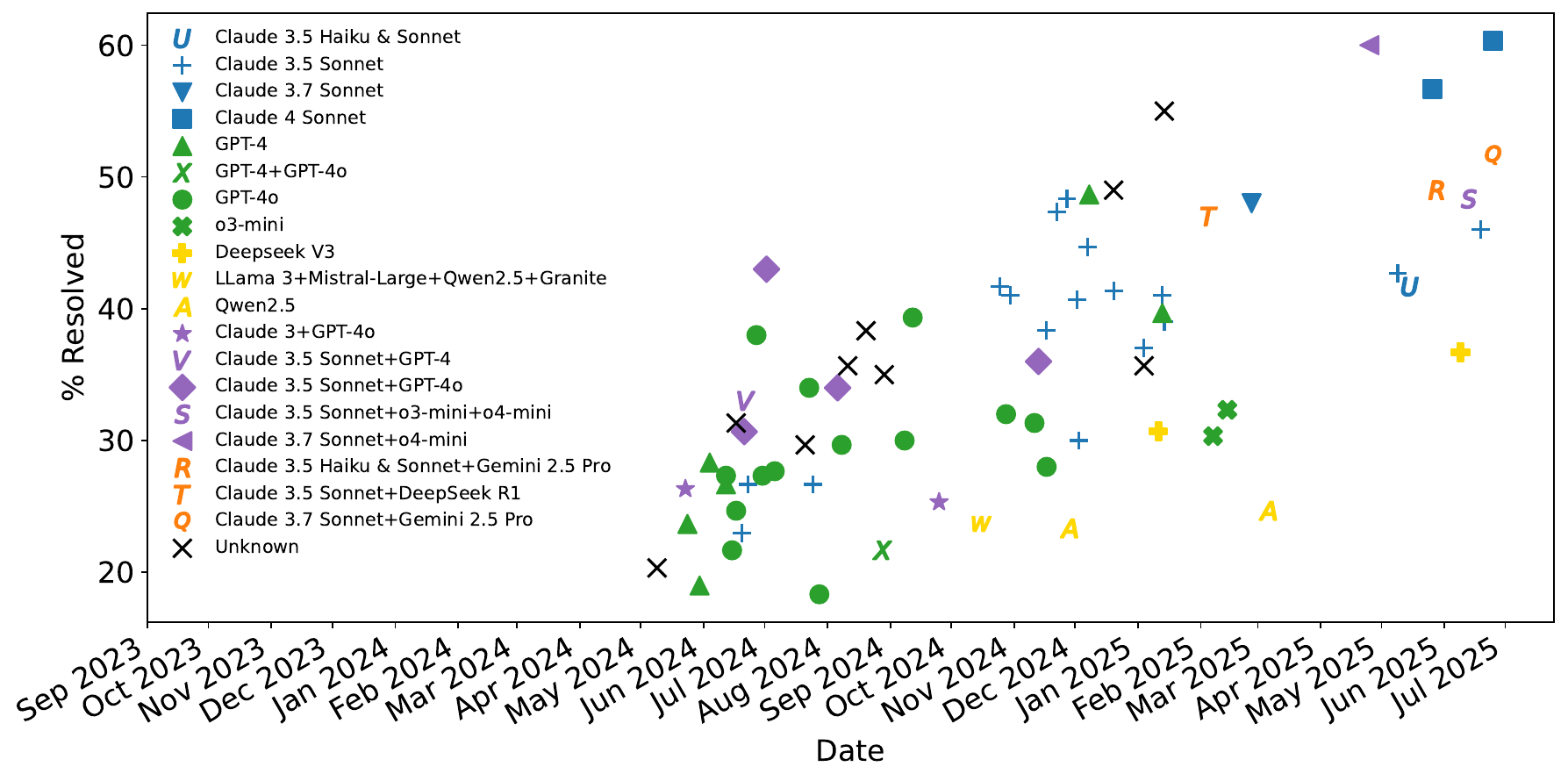}
        \caption{\swelite{}}
        \label{fig:evolLLMComboLite}
    \end{subfigure}%
     ~ 
      \begin{subfigure}[t]{0.5\textwidth}
        \centering
        \includegraphics[width=\textwidth]{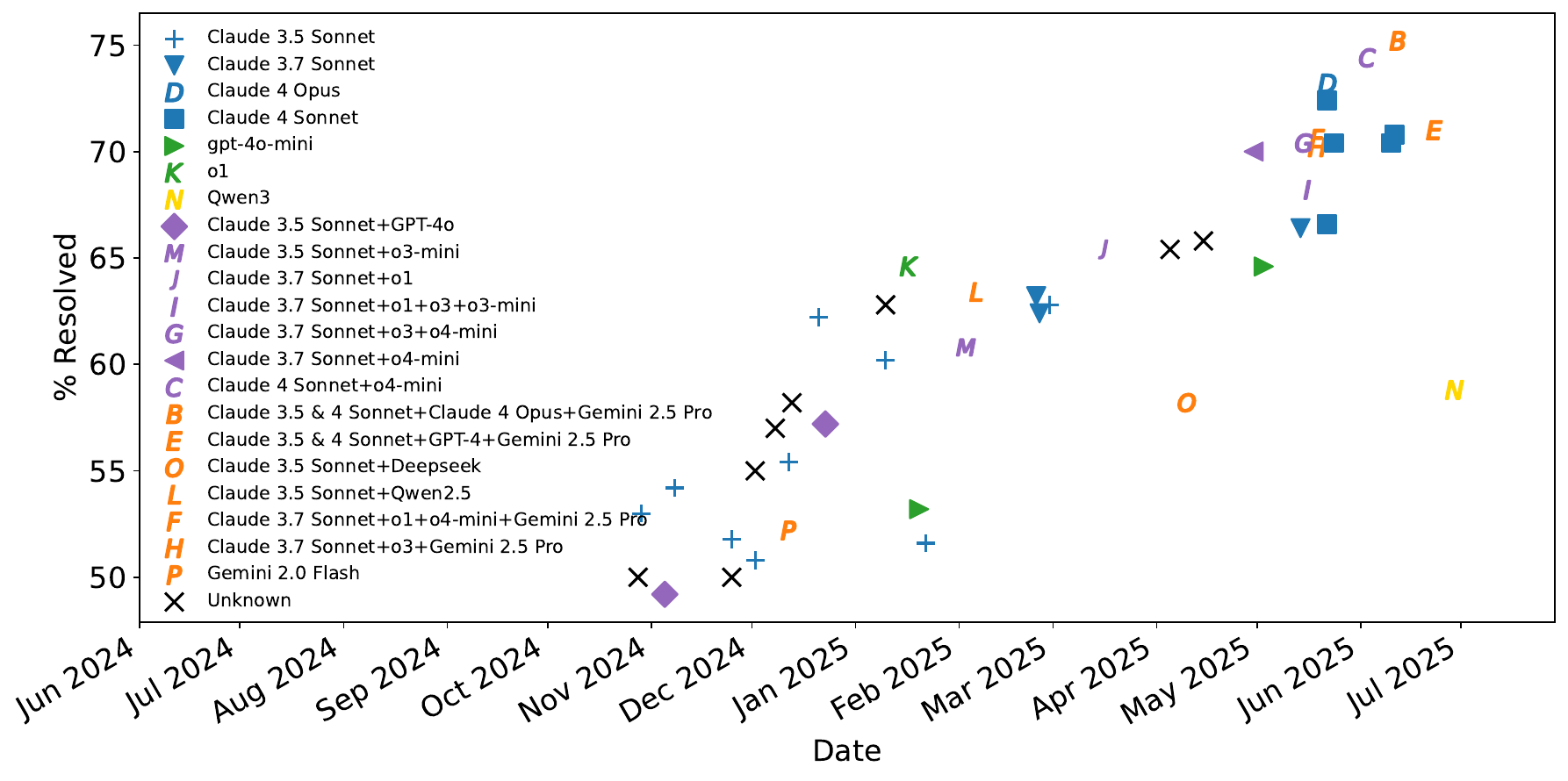}
        \caption{\sweverif{}}
        \label{fig:evolLLMComboVerified}
    \end{subfigure}%
    \caption{Evolution of \texttt{\% Resolved} by LLMs used.}
    \label{fig:evolLLMsCombo}
\end{figure*}

Figure~\ref{fig:evolLLMsCombo} shows the precision over time for each set of LLMs used together, based on their publication dates on the leaderboards.
For clarity, Figure~\ref{fig:evolLLMComboLite} displays the top 70 out of 79 entries from \swelite{}, while Figure~\ref{fig:evolLLMComboVerified} shows the top 45 out of 99 entries from \sweverif{}.
The colors represent:
\begin{inparaenum}
\item \ClaudeText{Blue}: submissions just using Anthropic Claude models,
  \item \OpenAIText{Green}: submissions just using OpenAI models,
  \item \ClaudeOpenAIText{Purple}: submissions using Anthropic and OpenAI, but not others,
  \item \OpenSourceText{Yellow}: submissions only using open-source models,
  \item \MultiVendorText{Orange}: other cases such as submissions using other vendors, or combinations between closed- and open-source models, and finally,
  \item \UnknownText{Black}: submissions without specifying the model used.
\end{inparaenum}

In \swelite{}, between June 2024 and March 2025, there is a considerable number of submissions  (represented with \OpenAIText{$\bullet$} and \OpenAIText{$\blacktriangle$}) using exclusively OpenAI's models. One of them, \approach{AbanteAI MentatBot}, achieved state-of-the-art results (38\%) with \llm{GPT-4o} (\OpenAIText{$\bullet$}) in June 2024. 
During the same period, there were also submissions using only \llm{Claude 3.5 Sonnet} (\ClaudeText{$+$}), though with lower performance.
Nevertheless, some competitive submissions  combined Anthropic Claude and OpenAI models (e.g. \ClaudeOpenAIText{$\blacklozenge$}, \ClaudeOpenAIText{$M$}), with one of them, \approach{CodeStory Aide} with	43\%, reaching top results using \llm{Claude 3.5 Sonnet} and \llm{GPT-4o} (\ClaudeOpenAIText{$\blacklozenge$}) in July 2024.
This submission held the top position for several months, until November 2024, when it was surpassed by solutions using a single LLM: \approach{Devlo} (47.3\%) and \approach{Globant Code Fix} (48.3\%), both relying solely on \llm{Claude 3.5 Sonnet} (\ClaudeText{$+$}). It was later outperformed by \approach{GRU}, which achieved 48.67\% using \llm{GPT-4}.
Between December 2024 and April 2025, submissions using only open-source models (e.g. \OpenSourceText{$N$}) began to appear, although their performance remained lower (<40\%).
This period also included submissions that achieved state-of-the-art results using unrevealed models (\UnknownText{X}).
Finally, from April to July 2025, new state-of-the-art results were achieved, first by combining \llm{Claude 3.7} with \llm{o4-mini} (\ClaudeOpenAIText{$\blacktriangleleft$}), and then by using \llm{Claude 4 Sonnet} alone (\ClaudeText{$\blacksquare$}).

In \sweverif{}, until January 2025, state-of-the-art submissions either used \llm{Claude 3.5 Sonnet} \ClaudeOpenAIText($+$)  or did not disclose the underlying models (\UnknownText{X}).
In January, the submission from \approach{W\&B} (\OpenAIText{$B$}) reached the top precision using \llm{o1} (64.6\%), holding the lead until the first submission combining \llm{Claude 3.7 Sonnet} and \llm{o1} (\ClaudeOpenAIText{$A$}).
Since May 2025, all leading submissions on the leaderboard have included models from the \llm{Claude 4} series. The first of these was a  submission  by Anthropic using only \llm{Claude 4 Opus}~(\ClaudeOpenAIText{$D$}), which achieved 73.20\%. 
Subsequent top submissions combined \llm{Claude 4} with other LLMs~(\MultiVendorText{$B$}), including the current leader by \approach{TRAE}, which reached 75.2\%

\begin{tcolorbox}
\underline{\bf{Answer to RQ 1 (LLMs):}}
Approaches leveraging proprietary models, first \llm{Claude 3.5 Sonnet} and more recently \llm{Claude 4} variants, have consistently achieved the highest precision on both leaderboards.
Solutions that combine closed and open-source LLMs—including those that incorporate fine-tuned models—also demonstrate competitive performance.
Approaches based solely on fine-tuned open-source LLMs tend to yield more modest precision but offer benefits in terms of transparency, reproducibility and adaptability.

\end{tcolorbox}

\subsection{RQ2: \rqcorecomponents}
\label{sec:results:approachDetails}


In this section, we classify the approaches according to three key dimensions that we presented in detail in \ref{sec:methodology:architecture}: 
\begin{inparaenum}[\it 1) ]
\item {\bf workflow authoring}, 
\item {\bf control flow autonomy}, and 
\item the {\bf total number of agents} that comprise the approach.
\end{inparaenum}
Rather than examining each dimension separately, we explore how these dimensions are combined within each solution's design.
This leads to the identification of seven architectural groups (G1-G7)\footnote{Some combinations are not possible, such as an agent performing a fixed execution.}, along with an additional group (G8) for submissions lacking sufficient information for classification.
These groups are: 
\begin{itemize}
  \item G1: Human-Workflow with Fixed execution - No Agent (Section \ref{sec:result:mainArchitecture:worflowFixedNoAgent}).
  \item G2: Human-Workflow with Fixed execution - with Single Agent  (Section \ref{sec:result:mainArchitecture:worflowFixedAgent}).
  \item G3: Human-Workflow with Fixed execution - with Multiple Agents  (Section \ref{sec:result:mainArchitecture:worflowFixedMultiplesAgents}).
  \item G4: Human-Workflow with Scaffolded Execution - Single Agent  (Section \ref{sec:result:mainArchitecture:worflowScaffExecSingleAgent}).
  \item G5: Human-Workflow with Scaffolded Execution - Multiple Agents  (Section \ref{sec:result:mainArchitecture:worflowScaffExecMultipleAgents}).
  \item G6: Workflow Emergent and Emergent Autonomy - Single Agent  (Section \ref{sec:result:mainArchitecture:worflowEmergentAutonomySingleAgent}),
  \item G7: Workflow Emergent and Emergent Autonomy - Multiple Agents  (Section \ref{sec:result:mainArchitecture:worflowEmergentAutonomyMultipleAgent}).
  \item G8: Unclassified Submissions  (Section \ref{sec:result:mainArchitecture:others}).
\end{itemize}

For each approach, we enumerate the different approaches, reporting their performance as listed in the leaderboards (column \precision{}). 
In cases where multiple submissions from the same approach exist (such as those using different LLMs), we just report the highest score.
Finally, in Section \ref{sec:results:architecture:precision}, we analyze the performance of each of these groups.

\subsubsection{\bf Human-Workflow with Fixed execution - No Agent (G1)}
\label{sec:result:mainArchitecture:worflowFixedNoAgent}

The entries listed in this section describe approaches that follow a predefined workflow, typically composed of two or three stages—such as localization, patch generation, and validation. These approaches execute the sequence of steps without dynamic decision-making or autonomous behavior. Moreover, none of the individual stages exhibit agentic behavior.

\approach{RAG + <model name>} refers to the initial set of entries on the \swebench{} leaderboard, introduced by its original creators as a baseline to evaluate the performance of LLMs in repairing issues from the benchmark they developed~\cite{jimenez2024SWEBenchLLMs}.
The approach follows a two-stage workflow: Localization, which uses BM25 retrieval~\cite{robertson2009probabilistic} to identify potentially buggy files, and Repair, where the selected LLM generates candidate patches.
In addition, they fine-tuned \llm{CodeLlama}~\cite{roziere2024codellama} on 19,000 issues to create the \llm{SWE-LLAMA} model, which is used during the repair stage.
Within this approach, the best results were obtained using \llm{Claude 3 Opus} \resultsBoth{4.33}{7}.

\approach{Agentless}~\cite{xia2024agentlessdemystifyingllmbasedsoftware} is a repair workflow based on LLMs, composed of three main stages: Localization, Repair and Validation~\resultsLite{27.33}.
This workflow is similar to traditional APR, such as those implemented in the Astor framework~\cite{martinez2016Astor} or the state-of-the-art of traditional APR TBar \cite{Liu2019Tbar}, with the novelty of using LLMs to perform each task. 
In contrast to LLM-based Agents, in \approach{Agentless}, LLMs are not used to decide how to proceed in the repair workflow.
Both leaderboards also include submissions of the posterior version: \approach{Agentless-1.5} \resultsBoth{40.67}{50.8}.

As a pioneering approach on \swebench{}, \approach{Agentless} has been extended, modified and improved, with the results of some of these variants also submitted to the leaderboards. 
For example:
\begin{itemize}
    \item 
\approach{Agentless - Lite} \resultsBoth{32.3}{42.4}, as the name suggests, is a simplification of the original approach:
\emph{``Use an embedding model to retrieve relevant files from the repository (Exclusively RAG-based localization). Query the LLM to generate a repair based on the top 5 retrieved files, retrying the generation until the model outputs a valid patch.''}.
\item \approach{Agentless + Repograph} \resultsLite{29.67} corresponds to the Agentless augmented with the plugin \approach{Repograph}. 
Repograph provides agents a structured way to analyze and interact with complex codebases, enabling detailed tracing of code dependencies, execution flow and structural relationships across the repository \cite{ouyang2024repographenhancingaisoftware}.
Repograph information is integrated across all stages of the Agentless workflow.

\item \approach{OrcaLoca + Agentless-1.5} \resultsLite{41} incorporates \approach{ORCALOCA}~\cite{yu2025orcaLocallmAgent} in the Localization stage from \approach{Agentless-1.5}, while retaining the Repair, Patch Validation and Patch Selection stages.
\approach{ORCALOCA} is an approach focused on file and function localization.
Instead of relying exclusively on LLMs to generate exploratory actions within the codebase, it introduces a dynamic action scheduling system that produces actions at varying levels of granularity and prioritizes them based on contextual relevance and urgency.

\item \approach{CodeFuse-AAIS} proposes two main improvements: 
\begin{inparaenum}[\it 1)]
    \item adaptive fault localization, and 
    \item expansion of the sampling space by combining multiples LLMs.
\end{inparaenum}. Nevertheless, we could not find more details about such improvements.

\item \approach{SWE-Fixer} \cite{xie2025SWEFixerTrainingopensourcellms} \resultsBoth{24.67}{32.8} is a further simplification of \approach{Agentless}, following a pipeline structure consisting of two primary stages:
\begin{inparaenum}[\it 1)]
    \item code file retrieval, and
    \item code editing.
\end{inparaenum}
In both stages, it uses fine-tuned open-source models (Qwen 2.5) trained using Chain-of-Thought (CoT) data \cite{wei2023chainofthought}.
\end{itemize}

There exist other workflow-based approaches beyond those that extend \approach{Agentless}.
For example, \approach{AppMap Navie}~\resultsBoth{36}{47.2} follows a structured workflow composed of three high-level phases: planning, generation, and validation. 
Each of these phases is further decomposed into sub-workflows that handle specific subtasks.
According to an official description, Navie is not agentic, it does not have the flexibility to choose or apply tools autonomously. Instead, it operates in a deterministic manner, executing a fixed sequence of stages.\footnote{\url{https://appmap.io/blog/2024/06/20/appmap-navie-swe-bench-leader/}}

\approach{Aider}~\resultsLite{26.33} employs an iterative process composed of the following main stages:
\begin{inparaenum}[\it 1)]
\item \textbf{File Filtering:} Aider builds a \emph{repository map}, a compact and expressive summary of the entire codebase, using static code analysis. 
This map helps the LLM understand the project structure and dependencies. The model then produces a natural language response indicating which files need to be edited and why.
\item \textbf{LLM-based Code Editing:} Code edits are performed using a set of carefully designed prompting strategies and configurable editing backends.
\item \textbf{Linting:} A static analysis phase is used to ensure that the modified code has no syntax or other fatal errors.
\item \textbf{Testing:} The final step consists of running all available tests to verify whether the generated patches are plausible.
\end{inparaenum}

The entry \approach{Patched.Codes Patchwork} \resultsLite{37} is based on the Patchwork framework. According to the official documentation, Patchwork is designed to support the construction of agentic workflows and provides several predefined ones. These include \emph{AutoFix}, which generates and applies fixes for code vulnerabilities, and \emph{ResolveIssue}, which identifies the relevant files to modify for resolving a given issue and automatically creates a pull request.
However, in a public announcement of their \swebench{} results, the CEO of the submitter company explicitly stated that the system \emph{``is not an `AI Engineer' or a `Coding Agent'—just a workflow engine that you can customize, self-host, and wrap around your LLM of choice''}.\footnote{Patched.Codes Patchwork founder on their \swebench{} submission: \url{https://x.com/rohan_sood15/status/1878698759632351277}}
This statement suggests a lack of autonomous agentic behavior, indicating a more deterministic, workflow-driven system rather than an agent-based one.

\subsubsection{\bf Human-Workflow with Fixed execution - with Single Agent (G2)}
\label{sec:result:mainArchitecture:worflowFixedAgent}

Similar to the previous group (G1), the entries listed in this section describe approaches that follow a predefined workflow and execute the sequence of steps without dynamic decision-making or autonomous control.
However, approaches in this group exhibit agentic behavior within one of the workflow stages.

Meta introduced the \approach{SWE-RL} \cite{wei2025SWE-RL} to train LLMs  —resulting in the model \approach{Llama3-SWE-RL}— with the goal of enabling them to autonomously recover and replicate a developer's reasoning process.
To evaluate the ability of the trained models to fix issues, the authors generated a pipeline-based scaffold named \approach{Agentless Mini} with include capabilities such as file-level fault localization, generating repair edits, reproduction, test generation, and regression test selection. 
\approach{Agentless Mini} enhances the original Agentless' pipeline with two
key improvements:
\begin{inparaenum}[\it 1)]
    \item test generation by considering other tests, and
    \item patch selection based on voting.
\end{inparaenum}
The evaluation of this scaffold using the trained model achieved a  41.2\% on \sweverif{}.


\approach{Nemotron-CORTEXA} \resultsVerified{68.2} introduces a two-stage workflow for automated software issue resolution.
First, the \emph{Localization Stage} operates in two substeps:
\begin{inparaenum}[\it a)]
    \item identifying the most relevant files related to the issue, and
    \item refining the granularity by selecting specific classes, functions, or methods.
\end{inparaenum}
A dedicated \emph{Localization Agent} is responsible for this second substep. To support the agent's reasoning and navigation, the system builds a graph representation of the repository that captures structural relationships among code entities.
Second,  there is a \emph{Repair Stage}, where multiple source code contexts and prompt formulations are used to generate a diverse set of candidate patches via LLMs. These patches are filtered based on both existing and LLM-generated tests. A final patch is selected through LLM-based reasoning over the filtered candidates.
Agentic behavior is primarily present in the \emph{entity-level localization}, where the Localization Agent autonomously filters and selects relevant code entities.

The company \approach{nFactorial} submitted four times their results on \sweverif{}. Only the first one, \resultsVerified{25.8}, follows a fixed pipeline consisting of three predefined phases (localization, fixing, and analysis) where the agent does not control the execution flow. 
In contrast, the subsequent entries delegate the control flow to the agent, allowing it to autonomously determine the execution path.

\subsubsection{\bf Human-Workflow with Fixed execution - with Multiple Agents (G3)}
\label{sec:result:mainArchitecture:worflowFixedMultiplesAgents}

Similar to the previous group (G2), the entries listed in this section describe approaches that follow a predefined workflow and execute the sequence of steps without dynamic decision-making or autonomous control.
However, the approaches in this group exhibit varied agentic behaviors across different workflow stages, often leveraging multiple agents with complementary roles.

An entry with multi-agents is \approach{AgentScope} \resultsVerified{63.4} which implements a workflow with three sequential stages (reproducing, fixing, and testing). 
Each stage is handled by a specialized agent which works in a reasoning-acting manner and is equipped with tools like Bash execution and file editor.

MASAI \cite{arora2024masaimodulararchitecturesSE}  \resultsBoth{27.33}{32.6} presents a high-level modular architecture composed of five distinct stages, including Edit Localizer, Fixer and Ranker.
Each stage is executed by a dedicated LLM-powered sub-agent, which is instantiated with defined objectives and tailored strategies designed to fulfill those objectives. In particular, each sub-agent is equipped with a set of actions and a corresponding \emph{strategy}, a problem-solving approach for using the LLM to address its assigned sub-task. This strategy can include methods such as vanilla completion, Chain-of-Thought\cite{wei2023chainofthought} or ReAct \cite{yao2023ReAct}.
Note that in MASAI, agentic behavior is confined to individual stages, with each sub-agent operating independently within its assigned role, rather than exercising control over the overall system workflow.

IBM submitted two solutions to \swebench{}, namely \approach{IBM Research Agent-101} \resultsLite{26.67} and \approach{IBM AI Agent SWE-1.0} \resultsLite{23.67}.
We were only found detailed information for the latter.
\approach{IBM AI Agent SWE-1.0} is a multi-agent system in which each agent is responsible for a specific task involved in the issue resolution process.
The system is structured around four main components that operate in a pipeline: \emph{Localization}, \emph{Editing}, \emph{Testing}, and \emph{Judging}.
Notably, the \emph{Localization} and \emph{Editing} stages follow the Observe-Think-Act loop~\cite{Balduccini2010AutonomiusAgentArch}, enabling agents to iteratively plan and execute actions based on observations from prior steps.

MarsCode Agent \cite{liu2024marscode} \resultsBoth{39.33}{50} introduces a multi-agent collaboration framework that dynamically allocates either static or adaptive solving pipelines depending on the nature of the bug, enabling flexibility across a range of software repair tasks.
The system defines six specialized agent roles (Searcher, Planner, Reproducer, Programmer, Tester, and Editor), each responsible for a specific software engineering subtask and equipped with tailored toolsets.
To enhance repository comprehension, MarsCode builds a multi-directional code knowledge graph from source code and documentation. This graph empowers agents with repository-level question answering (Q\&A) capabilities.

\subsubsection{\bf Human-Workflow with Scaffolded Execution - Single Agent (G4)}
\label{sec:result:mainArchitecture:worflowScaffExecSingleAgent}

Similar to the previous group (G3), these approaches follow a predefined, human-authored workflow. However, instead of rigid execution, they adopt a scaffolded model where a single agent—typically an LLM—has autonomy within and across workflow stages. This allows agentic behavior in localized decisions, while the overall control flow remains fixed.

\approach{PatchPilot}~\cite{li2025patchpilot} \resultsLite{41.33} introduces a human-designed, five-stage workflow consisting of: reproduction, localization, generation, validation, and refinement.
According to the authors, the refinement component distinguishes their approach from prior work. In this stage, partially correct patches, identified through validation feedback, are fed back into the generation component—along with the corresponding validation results— to produce a new batch of refined patches.

\approach{DARS} \cite{aggarwal2025darsdynamicactionresampling} \resultsLite{47} is an approach that extends \approach{SWE-Agent} with the primary goal of enhancing an agent’s ability to recover from and adapt to sub-optimal decisions. To achieve this, DARS identifies critical decision points during execution and selectively branches to explore alternative actions based on prior feedback. This reduces the number of full resets and accelerates convergence toward a valid patch. It also introduces a refined set of actions compared to previous work.
\approach{DARS} operates within a scaffolded workflow composed of two main phases:
\begin{inparaenum}[\it 1)]
\item \emph{Generation}, which encompasses reproduction, localization, and bug fixing; and
\item \emph{Evaluation}, where a reviewer LLM assigns scores to candidate patches and selects the best one for submission.
\end{inparaenum}

\approach{GRU} \resultsBoth{48.67}{57} is an agentic workflow which follows predefined tasks:
\begin{inparaenum}
    \item Analysis repository info;
    \item Explore code base;
    \item Analysis details (elaborates a change plan);
    \item Edit files and analysis; and
    \item Generate diff.
\end{inparaenum}
Note that the agent decides what to do in every task to achieve the result it desires.

\approach{Moatless Tools} \resultsLite{38.33} is a pipeline-based approach that includes predefined workflows (aka \emph{flows}).
It contains three main flows: 
\begin{inparaenum}[\it 1)]
   \item  \emph{tool coding}: Flow using tool calls to the LLM, recommended by the authors when using an underlying LLM with native support for tool calls (e.g., Claude 3.5 Sonnet).
   \item \emph{react coding}: Flow using ReACT \cite{yao2023ReAct} format instead of tool calls, recommenced when using an open source models without native support for tool calls (e.g., DeepSeek).
   \item \emph{tool thinking blocks}: Flow using tool calls with adjusted prompt for reasoning, recommended for using with reasoning models (e.g., DeepSeek R1, OpenAI o3).
\end{inparaenum}

Other approaches that report results on \swebench{} using agentic workflows, but with less publicly available detail, include  \approach{Aegis} \resultsLite{30.33}, \approach{Amazon Q Developer Agent} \resultsBoth{29.67}{65.4} and \approach{AbanteAI MentatBot} \resultsLite{38}.

\subsubsection{\bf Human-Workflow with Scaffolded Execution - Multiple Agents (G5)}
\label{sec:result:mainArchitecture:worflowScaffExecMultipleAgents}

The entries of this group adopt a human-defined scaffolded workflow executed by multiple agents, where each agent operates autonomously within its assigned stage. This design enables specialization across tasks —such as localization, patch generation, and validation— while preserving a globally structured execution flow.

\approach{AutoCodeRover} \cite{Zhang2024AutoCoderRover} \resultsBoth{19}{38.4} employs a predefined two-stage workflow consisting of context retrieval and patch generation.
In each stage, an LLM carries out a main task (e.g., find relevant code snippets and craft a patch), exhibiting agentic behavior.

\approach{SpecRover} \cite{ruan2024specrovercodeintentextraction} (also referred to as AutoCodeRover v2) \resultsBoth{30.67}{51.6} extends AutoCodeRover by introducing a broader spectrum of specification sources, including function-level summaries, program structure, and test cases. 
The agents iteratively infer specifications, where each agent can consume specifications produced by others and generate new ones in turn. This process continues until a patch is produced and deemed correct by one of the agents.
\approach{SpecRover} is organized as a multi-agent system with a human-defined workflow comprising five distinct roles, three of which are novel compared to AutoCodeRover: the Reproducer, Selection, and Reviewer agents. 
Two key contributions of \approach{SpecRover} lie in these new agents:
\begin{inparaenum}[\it 1)]
  \item  The \emph{Selection} agent chooses a patch based on the natural language issue description beyond test cases results. This a key contribution to avoid discarding correct patches due to incomplete test cases.
  \item The \emph{Reviewer} agent provide two types of feedback: a binary decision of whether the patch and the reproducer test are correct respectively, and an explanation for the decisions. This explanation is then used by the  \emph{Patching} agent and the \emph{Reproducer Test} agents to improve the patch.
\end{inparaenum}

\approach{CodeR} ~\cite{chen2024CoderRMultiAgent} \resultsLite{28.33} is  a multi-agent system designed for automated issue resolution.
It uses four predefined plans, each structured as a directed graph that encodes a standard resolution flow.
Each sub-task in a plan is executed by different a set of agents using the ReAct framework~\cite{yao2023ReAct}.
Given an issue to repair, the Manager agent selects the appropriate plan based on the issue description.
CodeR consists of five agents:
\begin{inparaenum}[\it a)]
\item \textbf{Manager}: selects the plan and interprets its execution summary;
\item \textbf{Reproducer}: generates test cases;
\item \textbf{Fault Localizer}: identifies the code regions that could cause the issue;
\item \textbf{Editor}: collects relevant context (leveraging AutoCodeRover’s search~\cite{Zhang2024AutoCoderRover}) and performs code edits;
\item \textbf{Verifier}: runs the reproduced or integration tests to assess patch correctness.
\end{inparaenum}
\approach{CodeR} defines 23 distinct actions that agents may execute depending on their role, some adopted from earlier frameworks like \approach{SWE-agent} and \approach{AutoCodeRover}, and others newly introduced, such as actions for generating new plans.

\approach{LingmaAgent} \cite{ma2025alibabalingmaagent} \resultsVerified{28.8}, in early versions named \emph{RepoUnderstander}, is an approach designed to improve automated issue resolution by enabling deeper comprehension of entire code repositories. It focuses on understanding the global context and intricate inter-dependencies among functions and classes, going beyond file-level reasoning. 
\approach{LingmaAgent}follows a modular three-stage workflow:
\begin{inparaenum}[\it 1)]
    \item \emph{Repository Knowledge Graph Construction:} A knowledge graph is constructed to represent semantic relationships between entities in the repository.
    \item \emph{MCTS-Enhanced Repository Understanding:} LLM-based agents dynamically explore the graph using Monte Carlo Tree Search (MCTS) to discover important code elements and structures that influence issue resolution. MCTS enables the simulation of multiple exploratory trajectories, guided by a reward function that leverages in-context learning (ICL) \cite{dong2024surveyincontextlearning} and Chain-of-Thought reasoning \cite{wei2023chainofthought}. This allows the system to adaptively narrow the search space.
    \item \emph{Information Utilization and Patch Generation:} Agents synthesize the most relevant information collected during exploration to locate faults and generate patches.
\end{inparaenum}

\approach{Globant Code Fixer Agent} \resultsLite{48.33} is a loop-based, multi-agent-driven approach composed by two main stages.
The \emph{Localization} stage is composed by a set of agents with access to tools navigates the codebase and searches for candidate files that may be causing the bug. 
The \emph{Fixing Stage} is composed by three specialized agents:
\begin{inparaenum}
\item Architect Agent: Generates solution proposals based on the bug localization report and contextual understanding of the codebase. 
Moreover, it drafts fixes aligned with coding best practices and project specifications.
\item Editor Agent: Implements the proposed solution by modifying the source code, and assures the syntactical and logical correctness of the solution.
\item Critic Agent: Evaluates the implemented solution for correctness. If it is incorrect, the agent triggers a retry loop, prompting the agents to refine the fix.    
\end{inparaenum}

As previously noted, the three most recent entries from \approach{nFactorial} \resultsVerified{49.2} adopt a scaffolded workflow that grants the agent full control over the execution flow. The initial implementation featured five distinct agents (\emph{Censorer}, \emph{Reproducer}, \emph{Localizer}, \emph{Fixer}, and \emph{Code Editor}) but the composition and roles of these agents have evolved across different submissions.

\approach{HyperAgent}\cite{phan2024hyperagent} \resultsLite{25.33} is a centralized multi-agent systems composed of four agents.
The code design principle of the approach centers on the centralization of advanced reasoning in a \emph{Planner} agent, while delegating computationally intensive yet conceptually simpler tasks to three specialized child agents: the \emph{Navigator}, \emph{Editor}, and \emph{Executor}.
The communication between the Planner and these specialized agents  is managed via an asynchronous message queue.
According to the authors, this architecture enables parallel processing of subtasks, dynamic load balancing, and efficient handling of complex software engineering challenges.

The submission \approach{SIMA} \resultsLite{27.67}, developed independently by Alex Sima, explores the concept of \emph{Agent Forest}, originally introduced by Li et al.~\cite{li2024MoreAgentsneed}. This method follows a two-phase process: 
\begin{inparaenum}[\it 1)]
  \item  \emph{sampling}, where a task input is iteratively processed by a single LLM or a collaborative group of LLM-based agents to generate multiple candidate outputs, and 
\item  \emph{voting}, where a majority decision determines the final result. 
\end{inparaenum}
To investigate how different organizational structures affect agent performance, Sima instantiated 13 variants of the Agent Forest framework, each inspired by the organizational hierarchy of major tech companies. For instance, the ``Amazon'' structure was modeled as a binary tree with a top-level manager agent,  the ``Apple'' structure consisted of numerous small, minimally structured, competing teams; and the ``Facebook'' structure represents a non-hierarchical, yet densely interconnected organization with web-like links among agents.
This experiment suggests that adopting multiple competing agent teams can increase the  diversity of problem-solving approaches, increasing the probability of resolving an issue.

Other approaches that report results on \swebench{} using agentic workflows with multiple-agents, but with less publicly available detail, include  \approach{Bracket.sh} \resultsVerified{53.2} and  \approach{Composio SWE-Kit} \resultsBoth{41}{48.6}.


\subsubsection{\bf Workflow Emergent and Emergent Autonomy - Single Agent (G6)}
\label{sec:result:mainArchitecture:worflowEmergentAutonomySingleAgent}

The entries in this group operate without a predefined workflow, unlike those in the previous groups. Here, a single agent autonomously determines the execution path, selecting and ordering steps based on goals, context, and feedback.

SWE-agent \cite{yang2024sweagentagent} \resultsBoth{23}{40.2} is a Agent-based system which follows the ReAct technique \cite{yao2023ReAct}: At each step, generates a thought and a command (that is, an action), then incorporates the feedback from the command's execution (file search, navigate, edit) in the environment.

Other entries build upon (or even fork) \approach{SWE-Agent}.
For instance, \approach{Nebius AI} \resultsVerified{40.6} introduces enhancements primarily targeting agent–environment communication and robust error handling.
The key contributions of their submissions~\cite{golubev2024search,zainullina2025guidedsearchstrategies} include:
\begin{inparaenum}[\it a)]
\item the training of an \emph{Action Generator} model using successful SWE-Agent trajectories on SWE-bench-like problems,
\item the development of \emph{Critic Models}, trained using data from the Action Generator, and
\item a \emph{trajectory selection strategy} that leverages the critic model to evaluate and choose the most promising repair trajectories for execution.
\end{inparaenum}
The authors trained open-source models, specifically \emph{Qwen2.5}.
Moreover, the authors shows that critic-based approach adds value not only on top of open weight models but also when paired with frontier LLMs such as \texttt{GPT-4o}.

\approach{OpenHands} (formerly known as \approach{OpenDevin}) \resultsBoth{41.67}{65.8} is a platform designed for developing AI agents that perform software development tasks, including code writing, command-line interaction, and web browsing. 
OpenHands provides modular implementations of agents, environments, and evaluation tools.
The authors implemented \approach{CodeActAgent} using the agent abstraction offered by OpenHands. This default generalist agent is based on the CodeAct framework~\cite{wang2024CodeAct} and serves as an example of a single-agent system. According to the authors, such a configuration involves a single LLM, a unified action space, and a single prompting technique.

The company Anthropic submitted the entries \approach{Tools + Claude} \resultsVerified{63.2}, which employ a minimal scaffolding framework specifically optimized for their models, such as Claude 3.5 Sonnet. 
According to the submitters, the primary objective of this design is to grant the language model  as much control as possible in the repair process. 
With this minimal scaffolding, the model is free to determine how it progresses from one step to the next, relying on its own judgment rather than following strict, discrete transitions or being hardcoded into a fixed workflow or pattern.
Other entries, such as \approach{Augment Agent}\resultsVerified{65.4}, build upon Anthropic's minimal scaffolding and incorporate additional enhancements, for instance, leveraging the Sequential Thinking MCP Server\footnote{\url{https://github.com/modelcontextprotocol/servers/tree/main/src/sequentialthinking}} framework as the planning tool (not released by Anthropic). This framework facilitates dynamic and reflective problem-solving through a structured reasoning process.

A similar approach, but using different LLMs, corresponds to the entries \approach{W\&B Programmer O1}~\resultsVerified{64.6}, which was the first on \swebench{} to evaluate the reasoning capabilities of \texttt{OpenAI's GPT-o1}, and \approach{Google Jules}~\resultsVerified{52.2}, which uses Google's Gemini model.

\subsubsection{\bf Workflow Emergent and Emergent Autonomy - Multiple Agents (G7)}
\label{sec:result:mainArchitecture:worflowEmergentAutonomyMultipleAgent}

The entries of this group adopt an emergent workflow executed by multiple agents with predefined roles, where the coordination and execution flow are not scripted but emerge dynamically through inter-agent interaction.

\approach{InfantAgent} \cite{lei2024InfantAgent} \resultsLite{30} is a hierarchical multi-agent collaboration system designed to solve software engineering tasks through logical reasoning and self-reflection. The agents are organized into two levels: \emph{brain-level} agents, responsible for reasoning and coordination, and \emph{hand-level} agents, which execute tasks by invoking tools. Hand-level agents are designed using prompt engineering and can be further trained on carefully curated datasets.
The system operates iteratively through a structured sequence of steps until the task is fulfilled: 
\begin{inparaenum}[\it a)]
   \item  reasoning and analysis (employing Chain-of-Thoughts~\cite{wei2023chainofthought}), 
    \item  task scheduling (by the brain-level agent),
    \item task execution (by the assigned hand-level agent), and 
    \item results evaluation and summarization (performed by the brain-level agent).
\end{inparaenum}

The \approach{CodeStory Aide} \resultsBoth{43}{62.2}, introduces a novel design based on a population of $N$ agents, where each agent is assigned to operate within the scope of a specific \emph{code symbol},  such as a Python class or function. 
This architecture grounds each agent to a well-defined unit of work, enabling fine-grained code comprehension and modification, while minimizing conflicts and overlaps in responsibilities across agents.

The entries \approach{Emergent E1} \resultsVerified{57.2} correspond to multi-agent architecture, including Search Agent, Edit agent and a Patch `Adjudicator' Agent (a judge).

\subsubsection{\bf Unclassified Submissions Due to Limited Evidence on Architectural Aspects (G8)}
\label{sec:result:mainArchitecture:others}

There are additional entries for which limited or no publicly available information about the architecture is provided. As a result, we do not have sufficient evidence to classify them within the categories previously described.
Examples of these entries are:
\begin{inparaenum}[\it a) ]
\item \approach{Artemis Agent},
\item \approach{Blackbox AI Agent},
\item \approach{devlo},
\item \approach{EPAM AI/Run Developer Agent},
\item \approach{Honeycomb},
\item \approach{Kodu-v1},
\item \approach{Isoform},
\item \approach{Kortix AI},
\item \approach{Learn-by-interact},
\item \approach{OpenCSG Starship Agentic Coder},
\item \approach{Solver},
\item \approach{ugaiforge},
and 
\item \approach{Aime-coder v1}.
\end{inparaenum}

\subsubsection{Architecture of solutions and \precision{}}
\label{sec:results:architecture:precision}

\begin{table}[t!]
\centering
\begin{tabular}{|l|rll|rll|}
\hline
&\multicolumn{3}{c|}{\swelite{}} & \multicolumn{3}{c|}{\sweverif{}} \\ 
\cline{2-7}
Architecture&  & \multicolumn{2}{c|}{\precision{}}
&  & \multicolumn{2}{c|}{\precision{}}\\
\cline{2-7}

&\#E   & Median & Max & 
\#E   & Median & Max\\
\hline

G1 Human-Workflow with Fixed execution - No Agent & 23 & \ccb{27.33} & \ccg{46} & 14 & \ccb{24.7} & \ccg{50.8} \\
G2 Human-Workflow with Fixed execution - with Single Agent & - & - & - & 4 & \ccb{53.7} & \ccg{68.2} \\
G3 Human-Workflow with Fixed execution - with Multiple Agents & 8 & \ccb{36.66} & \ccg{60.33} & 5 & \ccb{63.4} & \ccg{75.2} \\
G4 Human-Workflow with Scaffolded Execution - Single Agent & 16 & \ccb{38.16} & \ccg{60} & 10 & \ccb{56} & \ccg{70.8} \\
G5 Human-Workflow with Scaffolded Execution - Multiple Agents & 10 & \ccb{31.84} & \ccg{48.33} & 15 & \ccb{40.6} & \ccg{74.4} \\
G6 Workflow Emergent and Emergent Autonomy - Single Agent & 10 & \ccb{24.84} & \ccg{56.67} & 31 & \ccb{54.2} & \ccg{73.2} \\
G7 Workflow Emergent and Emergent Autonomy - Multiple Agents & 2 & \ccb{36.5} & \ccg{43} & 4 & \ccb{54.5} & \ccg{62.2} \\
G8 Unclassified Submissions & 10 & \ccb{36.66} & \ccg{55} & 16 & \ccb{43.9} & \ccg{66.4} \\

\hline
\end{tabular}

\caption{Total (\#E), Median and Max precision (\precision{}) of submissions grouped by the architecture.} 
\label{tab:resultsArchitecture}
\end{table}

Table \ref{tab:resultsArchitecture} groups the submission according to the classification of the architectures discussed previously. 
For each group, it shows the total number of submissions, and the median and max precision (\precision{}).

In \swelite{}, the highest precision scores in \swelite{} were attained by submissions falling under G3 (Human-Workflow with Fixed execution - with Multiple Agents) and G4 (G4 Human-Workflow with Scaffolded Execution - Single Agent), with values of 60.33\% and 60\%, respectively.
These two groups also stand out for the highest median precision observed among the submissions assigned to them (33.66\% and 38.16\%), despite similar values in other groups.
This parity is confirmed by the Kruskal–Wallis test, which did not find statistically significant differences between the groups (H = 12.1876, p = 0.0579).
These results indicate that well-structured, human-defined workflows, whether fixed (scripted execution) or scaffolded (allowing local autonomy), can achieve competitive performance.
Moreover, the table also shows that  G1 (Human-Workflow with Fixed execution - No Agent) and the mentioned  G4 have the highest number of entries, with 23 and 16 submissions respectively.

In \sweverif{}, the top-performing submission, classified under G3 (Human-Workflow with Fixed execution - with Multiple Agents), achieved 75.2\%, though submissions from other groups recorded comparable performance levels.
This group also achieved the highest overall performance, despite having a relatively small number of submissions (5).
In contrast, G6 (Workflow Emergent and Emergent Autonomy – Single Agent) includes the largest number of approaches (31), with submissions reaching competitive results (up to 73.2\%) and a median score that contends for second place after G3.
This suggests that autonomous single-agent systems, such as those in G6, can achieve competitive repair performance, which may be attributed to the improved reasoning abilities of modern LLMs —such as Claude 4 Opus and Sonnet— that reduce the need for explicit complex scaffolding.
The Kruskal–Wallis test reveals a significant difference between submitter groups (H = 19.41, p = 0.0070).
Dunn’s post-hoc test shows that G1 differs significantly from G3 ($p$ = 0.0207), G4 ($p$ = 0.0125), and G6 ($p$ = 0.0003).

\begin{tcolorbox}
\underline{\bf{Answer to RQ 2 :}}
Submissions classified under G3 (Human-Workflow with Fixed Execution – with Multiple Agents) and G4 (Human-Workflow with Scaffolded Execution – Single Agent) consistently achieved the highest performance on both \swelite{} and \sweverif{}.
Additionally, G6 (Workflow Emergent and Emergent Autonomy – Single Agent), which includes the largest number of submissions (31) on \sweverif{}, demonstrated that autonomous single-agent approaches —empowered by the reasoning capabilities of state-of-the-art LLMs such as Claude 4 Sonnet— can also deliver competitive results without requiring complex scaffolding.
Nevertheless, submissions without agents (G1) tend to show lower precision.

\end{tcolorbox}


\subsection{RQ 3 \rqpipeline}
\label{sec:results:phasesE2EMaintenance}

We describe the end-to-end software maintenance pipeline proposed in Liu et al.' survey~\cite{liu2024SurveyAgentsLLM4SE}, which also classifies six agent-based systems that we examine in this paper:
\approach{Agentless}~\cite{xia2024agentlessdemystifyingllmbasedsoftware},~\approach{AutoCodeRover}\cite{Zhang2024AutoCoderRover}, \approach{RepoUnderstander} (subsequently renamed \approach{LingmaAgent})~\cite{OLDma2024understandRepoUnderstander}, \approach{MASAI}~\cite{arora2024masaimodulararchitecturesSE}, \approach{SWE-Agent}\cite{yang2024sweagentagent} and \approach{CodeR}~\cite{xia2024dcoder}.

\subsubsection{\bf Preprocessing}
\label{sec:results:phases:preprocessing}

One of the initial phases commonly employed across approaches is the preprocessing of the codebase to construct a condensed representation, such as a graph or an index, that summarizes key information from the repository, including source code and, in some cases, documentation. This representation enables more efficient code navigation and supports contextual understanding. We identify different techniques used in this preprocessing phase.
     

Several approaches evaluated on \swebench{} adopt \emph{knowledge graph representations} of the codebase.
For instance, the \approach{Bytedance MarsCode} Agent \cite{liu2024marscode} \resultsBoth{39.33}{50} analyzes and organizes the code and documentation within a repository to generate a multi-directional graph.
Similarly, \approach{Nemotron-CORTEXA} \cite{sohrabizadeh2025nemotroncortexa} \resultsVerified{68.2} constructs a graph-based representation of a software repository, where nodes correspond to code files and entities —such as functions, classes, and methods— and edges capture \emph{contain} and \emph{use} relationships. Some approaches integrate off-the-shelf components in their solutions. For instance, 
\approach{Orcaloca}~\cite{yu2025orcaLocallmAgent} \resultsLite{41} employs a graph-based representation, called \emph{CodeGraph}, to index code repositories and support code entity search. This graph captures hierarchical relationships —such as methods within classes— as well as reference relationships, including function calls between entities. 
Similarly, \approach{DARS}~\cite{aggarwal2025darsdynamicactionresampling} \resultsLite{47} utilizes \emph{RepoGraph}~\cite{ouyang2024repographenhancingaisoftware}, a hierarchical structure in which nodes represent code definitions and edges encode dependencies among them. \emph{RepoGraph} is also integrated in one of \approach{Agentless}' entries \resultsLite{29.67}.
As a last example, \approach{KGCompass}~\cite{yang2025enhancing} \resultsLite{46} incorporates a repository-aware knowledge graph designed to accurately link repository artifacts (such as issues and pull requests) with codebase entities (including files, classes, and functions). This graph is used both to narrow the search space during issue localization and to augment LLMs with relevant contextual information during patch generation. 
Other approaches using knowledge graphs are \approach{Alibaba Lingma Agent} \resultsBoth{33}{28.8} and Bytedance's \approach{TRAE} \resultsVerified{75.2}.

\emph{Repository mapping} is a variant of knowledge graph, also adopted as a strategy to represent the codebase structure, as implemented by \approach{Aider} \resultsLite{26.33} and \approach{CodeStory Midwit Agent} \resultsVerified{62.2}.
In the case of \approach{Aider}, the map provides a concise summary of the codebase —highlighting key classes and functions— and is included with each request sent to the LLM.\footnote{Aider RepoMap: \url{https://aider.chat/2023/10/22/repomap.html}}

\emph{Vector stores} are employed to index the codebase in submissions such as \approach{Moatless Tool} \resultsBoth{38.33}{70.8}, \approach{GRU} \resultsBoth{48.67}{57}, and \approach{SuperCoder}~\resultsLite{34}.
Notably, \approach{Patched.Codes Patchwork} \resultsLite{37} embeds the entire repository into a vector database using a Sentence-Transformers based embedding model. \approach{Aegis} \resultsLite{30.33} encodes code snippets with Cohere Embed v3 and stores them in a local FAISS vector store. In the case of \approach{Bracket.sh} \resultsVerified{53.2}, the system constructs an intermediate representation of large codebases, enabling agents to operate at higher levels of layered abstraction.
Other submissions, such as \approach{AppMap Navie} \resultsBoth{36}{47.2}, do not rely on embeddings or pre-built indexes but instead generate an index of the entire codebase.

Some approaches utilize \emph{Repository Trees} to facilitate navigation. For example, \approach{Amazon Q Developer Agent} \resultsBoth{29.67}{65.4} uses a structured XML representation of the repository’s file system, while most of \approach{Agentless}' entries convert the project codebase into a tree-like structure.
\approach{Factory Code Droid} \resultsBoth{
31.33}{37} models the codebase using a multi-resolution representation that combines explicit relationships, captured through graphs, and implicit relationships, encoded via embeddings in latent space, enabling reasoning across different levels of abstraction and a deeper understanding of the code. Repository trees are also used by \approach{SWE-Agent} \resultsBoth{56.67}{66.6} and \approach{Composio SWE-Kit} \resultsBoth{41}{48.6}.

Other submissions combine different techniques. 
For example, \approach{Zencoder} \resultsVerified{70} relies on \approach{Repo Grokking}, a foundational engine that continuously indexes the entire codebase, capturing structural and semantic relationships through embeddings and dependency graphs. It supports advanced search capabilities—including embedding search, graph traversal, and full-text retrieval, all enhanced by LLMs for deeper contextual understanding.

It must be remarked that some submissions either do not perform preprocessing or do not report it explicitly. For instance, \approach{AbanteAI MentatBot} \resultsLite{38} uses a Library Context tool to select relevant files, whereas \approach{Globant Code Fixer Agent} \resultsLite{48.33}, \approach{AutoCodeRover} \resultsBoth{38.4}{30.67} and \approach{SpecRover} \resultsVerified{51.6} access the file system directly via specialized tools and APIs.

\subsubsection{\bf Issue Reproduction}
\label{sec:results:phases:issueReproduction}

By design, \swebench{} does not provide submitters with the tests that expose and reproduce the benchmark bugs. These tests are only available during the evaluation phase of a submission. This contrasts with other benchmarks in the APR community, such as Defects4J, which include \emph{failing} tests that are used not only to reproduce bugs but also to verify the plausibility of candidate patches. As a result, in the absence of such tests, some solutions attempt to create them manually or generate scripts to reproduce the bugs.

Submissions such as \approach{SWE-Agent} \resultsBoth{56.67}{66.6} and \approach{CodeR} \resultsLite{28.33} leverage LLMs to generate reproduction tests that aim to replicate the original issue directly from issue descriptions. \approach{Agentless} \cite{xia2024agentlessdemystifyingllmbasedsoftware} \resultsBoth{40.67}{50.8} generates multiple candidate reproduction tests and selects the optimal one based on actual execution results on the original codebase. \approach{SWE-RL} \cite{wei2025SWE-RL} \resultsVerified{41.2} builds upon this approach through the creation of the \approach{Agentless Mini} agent scaffold, introducing two key improvements:
\begin{inparaenum}[\it 1)]
\item Instead of relying solely on the issue description, it retrieves a relevant existing test file to guide test generation; and
\item Rather than selecting a single majority-voted test, it allows the selection of multiple top-ranked test samples based on voting outcomes.
\end{inparaenum}
\approach{PatchPilot} \resultsBoth{41.33}{64.6} extends \approach{Agentless} with a \emph{self-reflection-based PoC reproduction} mechanism, inspired by the \emph{Reflexion mechanism}~\cite{shinn2023reflexion}. To address common issues at this stage —such as misconfigurations and missing dependencies— the approach leverages an LLM to iteratively generate and refine scripts that reproduce the issue (referred to as a Proof-of-Concept, or PoC). It uses carefully crafted prompts to ensure the inclusion of necessary dependencies and configurations.

Agents specialized in issue reproduction are incorporated in various multi-agent approaches.
For instance, \approach{Bytedance MarsCode Agent} \resultsBoth{39.33}{50} features a \agent{Reproducer Agent} that generates reproduction scripts based on the relevant code and issue description, and performs dynamic debugging in a sandbox to confirm successful reproduction \cite{liu2024marscode}. 
\approach{SpecRover} \resultsVerified{51.6}  also includes a \agent{Reproducer Agent} that writes a test to reproduce the program fault described in the issue, which is also used to verify the patch. Since the test is generated by an LLM and may vary due to its non-deterministic nature, \approach{SpecRover} further employs a \emph{Reviewer Agent} to assess the correctness of the test based on the issue description.

One of the primary obstacles in issue reproduction is the creation of a runnable environment, as emphasized by \approach{SWE-RL}~\cite{wei2025SWE-RL} \resultsVerified{41.2}.
To address this, the multi-agent system \approach{MASAI} \cite{arora2024masaimodulararchitecturesSE} \resultsBoth{27.33}{32.6}  defines two specialized sub-agents for this task:
\begin{inparaenum}[\it 1)]
\item the \emph{Test Template Generator}, which analyzes documentation, existing tests, and repository setup to determine how to write and execute new tests. It then produces a repository-specific, issue-independent test template along with the command required to run it; and the \emph{Issue Reproducer}, which uses the generated test template to write a test that replicates the behavior described in the issue.
\end{inparaenum}
\approach{nFactorial} \resultsVerified{49.2} defines a \agent{Reproducer Agent} tasked with generating a standalone Python script to replicate the issue, using tools for executing bash scripts and editing files, and ensuring the result is reliably reproducible across multiple executions.
DARS~\cite{aggarwal2025darsdynamicactionresampling} \resultsLite{47} equips agents with actions for creating and improving reproduction scripts based on previous attempts.

Some approaches such as \approach{Alibaba Lingma Agent} \resultsBoth{33}{28.8} and \approach{SWE-fixer} \resultsBoth{24.67}{32.8} do not offer issue reproduction. 
Notably, some others have incorporated issue reproduction progressively over time. For example, \approach{GRU} has two submissions on each leaderboard, and one key difference between them is the addition of a new task in the workflow aimed at a reproduction script. 


%
%
%

\subsubsection{\bf Issue Localization}
\label{sec:results:phases:issueLoc}

The goal of issue localization is to identify code elements —at a specific granularity, such as line— that are suspected to be related to the issue and may require modification. 
We adopt the four-category taxonomy proposed by Liu et al.~\cite{liu2024SurveyAgentsLLM4SE}, which classifies seven agent-based approaches according to their localization strategies.

\paragraph{Retrieval-based Localization:}

The use of retrieval-based strategies enables approaches to identify relevant code elements by measuring their similarity to inputs such as issue descriptions. Several techniques are used:

Our analysis reveals that multiple submissions rely on large language models (LLMs) to perform at least some steps of the localization process. For example, \approach{SWE-Fixer}~\cite{xie2025SWEFixerTrainingopensourcellms} \resultsBoth{24.67}{32.8} adopts a coarse-to-fine strategy, first applying BM25~\cite{RobertsonBM25} for initial file retrieval, and then using a model to identify the defective files from the retrieved set.

Repository representations created during the preprocessing phase (Section~\ref{sec:results:phases:preprocessing}) are leveraged in the localization phase alongside LLMs. 
For instance, 
\approach{DARS}  \resultsLite{47} employs a keyword-based subgraph retrieval technique (ego-graphs~\cite{hu2024graggraphretrievalaugmentedgeneration}) over its \emph{RepoGraph}, enabling it to identify files and specific line numbers relevant to the issue.
\approach{Composio SWE-Kit} \resultsBoth{41}{48.6} includes a tool that analyzes the codebase and create a structured code map. This map allows the agent to efficiently locate, retrieve, and understand relevant code sections without -according to the authors- overloading the LLM's context.
A similar strategy, combining a repository map with keyword-based search, is employed by \approach{CodeStory}.

\approach{KGCompass}~\cite{yang2025enhancing} \resultsLite{46}  employs a hybrid approach that combines knowledge graph analysis with LLM-based reasoning to identify potential bug locations. 
It first ranks function entities based on their relevance to the issue description, using a score that integrates semantic and textual similarity to the issue text as well as structural proximity in the graph; then it selects the top 15 candidates. An LLM further analyzes the problem description to identify up to 5 additional locations.

\approach{Amazon Q Developer Agent} \resultsBoth{29.67}{65.4} leverages a structured XML representation of the repository, along with additional context, to guide an LLM in identifying relevant files for retrieval.
\approach{SuperCoder} ~\resultsLite{34} employs a hierarchical search space reduction strategy, using an LLM at each step to progressively narrow down from candidate files to specific locations within those files, via RAG and repository maps.
\approach{SWE-RL} \cite{wei2025SWE-RL} \resultsVerified{41.2}, through its scaffold \approach{Agentless Mini}, uses an LLM to predict relevant file paths based on the given issue and a repository structure mapped to file paths. 
\approach{HyperAgent} \resultsLite{25.33} incorporates an LLM-based agent named \agent{Navigator}, specialized in information retrieval within the repository.

In contrast, some approaches do not rely on LLMs for localization. For example, \approach{AppMap Navie} \resultsBoth{36}{47.2} defines a method to retrieve the relevant code context based on the issue text. According to the authors, this method operates locally without invoking an LLM, making it fast, predictable, and efficient.


\paragraph{Navigation-based Localization:}

The navigation strategy allows agents to perform code search actions across the entire repository structure \cite{liu2024SurveyAgentsLLM4SE}.
\approach{Agentless} \cite{xia2024agentlessdemystifyingllmbasedsoftware} \resultsBoth{40.67}{50.8}  employs a hierarchical localization process that first narrows down the fault to specific files, then to relevant classes or functions, and finally to fine-grained edit locations. This is achieved through a combination of LLM-based techniques and classic information retrieval-based localization methods \cite{Zhou2012RetrievalBug}.
Both \approach{SWE-Agent} \cite{yang2024sweagentagent}  \resultsBoth{56.67}{66.6} and \approach{MASAI} \cite{arora2024masaimodulararchitecturesSE}  \resultsBoth{27.33}{32.6} include tools that enable their agents to navigate the repository to locate relevant files or to visualize files to extract code elements.
    
\approach{Co-PatcherR}~\cite{tang2025copatchercollaborativesoftwarepatching} \resultsVerified{46}, build on top of \approach{PatchPilot}~\cite{li2025patchpilot}, improves localization by dividing it into two \emph{small}-LLM-driven sub-tasks: \emph{file localization}, where an LLM receives the issue description and repository file structure to identify relevant files, and \emph{line localization}, where the LLM analyzes code from each selected file along with the issue description to pinpoint faulty lines. Using \llm{Qwen-2.5-Coder} as the base model, the localization LLM is fine-tuned on distillation data with reasoning chains generated by \llm{Claude-3.7-Sonnet}.

For instance, \approach{IBM AI Agent SWE-1.0} \resultsLite{23.67} employs multiple localization agents, 
each potentially backed by a different LLM, that collaborate with a judge to identify the bug location (file, function, and line) by following a \emph{Think-Act-Observe} loop and utilizing tools based on BM25, LLM-based search and syntactic search.
\approach{AutoCodeRover}~\cite{Zhang2024AutoCoderRover} \resultsBoth{38.4}{30.67} uses an LLM-based agent, \agent{Context Retrieval}, to navigate the codebase and extract relevant snippets, leveraging context retrieval APIs such as \texttt{search\_method\_in\_class} which works at the AST-level.
\approach{Nemotron-CORTEXA} \cite{sohrabizadeh2025nemotroncortexa} \resultsVerified{68.2} employs a \agent{Localization Agent} in a two-step process: first, it identifies relevant files using a custom-trained code embedding model aligned with issue descriptions; then, through candidate entity filtering, the agent iteratively traverses the code graph to refine results at the level of functions, classes, or methods.
\approach{nFactorial} \resultsVerified{49.2} includes a \agent{Localizer} agent designed to navigate a custom file tree structure and identify code segments relevant to the problem statement. Similarly, \approach{Augment Agent} \resultsVerified{70.4} equips its agents with tools such as \texttt{grep} and \texttt{find} for repository navigation. 
\approach{Globant Code Fixer Agent} \resultsLite{48.33} features a \emph{Localization stage}, where a set of agents explores the codebase to identify candidate buggy.
\approach{ORCALOCA} \cite{yu2025orcaLocallmAgent} \resultsLite{41} enhances issue localization with LLM-based agents by combining dynamic action scheduling, hierarchical action decomposition, and a context manager that prunes irrelevant information to ensure focused and efficient codebase exploration.

The structured representation of a repository  (Section~\ref{sec:results:phases:preprocessing}) is used to support navigation.
For example,
\approach{Agentless + RepoGraph} \resultsLite{29.67} extends \approach{Agentless} by integrating RepoGraph, a structured repository graph that supports keyword-based queries (e.g., issue derived) through the \texttt{search\_repograph()} method.
which searches on the build graph given as input key terms, such as some extracted from the issue description.
\approach{Aider} \resultsLite{26.33}  uses the built repository map, compact and powerful summary of the entire codebase, to help the LLM to decide which file to edit, and a justification of that decision. This map is constantly tailored show repository context that is relevant to the current state of the problem.



\paragraph{Spectrum-based Localization.}

Spectrum-based fault localization (SBFL) techniques have been widely used in traditional APR systems, such as GenProg \cite{LeGoues2011genprog}, to compute the suspiciousness of code elements (e.g., lines) based on their coverage in failing and passing test cases.
\approach{AutoCodeRover} \resultsBoth{38.4}{30.67}  complements the technique mentioned above with SBFL in order to augment the problem statement (i.e., the issue description) with additional relevant classes and methods. 
\approach{CodeR} \resultsLite{28.33}  adopts a hybrid approach that linearly combines suspiciousness scores from SBFL (based on tests generated by its \emph{Reproducer} agents; see Section~\ref{sec:results:phases:issueReproduction}) and BM25-based file-level localization using issue similarity.
Unlike SBFL-based methods, \approach{PatchPilot} \cite{li2025patchpilot} adopts a runtime-driven hybrid approach that executes the issue to collect execution traces, uses LLMs to filter relevant files and perform fine-grained searches, and finally applies an LLM-based review step to retrieve code snippets linked to the root cause.

\paragraph{Simulation.}

\approach{Alibaba Lingma Agent}\cite{ma2024lingmaswegpt}  \resultsBoth{33}{28.8} performs issue localization through simulation. It first constructs a Repository Knowledge Graph (see Section~\ref{sec:results:phases:preprocessing}) and then applies a Monte Carlo Tree Search (MCTS) strategy to explore the graph. By simulating multiple exploration trajectories and assigning reward values to graph nodes (i.e., code elements), the approach narrows the search space to localize faults more precisely. 
MCTS has also been explored by other approaches, such as \approach{CodeStory Midwit Agent} \resultsVerified{62.2}. However, according to the submission authors, they dropped MCTS from their framework due to long execution times.

\paragraph{Combining techniques}

\approach{Codev} identifies relevant files using a multi-source approach that combines static analysis, semantic search, and LLM-based reasoning. 
Six complementary sources are used to retrieve potentially relevant code blocks: a code graph, two types of string search, a semantic embedding model, an LLM-based analyst, and an agentic top-down search.
Each retrieved block is evaluated and scored by a lightweight LLM, which classifies them as \emph{Solution Target}, \emph{Supporting}, \emph{Interesting}, or \emph{Not Relevant}. 
These scores are aggregated into a confidence measure per file, guiding the selection of files for subsequent development planning.



\subsubsection{\bf Task Decomposition - Planning}
\label{sec:results:phases:taskDecompPlan}

There are approaches that decompose the main task (in this case, resolve an issue) into more fine-grain sub-tasks.
The \approach{Tools + Claude} entries by Anthropic \resultsVerified{73.2} furnish an LLM (Claude 3.5/3.7/4 Sonnet) with a set of tools for planning. In addition to the \texttt{Edit} and \texttt{Bash} tools, they introduces a ``planning'' tool that prompts the LLM to write down its thoughts as it solves the problem.
This tool leverages the inherent reasoning capabilities of Claude Sonnet’s default thinking mode.
As the details of the planning tool were not publicly disclosed, the submission \emph{Augment Agent} \resultsVerified{70.4} replaces it with the Sequential Thinking MCP framework, which enables dynamic and reflective problem-solving through a structured reasoning process.

Agents for plan elaboration are also employed.
For example, \approach{Alibaba Lingma Agent} \cite{ma2025alibabalingmaagent} \resultsBoth{33}{28.8} includes an LLM-based agent, \agent{Summary Agent}, which takes as input the relevant code fragments identified through \emph{MCTS-based repository understanding} along with the issue description, and produces a sequential summary of the relevant code and plans a solution.
\approach{CodeR} \cite{xia2024dcoder} \resultsLite{28.33}  is a multi-agent system featuring a \agent{Manager Agent} responsible for selecting one of four predefined plans based on the issue description and interpreting the execution summary after the plan is carried out. 
A similar approach is adopted by \approach{Bytedance MarsCode}~\cite{liu2024marscode} \resultsBoth{39.33}{50}, which includes a \agent{Planner} that analyzes the collected code snippets and classifies the issue into one of two workflows: dynamic debugging repair or static repair.
\approach{SuperCoder} \cite{gautam2024supercoder}~\resultsLite{34} includes a \agent{PlannerAgent} responsible for navigating code files, identifying buggy locations, and determining the changes needed to resolve the issues.
\approach{HyperAgent} \cite{phan2024hyperagent} \resultsLite{25.33} is a multi-agent system centered around a \agent{Planner} agent that processes issues, generates resolution strategies, and coordinates specialized child agents (Navigator, Editor, Executor) via an asynchronous message queue in an iterative planning and feedback loop.

Some approaches do not detail the use of LLMs or agents. \approach{AppMap Navie} performs a planning step by combining the issue description with context to produce a structured plan outlining the problem, a high-level solution, target files, and file-specific change descriptions without code.
Similarly, \approach{Gru} \resultsBoth{48.67}{57} takes the list of relevant files from localization, analyzes the corresponding code, and decides which files to modify and how the modifications should be made. In the extreme case, approaches such as \approach{Aegis} \resultsLite{30.33}, \approach{Amazon Q Developer Agent} \resultsBoth{29.67}{65.4}, and \approach{Google Jules} \resultsVerified{52.2}, mention the inclusion of a planning phase or the generation of a plan, but they do not provide further implementation details.


\subsubsection{\bf Patch Generation}
\label{sec:results:phases:patchGeneration}

In the patch generation phase, approaches produce patches based on suspicious code elements identified during the localization phase (Section~\ref{sec:results:phases:issueLoc}) or derived from a modification plan (Section~\ref{sec:results:phases:taskDecompPlan}).
Patch generation is typically included as a distinct phase in approaches with a crafted workflow such as \approach{Agentless}~\cite{xia2024agentlessdemystifyingllmbasedsoftware} \resultsBoth{40.67}{50.8}, \approach{PatchPilot}~\cite{li2025patchpilot} \resultsBoth{41.33}{64.6}, and \approach{DARS}~\cite{aggarwal2025darsdynamicactionresampling}  \resultsLite{47}, consistent with the design of traditional APR techniques.
In this phase, submissions either rely on foundation models —as in the mentioned approaches— or use fine-tuned models. 
This is the case for
\approach{Co-PatcherR}~\cite{tang2025copatchercollaborativesoftwarepatching}  \resultsVerified{46}, which enhances generation by combining two LLM-driven sub-tasks: patch generation, where the model receives the issue description and identified root causes to produce a candidate patch, and patch critique, where the model evaluates a given patch and suggests corrections if needed. Using \llm{Qwen-2.5-Coder-14B} as the base model, the generation LLM is finetuned on reasoning-chain distillation data produced by \llm{Claude-3.7-Sonnet}. 
Notably, in \approach{Co-PatcherR},  a single model is trained for both localization (See \ref{sec:results:phases:issueLoc}) and generation tasks, as both require similar capabilities—interpreting the issue and understanding the codebase.

A number of solutions use specialized agents tasked with generating patches. These agents are referred to using different names; for example, \agent{Editor (Editing) Agent} in  \approach{Globant Code Fixer Agent}, \approach{MarsCode} \cite{liu2024marscode}, \approach{CodeR} \cite{xia2024dcoder}, \approach{SWE-Fixer} \cite{xie2025SWEFixerTrainingopensourcellms}, \approach{HyperAgent} \cite{phan2024hyperagent}, \approach{IBM AI Agent SWE-1.0};
\agent{Patching Agent} in \approach{SpecRover} \cite{ruan2024specrovercodeintentextraction}; and \agent{Fixer Agent} in \approach{nFactorial}.
These agents typically receive, in addition to suspicious code elements, the problem statement (e.g., issue description) and additional context, which may include other files, relevant code snippets, failing test cases, or —such as in \approach{AutoCodeRover}~\cite{Zhang2024AutoCoderRover}— a history of context retrieval, including invoked APIs, their results, and prior code analysis performed by other agents.
In some submission, patching agents follow a retry-loop strategy for patch generation. For instance, in \approach{AutoCodeRover}\cite{Zhang2024AutoCoderRover} and \approach{Alibaba LingmaAgent}\cite{ma2025alibabalingmaagent}, if a generated patch fails to conform to the required format (e.g., an invalid diff) or cannot be syntactically applied to the original program, the agent is prompted to retry.

Authors from some submissions, such as   \approach{Aider} \resultsLite{26.33}  and \approach{TOOLS + Claude} \resultsVerified{73.2},  designed and developed specialized tools for editing files and generating patches, or as the case of  \approach{IBM Research Agent-101} to prevent common errors such as incorrect indentation, missing imports, and imprecise line numbers.

Similarly to some traditional APR approaches such as \cite{Bach2016History}, which leverage historical repair data, \approach{ExpeRepair}~\cite{mu2025experepairdualmemoryenhancedllmbased} \resultsLite{60.33} adopts a multi-agent architecture enhanced with a dual-memory mechanism:
\begin{inparaenum}[\it 1)]
\item \emph{episodic memory}, which stores concrete and reusable demonstrations of past repairs, and
\item \emph{semantic memory}, which captures high-level, generalizable insights (e.g., repair strategies).
\end{inparaenum}
Two of \approach{ExpeRepair}'s agents, the \agent{Patch Agent} and the \agent{Test Agent}, leverage this memory system when generating patches and tests, respectively. They retrieve relevant demonstrations (patches or tests) from the episodic memory, and use summarized natural language insights from the semantic memory to guide their reasoning with general repair strategies.

The patch generation phase may produce one or multiple candidate patches, which are subsequently evaluated and ranked in later phases.
For instance, \approach{Agentless}~\cite{xia2024agentlessdemystifyingllmbasedsoftware} and \approach{OpenHands} \cite{wang2024openhandsopenplatformai} sample multiple candidate patches per issue to maximize the chance of generating a correct fix. 
\approach{W\&B Programmer O1} \resultsVerified{64.6} employs a parallelization strategy by generating five candidate patches simultaneously for each issue.
Other submissions such as \approach{TRAE} \resultsVerified{75.2}, \approach{CodeFuse-AAIS} \resultsLite{35.67} and one particular from OpenHands -\approach{4x Scaled (2024-02-03)}- \resultsVerified{60.8} combine multiple large language models to expand the sampling space.
\approach{Aider}  \resultsLite{26.33} employs a cascade strategy, beginning with one LLM and, if no plausible solution is found, proceeding with another. 
\approach{IBM AI Agent SWE-1.0} \resultsLite{23.67} uses multiple \agent{Editing} agents —each powered by a different LLM— and a \agent{Judge} agent responsible for analyzing and selecting among the resulting patches.
\approach{Devlo} \resultsVerified{70.2} employs multiple LLMs with ``\emph{diverse training data, inductive biases, and reasoning habits}''. According to the authors, this diversity allows \approach{Devlo} to surface what one model might miss, triangulate more accurate solutions, and gain confidence when models independently propose similar fixes.


Multiple patches can result from varying the configuration, the application of different models and strategies.
For example, \approach{SuperCoder}~\resultsLite{34} produces different solutions by sampling with different temperature settings. 
In \approach{ExpeRepair} \resultsLite{60.33} , the \agent{Patch Agent} performs multiple samplings per iteration, with the temperature increasing at each step to promote greater diversity among candidate patches.
KGCompass~\cite{yang2025enhancing} \resultsLite{46} employs a hybrid sampling strategy, combining deterministic (temp = 0) and exploratory (temp > 0) sampling methods.

Similarly, \approach{SIMA} \resultsLite{27.67} submission deploys multiple agents, each configured with a distinct temperature value.
Moreover, multiple patches can also result from the creation and evaluation of different execution trajectories. For example, \approach{DARS}~\cite{aggarwal2025darsdynamicactionresampling} \resultsLite{47} generates multiple patches through its \emph{Expansion mechanism}, which branches the trajectory at key decision points based on prior trajectory history and execution feedback from previous attempts.
Nevertheless, the generation of multiple patches can be fostered in the earlier stages of the repair pipeline. For example, \approach{PatchPilot}~\cite{li2025patchpilot} \resultsBoth{41.33}{64.6} encourages patch diversity at the planning stage (Section~\ref{sec:results:phases:taskDecompPlan}) by designing three types of prompts for plan generation. These prompts explicitly guide the LLM to produce patching plans with different focuses, for instance, generating a minimal patch with the smallest possible modifications.
\approach{Codev} \resultsLite{49} launches multiple rollouts based on the number of files identified as `interesting' by the fault localization stage (see Section \ref{sec:results:phases:issueLoc}). For each rollout, it generates a distinct development plan to introduce variance and explore alternative solutions.
\approach{Zencoder} \resultsVerified{70} tackles each SWE-bench task by running four agents in parallel, each equipped with a distinct LLM (\llm{Claude Sonnet 3.7} or \llm{o4-mini}) and a unique set of diagnostic, search, editing, and reasoning tools.

\subsubsection{\bf Patch Verification}
 \label{sec:results:phases:patchvalidation}

This phase aims to verify both the syntactic and semantic correctness of the candidate patches generated in the previous phase.

\emph{Syntax checkers} or \emph{linters} —static analysis tools that detect stylistic issues, potential bugs, and suspicious code constructs—  are used in several submissions, including \approach{Alibaba Lingma Agent} \resultsBoth{33}{28.8}, \approach{Amazon Q Developer Agent} \resultsBoth{29.67}{65.4}, and \approach{Nemotron-CORTEXA} \resultsVerified{68.2}.
Different types of tools are used across submissions to assist in patch validation. 
For example, \approach{Aegis} \resultsLite{30.33} and \approach{PatchPilot} \resultsBoth{41.33}{64.6} execute Flake8\footnote{https://flake8.pycqa.org/en/latest/}, a command-line tool that checks Python code against coding style (PEP 8), detects programming errors, and flags overly complex constructs.
\approach{Artemis Agent} \resultsVerified{32} invokes external utilities such as profilers (e.g., VTune\footnote{https://www.intel.com/content/www/us/en/developer/tools/oneapi/vtune-profiler.html}, Speedscope\footnote{https://github.com/jlfwong/speedscope}) and security analyzers (e.g., SonarQube\footnote{https://github.com/SonarSource/sonarqube}).
\approach{CodeStory Midwit Agent} \resultsVerified{62.2} relies on Pyright\footnote{https://github.com/microsoft/pyright} to gather diagnostics such as missing imports.
\approach{Factory Code Droid} \resultsBoth{
31.33}{37} extends validation further by employing not only linters and static analyzers, but also debuggers.
\approach{SWE-Agent} \resultsBoth{56.67}{66.6}  integrates a linter into the \texttt{edit} function to automatically alert the agent of any mistakes introduced during file editing \cite{yang2024sweagentagent}

Syntax or linter errors can be fed back into the generation phase to produce revised code that avoids the detected issues, as implemented in submissions such as \approach{SWE-agent}, \approach{AppMap Navie}, and \approach{CodeStory Midwit Agent}. Notably, \approach{Aider} \resultsLite{26.33} prompts the LLM with linting errors represented in abstract-syntax-tree format, allowing it to display the relevant code context for each error. This enables the model to better understand the problem and make the appropriate corrections.

\emph{Dynamic verification} is also used to assess the plausibility of candidate patches. 
Traditional APRs, such as GenProg~\cite{LeGoues2011genprog} pioneered on the usage of test cases for semantically candidate validating patches.
The majority of the submissions to \swebench{} execute regression test and, when available, also run the issue reproduction test or script (Section~\ref{sec:results:phases:issueReproduction}).
Submissions vary in their test execution strategies during patch validation. While some, like \approach{AppMap Navie} or \approach{Devlo}, execute all the test cases from the repository, others, such as \approach{AgentScope} \resultsVerified{63.4}, limit execution to a selected subset.
\approach{Aegis} \resultsLite{30.33}, for example, filters tests based on filename similarity between test files and the modified files. 
In contrast, \approach{Agentless} \resultsBoth{40.67}{50.8} begins with a full test run and subsequently narrows the test set using LLM-guided selection. 
Additionally, discussed in Section \ref{sec:results:phases:issueReproduction}, certain approaches (e.g., \approach{CodeShellTester} \resultsLite{31.33}) enhance patch validation by synthesizing new test cases.

\emph{Patch refinement} is an additional phase included in some approaches following validation.
For example, \approach{PatchPilot}~\cite{li2025patchpilot} \resultsBoth{41.33}{64.6} explicitly incorporates a \agent{Refinement stage} in its workflow, where the current patch is iteratively improved based on validation feedback.
\approach{SpecRover}~\cite{ruan2024specrovercodeintentextraction} \resultsVerified{51.6} introduces a \agent{Review} agent that generates Reviewer Feedback —a summary derived from inferred specifications explaining the rationale behind the patch correctness judgment. This feedback is then passed to the \agent{Patching} agent to guide the generation of an improved patch.
In \approach{ExpeRepair} \resultsLite{60.33}, patch refinement is performed during the Validation phase. In this phase, the \agent{Patch Agent} improves the solution by addressing edge case handling, regression risks, and adherence to language-specific best practices, while the \agent{Test Agent} enhances the test suite with additional cases targeting boundary conditions and corner cases.

\emph{LLM-as-judge} have be also employed by submissions, eventually after other validations.
For instance,  \approach{CoDev} \resultsLite{49}, once a solution passes the test script, it undergoes regression testing. If a regression test fails, a reasoning model assesses whether the failure is still relevant or if the test should be updated in order to ensure that valid solutions are not wrongly discarded due to outdated tests.



\subsubsection{\bf Patch Ranking and Selection}
\label{sec:results:phases:ranking}

The last phase is usually the selection of a single patch to be presented to the developer, and in the case of experimentation on \swebench{}, to be submitted for further evaluation (e.g., using the bug-revealing tests not available during repair).
For selecting the best patches, some approaches apply patch ranking.

\emph{Majority voting} is adopted by several submissions as a strategy for selecting the final patch.
For example, \approach{Agentless}~\cite{xia2024agentlessdemystifyingllmbasedsoftware} \resultsBoth{40.67}{50.8} normalizes candidate patches to eliminate superficial syntactic differences, then selects the one with the highest frequency.
A similar majority-based strategy is adopted by \approach{Nemotron-CORTEXA} \cite{sohrabizadeh2025nemotroncortexa} \resultsVerified{68.2}.
\approach{Bytedance MarsCode Agent} \resultsBoth{39.33}{50} normalizes patches at the AST level and leverages an LLM to vote across all candidates, selecting the patch receiving the most votes.
\approach{Augment Agent}  \resultsVerified{70.4} also employs a simple majority voting strategy using the \llm{OpenAI o1} model: it presents the model with a list of candidate diffs and the problem statement, prompting it to pick the majority vote solution.
\approach{TRAE} \resultsVerified{75.2} initially adopted the majority voting from \approach{Augment Agent}, but discontinued its use after observing a decline in performance as the sampling space increased.
Instead, \approach{TRAE} is equipped with a  \emph{Selector agent} with two stages: 
\begin{inparaenum}
    \item \emph{Syntax-Based Voting}: the agent first performs syntax-based voting by clustering patches based on AST equivalence and selecting the most frequent variant.
    Then, the agent verifies whether the syntax-voted patch behaves as expected using the issue description and a set of tools (e.g., bash, file editor)
   \item \emph{Multi-Selection Voting}: If the agent in the previous phase cannot determine the correctness of the selected patch, the process proceeds by first deduplicating the candidate patches. Then, one or more Selector agents are tasked with selecting the most likely correct patch, and the final output is determined by aggregating their votes.
\end{inparaenum}

\emph{Measuring similarity} among candidate patches guides patch selection in certain submission
For example, \approach{SIMA} \resultsLite{27.67} applies a \emph{vector similarity} voting method that selects the patch most semantically aligned with the others based on embedding distance.
\approach{CodeShellAgent} \resultsVerified{44.2} clusters vectorized candidate patches using an embedding LLM and k-means, selecting the patch closest to the centroid of the largest cluster.

\emph{Prompting LLMs} is also a strategy for selecting a single patch.
For example, \approach{W\&B Programmer O1} \resultsVerified{64.6}, which parallelizes patch generation, uses \llm{OpenAI o1} as a tie-breaker to select the best patch.
\approach{Zencoder} \resultsVerified{70}  employs \llm{o3-mini} to select the best patch from the candidates generated by four agents running in parallel (See \ref{sec:results:phases:patchGeneration}). 
\approach{MASAI} \cite{arora2024masaimodulararchitecturesSE} \resultsBoth{27.33}{32.6} employs an LLM to rank candidate patches based on the issue description and, when available, the results of repository test executions.
In the initial submission (\texttt{2025-04-25}) \approach{Refact.ai Agent} included a specialized tool, \texttt{deep\_analysis}, which can be invoked during the repair workflow. This tool follows a three-step process: after generating an initial patch, it performs a \emph{Critique} step that identifies weaknesses, limitations, or bugs in the generated code. Finally, in the \emph{Refinement} step, the patch is improved based on the feedback from the \emph{Critique} phase. This tool is backed by a reasoning model \llm{o4-mini} that handles the cognitive load of problem.
However, in a later submission, the authors removed the \emph{Critique} step, and the tool now only generates a candidate patch. They found that the approach performs better when it simply runs tests and determines the next steps based on the results.
\approach{ExpeRepair} \resultsLite{60.33} has a dedicated review agent, which selects the final patch based on criteria
such as correctness, code style, and adherence to best practices.

\emph{Scoring the quality} of patches is also employed by approaches such as \approach{Aider} and \approach{AppMap Navie} to support patch selection. For instance, \approach{Aider} \resultsLite{26.33} computes a score based on factors such as the number of edits, linting and testing error.
\approach{DARS}~\cite{aggarwal2025darsdynamicactionresampling} \resultsLite{47} uses a fine-tuned \llm{Deepseek R1} as a \agent{Reviewer LLM} to assign patch scores based on three predefined rubrics. These include criteria such as correctness of the fix and regression risk, each rated on a three-point scale.

A combination of \emph{scoring and voting} is also employed.
For instance, \approach{SWE-RL} \resultsVerified{41.2}, through its scaffold \approach{Agentless-Mini}, first clusters patches into \emph{concept groups}, where each group consists of patches that pass the same set of reproduction tests —a strategy shown to effectively reduce the candidate space~\cite{martinez2024xTestCluster}.
Next, the approach assigns a score to each group based on the number of passing tests and the number of patches it contains.
Finally, it selects the consensus group with the highest score and applies majority voting within the group to determine the best patch.
Similarly, \approach{Codev} \resultsLite{49} applies a cross-validation algorithm, inspired the technique \texttt{xTestCluster}~\cite{martinez2024TestCluster}: given patches and test scripts generated by different trajectories, the technique validates a patch from one trajectory against the generated test from the other trajectories.
\approach{Codev} then assigns scores based primarily on the results of the generated test scripts and regression tests.


\emph{Training or fine-tuning} models to evaluate candidate patches has also been explored.
\approach{AgentScope} \resultsVerified{63.4} employs a reward model during its voting stage, which takes as input a candidate patch and the summarized repair trajectory, assigning a score used to select the best patch.
This reward model is obtained by fine-tuning an open-source model using data from benchmarks such as \texttt{SWE-Gym}\cite{pan2024SWEGym} and \texttt{SWE-bench-extra}\footnote{\url{https://huggingface.co/datasets/nebius/SWE-bench-extra}, a preliminary benchmark for \texttt{SWE-re-bench}\cite{badertdinov2025swerebenchautomatedpipelinetask}}.
\approach{OpenHands} trains a dedicated critic model to select the best solution~\cite{wang2025OInferenceTimeScalingAndCritic}, using agent trajectories from SWE-Gym~\cite{pan2024SWEGym}. The model applies temporal difference (TD) learning to propagate success signals and incorporates a regression head to estimate reward values.
\approach{Nebius} \resultsVerified{40.6} employs a \emph{critic-guided search} strategy~\cite{zainullina2025guidedsearchstrategies} that supports two forms of supervision:
\begin{inparaenum}[\it a)]
\item \emph{process supervision}, which assesses the quality of individual actions and intermediate states, and
\item \emph{outcome supervision}, which evaluates the correctness of the complete trajectory leading to a patch.
\end{inparaenum}
The critic is based on a fine-tuned LLaMA 3 model, trained on trajectories from \approach{SWE-agent} using the \texttt{SWE-bench-extra} dataset developed by \company{Nebius}\footnote{\url{https://huggingface.co/datasets/nebius/SWE-agent-trajectories}}.

\emph{Agents} specialized in patch selection and ranking are also featured in several multi-agent submissions.
For instance, \approach{SpecRover}~\cite{casper2025aiagentindex} \resultsVerified{51.6} employs a \agent{Selection Agent} that selects a patch based on the issue description and provides a justification for its choice.
\approach{IBM AI Agent SWE-1.0} \resultsLite{23.67} integrates a \agent{Judge Agent} responsible for selecting the final patch from those proposed by multiple \agent{Editing Agents}, each potentially powered by a different foundation LLM.
\approach{Emergent E1} \resultsVerified{57.2} includes a \agent{Patch Adjudicator}, an LLM-based agent that evaluates candidate patches and their associated trajectories to select the most suitable one based on the problem statement.
Other submissions incorporating similar agents include \approach{Globant Code Fixer Agent} (\agent{Critic Agent}), \approach{nFactorial} (\agent{Scorer Agent}), and \approach{Bracket.sh} (\agent{Judge Agent}).







\section{Discussion}
\label{sec:discussion}

\subsection{Patch Overfitting on Program Repair and \swebench{}}


An overfitting patch refers to a fix that passes all available test cases, including the bug-revealing one,
yet remains incorrect  incorrect because of weaknesses in the validation oracle, which in practice consists of the test suite.
The software engineering community has a long history of studying patch overfitting, beginning with the seminal works by Smith et al.\cite{smith2015cure}, who coined the term \emph{overfitting patch}, and Qi et al.\cite{qi2015analysis} et al., who introduced the concept of \emph{plausible patches}, patches that pass all test cases yet incorrect.

Nevertheless, patch overfitting affects not only early APR systems but also recent state-of-the-art approaches, including Passerine from Google~\cite{rondon2025AgentGoogle}, Agentless~\cite{xia2024agentlessdemystifyingllmbasedsoftware} and AutoCoderRover~\cite{Zhang2024AutoCoderRover}.
In these works, the authors conducted manual analyses to identify and exclude overfitting patches from their reported results on \swebench{}, following a practice commonly adopted in earlier APR research evaluated on other benchmarks, such as Defects4J (e.g., \cite{martinez2017automatic, Liu2020:ontheefficiency}).
However, it remains unclear to what extent all submitters \swebench{} are aware of or account for the overfitting problem.
Recent studies of \swebench{} results, such as~\cite{wang2025solvedSWEBench}, have analyzed patch correctness and found a substantial number of overfitting patches, leading to an average overstatement of resolution rates by 6.2 absolute percentage points.

Academic or research-oriented submissions -specially those that have been involved in the software engineering community (e.g. \cite{xia2024agentlessdemystifyingllmbasedsoftware,ruan2024specrovercodeintentextraction,Zhang2024AutoCoderRover,rondon2025AgentGoogle,ma2025alibabalingmaagent}) are likely to be aware of the quality standards expected by the research community, for example, validating patch correctness beyond plausibility, often through manual inspection or even eventually with automated techniques such as DDR~\cite{Ye2021:DDR}. 
In their papers, these authors typically report the ratio of bugs repaired with correct patches as well as those with merely plausible ones.
However, we have not observe such analyzes in articles that targets the artificial intelligence community (e.g. \cite{yang2024sweagentagent,li2025patchpilot,zainullina2025guidedsearchstrategies}).
This highlights a misalignment between the software engineering and AI communities in evaluation standards: while the former emphasizes semantic correctness, the latter often relies solely on test-suite-based validation, potentially overlooking incorrect yet test-passing patches.
Additionally, some submissions —particularly from industry— may not be fully aware of the underlying issues associated with test-suite-based evaluation, such as patch overfitting.
These observations highlight the need for the software engineering community to more effectively disseminate established evaluation practices and engage in cross-community dialogue to help align standards more broadly.

The \swebench{} framework and leaderboards provide a valuable foundation for reproducible experimentation and progress tracking in automated issue repair.
However, it currently lacks mechanisms for deeper correctness validation beyond passing test cases. As a result, it remains unclear whether submitters—particularly those outside academia—perform additional analyses to verify the semantic correctness of their patches. This underscores the need for more rigorous evaluation practices to complement test-based validation. Future work could enhance the \swebench{} platform by incorporating optional or standardized procedures for post-submission correctness assessment, helping to ensure the reliability of reported results.





\subsection{Variants of SWE-Bench Beyond Lite and Verified}


Both \swelite{} and \sweverif{} are derived from the same benchmark, \swebench{}, which originally contains 2,294 issues.
The reduction in the number of issues in each subset is primarily due to two factors: benchmark quality control and optimization for experimentation.

Each subset was constructed based on distinct selection criteria:
\swelite{} excludes issues involving complex edits or vague descriptions, resulting in a subset of 300 issues.
\sweverif{} filters out issues with insufficient test coverage, vague descriptions, and other limitations, yielding 500 issues.

Some other sub-benchmarks have emerged from \swelite{} and \sweverif{}.
For instance, Xia et al.~\cite{xia2024agentlessdemystifyingllmbasedsoftware} constructed \texttt{SWE-bench Lite-S} by filtering out problematic issues from the full \swebench{} dataset to enable more rigorous evaluation and comparison—similar in spirit to the goals behind \sweverif{} developed by OpenAI.
Ma et al. proposed \texttt{SWE-bench-Lite-FIX}~\cite{OLDma2024understandRepoUnderstander}, an evolution of  \swelite{} where 45 issues out of the 300 from Lite were improved (fixed) by augmenting the issue description.
\sweverif{} has also been forked into smaller subsets. One example is SWEBench-verified-mini\footnote{\url{https://github.com/mariushobbhahn/SWEBench-verified-mini}}, a reduced version containing 50 issues. 
According to the authors, this subset maintains a similar distribution of performance metrics, test pass rates, and difficulty. Its issues are sourced from only two projects (\swebench{} has 13).
Another subset, named \texttt{lite\_and\_verified\_solvable}
\footnote{\url{https://github.com/aorwall/moatless-tools?tab=readme-ov-file\#run-evaluation}}, contains 84 issues that appear in both the Lite and Verified datasets and have at least one successful submission on \swebench{}.

The evolution of benchmarks often aims to improve data quality, address inconsistencies, and better support evaluation needs. This trend is not unique to \swebench{}; for example, other widely used bug benchmarks such as Defects4J \cite{Just2014Defects4J} have undergone significant revisions, resulting in distinct versions (e.g., v1 and v2). However, while such evolution can enhance a benchmark's reliability and applicability, it also introduces risks.

Although the \swebench{} leaderboard includes both the \swelite{} and \sweverif{} subsets, it does not currently track or evaluate submissions on other emerging forks and derivatives (e.g., \texttt{Lite-S}, \texttt{verified-mini}). 
Without controlled versioning and official support -such as recognition or integration by the leaderboard maintainers, as happened with \sweverif{}- these new subsets or evolutions of existing ones may remain underutilized and have limited influence within the research community.

\subsection{The debate: Single-Agent or Multiple- Agent architecture?}
\label{sec:discussion:singlemultiples}

In our study, we found that the majority of submissions correspond to the Agentic-based approach, and these type of solutions can be Single Agent or Multiple Agents.
From observing the results from both Lite and Verified leaderboards, we cannot determine that one architecture achieves better results than the other.
However, different players have publicly endorsed one approach over the other.

For instance, Cognition, creators of one of the \approach{Devin} pioneering AI software engineer agent, argued against building multi-agent systems. They consider such systems fragile because agents often lack access to a shared global context, which is crucial for making coherent, high-quality decisions.
At the same time, researchers have also focused on failures of multi-agent systems. For instance, Cemri et al.\cite{cemri2025multiagentllmsystemsfail} examined the trajectories of six multi-agent systems, including HyperAgent~\cite{phan2024hyperagent}, a system with results submitted to the \swelite{} dashboard. 
Based on 200 task executions, the authors identified 14 distinct failure modes, grouped into three overarching categories:
\begin{inparaenum}[\it 1)]
\item Specification issues (failures stemming from poor system design or ambiguous prompt specifications),
\item Inter-agent misalignment (failures caused by breakdowns in coordination and communication between agents during execution), and
\item Task verification (failures due to inadequate validation or premature task termination).
\end{inparaenum}
While not all failures were directly attributable to the multi-agent architecture, the most prevalent ones fell under the inter-agent misalignment category —highlighting issues that are inherently linked to the challenges of coordinating multiple agents.

In contrast, just one day later of that post from Cognition, Anthropic published a post outlining lessons learned from building their research system based on a multi-agent architecture.\footnote{\url{https://www.anthropic.com/engineering/built-multi-agent-research-system}}
They argue that, for their use case, sub-agents are beneficial because they operate in parallel with separate context windows, enabling better compression. Additionally, sub-agents support separation of concerns by using distinct tools, prompts, and exploration paths —reducing path dependency and allowing for more thorough and independent investigations.
The authors also acknowledge that certain domains, such as most coding tasks, are not well suited for multi-agent systems with current technologies, as they typically require all agents to share the same context or involve complex interdependencies between agents.
This reveals a partial alignment between the positions of both companies, despite their differing architectural choices.

We also found discussions about this architectural issue on the meta-data from \swebench{} entries.
For example, the entry \approach{Lingxi} focuses on a multi-agent approach, which offers a way to split the repair workflow into compact, purpose-built agents that each receive only the information they need. That offered an option to single agent pipelines where keep the entire dialogue in one context inevitably suffer from \emph{context dilution}, where the prompt is dominated by earlier discussion tokens.
On the contrary, OpenHands fosters a single-agent architecture \cite{neubig2024openhandsmultiagent}
characterized by a single 
\begin{inparaenum}[\it a)]
\item LLM,
\item action space, and
\item prompting technique.
\end{inparaenum}
The authors identify several challenges with multi-agent architectures, including:
\begin{inparaenum}[\it 1)]
\item designing an effective agentic structure, where each agent has clear responsibilities and dedicated tools;
\item preserving context across multiple agents; and
\item maintaining the agents' codebases and prompt.
\end{inparaenum}

Other submitters to \swebench{} have evolved the architecture of their approaches across submissions to address issues arising from their initial choice of single-agent or multi-agent design.
For instance, 
\approach{nFactorial} has submitted four entries to \sweverif{}, each featuring a distinct system architecture.
Their initial submission followed a non-agentic design. In the second submission, they transitioned to a multi-agent architecture incorporating specialized agents such as the Localizer, Fixer, and Code Editor.
However, in the third submission, they merged the Localizer and Fixer into a single agent. 
According to the authors, this change was motivated by the observation that passing context between separate agents sometimes resulted in the loss of critical information.
According to the authors of Warp \resultsVerified{71}, they initially explored multi-agent approaches —including dedicated testing, planning, and reasoning agents— but ultimately found that a single primary agent yielded the most consistent and reliable results.

In conclusion, there is no one-size-fits-all solution, no single architecture prevails, as submitters navigate trade-offs between the modularity of multi-agent systems and the simplicity and context integrity of single-agent designs.
Foundational frameworks and practical toolsets like MAST, proposed by Cemri et al.~\cite{cemri2025multiagentllmsystemsfail}, offer valuable means for diagnosing and improving both single- and multi-agent systems. These tools can help identify and mitigate the types of failures and weaknesses highlighted in the mentioned articles and studies.

\subsection{Known Limitations of \swebench{}}

Despite the success of \swebench{} —adopted by researchers across diverse communities, including Software Engineering and Artificial Intelligence, and having a notable impact on industry, as discussed in Section~\ref{sec:results:submitter}— it still shares some limitations common to other bug and issue benchmarks. These limitations have even been acknowledged and discussed by several submitters.
For example, Zencoder\footnote{https://zencoder.ai/blog/demystifying-swe-bench} summarizes the following.
\begin{inparaenum}
 \item {\bf Data Contamination}: Many issues in \swebench{} were created before the training cutoff dates of several LLMs. This raises the risk of data leakage, where models may have been exposed to issue descriptions or solutions during training, potentially inflating performance scores.  
    To address this, researchers have developed infrastructures to ensure freshness and isolation of benchmark data. For instance, Zhang et al.~\cite{zhang2025swebenchgoeslive} proposed \texttt{SWE-bench-Live}, an automated curation pipeline that streamlines the full process, from instance creation to environment setup, allowing the benchmark to be continuously updated with recent data.
\item {\bf Lack of Real-World Representation}: \swebench{} focuses exclusively on Python repositories, which may limit its ability to represent the full diversity of programming languages and software engineering tasks encountered in real-world development.  
    For example, Zeng et al.~\cite{zeng2025skyworkswe} introduced \textit{Skywork-SWE}, a dataset comprising 10,169 real-world Python task instances. Their evaluation using the \approach{OpenHands} scaffold showed significantly lower resolve rates than those observed on \swebench{} —e.g., 20.23\% with \llm{Gemini-2.5-Pro}, 18.54\% with \llm{GPT-4.1}, and 15.94\% with \llm{o3-mini}— suggesting potential overestimation of model performance on \swebench{}.
     \item {\bf Benchmark Specificity}: Scores on \swebench{} may reflect a model’s performance on known or curated datasets rather than its ability to generalize to novel, diverse, or noisy real-world scenarios. 
\end{inparaenum}

\subsection{Approaching Saturation: Observations from \swebench{} Trends}

In July 2025, submissions to \swebench{} achieved up to 75\% precision—an increase of more than 20 percentage points compared to the previous year.
This progress coincides with the release of increasingly capable LLMs. As depicted in Figure~\ref{fig:evolLLMComboVerified}, all systems exceeding 70\% precision relied on state-of-the-art Claude 4 models, either alone or in combination with others.
If such trends persist,  upcoming LLM releases could surpass 80\% or even  90\%, suggesting that \swebench{} may be approaching a saturation point, as previously observed with benchmarks like HumanEval for the code generation task.\footnote{\url{https://paperswithcode.com/sota/code-generation-on-humaneval}}

\subsection{The importance of Open-source frameworks for experimentation}

\swebench{} serves as an excellent example of the value of open-source tools in enabling rapid experimentation and research progress.
Beyond the benchmark infrastructure is open-source, one of the key factors contributing to its early success was the availability of open-source implementations of several initial submissions—particularly the tools or scaffolds used in approaches such as \approach{Agentless}, \approach{AutoCodeRover}, \approach{SWE-Agent}, \approach{Moatless}, \approach{OpenHands}, and \approach{PatchPilot}.
This openness has significantly accelerated the pace of research by providing a variety of accessible frameworks on which researchers, developers, and industry practitioners can build, extend, and evaluate new ideas.
As discussed in Section~\ref{sec:result:mainArchitecture:worflowFixedNoAgent}, numerous contributions have extended and adapted \approach{Agentless}, demonstrating the benefits of having a reusable and modifiable open-source foundation.
A notable example of this openness is the \approach{SIMA} submission \resultsLite{27.67}, where a single developer experimentally evaluated the ideas proposed in a paper on multi-agent architectures~\cite{li2024MoreAgentsneed} by implementing them on top of the open-source \approach{Moatless} tool.
This highlights how open-source frameworks lower the barrier to entry and empower individual contributors to engage in impactful research and benchmarking.

Surprisingly, some submitters from industry have also embraced open-source practices.
For instance, the small company \texttt{Augment Code}, in addition to offering commercial products (e.g., development plugins), has open-sourced the agent used in their \swebench{} evaluation.\footnote{Augment Code Agent: \url{https://github.com/augmentcode/augment-swebench-agent/tree/main}}
Notably, this agent held the highest precision on the Verified leaderboard (65.4\%) as of March 2025.
This open-source strategy not only enhances the transparency and reproducibility of their results but also contributes to the broader advancement of the community—including other industrial players—as demonstrated by its influence on ByteDance’s submission \approach{TRAE}, which adopted Augment Code’s open-sourced patch selector.

\section{Threats to Validity}
\label{sec:ttv}


\paragraph{\textbf{External Validity.}}
In this study, we focused on two \swebench{} leaderboards, which we selected due to the SWE-bench's substantial impact on both academia and industry, as demonstrated both by the companies involved (most industry big players) and the scientific impact, e.g. with Jimenez et al.'s seminal paper reaching over 780 citations on Google Scholar as of July 2025. Other benchmarks may be equally representative of the broader issue-fixing landscape, but we do not claim that our findings can be applied to them. 

\paragraph{\textbf{Internal Validity.}} We followed a systematic procedure to collect information from the leaderboard submissions, enabling us to characterize both the submitters and the approaches. However, there is a risk that we may have missed documents describing some approaches, resulting in certain submissions being left uncharacterized. To mitigate this, we extended our search beyond the official leaderboard metadata, consulting multiple sources including Google searches, LinkedIn profiles, and scientific publications indexed on Google Scholar and arXiv. Furthermore, the entire information-gathering process was repeated twice to ensure coverage and consistency.

\paragraph{\textbf{Construct Validity.}}

Our classification was based on a published taxonomy \cite{liu2024SurveyAgentsLLM4SE} and three architectural dimensions (workflow authoring, control flow autonomy and presence of agents). While these constructs were designed to capture core aspects of LLM-based repair systems, applying them consistently was challenging  because many submissions lacked detailed or standardized descriptions of their solutions. 
To mitigate potential misclassification, we used explicit categories for unknown cases and cross-validated all annotations among the authors.

Some potentially important dimensions and variables were not analyzed in this study.
First, we did not account for the number of parameters or the specific minor version of each LLM. 
This decision was made to simplify the analysis and focus on identifying broader trends across model families and major releases, rather than fine-grained differences between specific versions.
Second, we did not evaluate the monetary cost associated with the different submissions, despite some submitters reporting this information.
While cost is a relevant factor for assessing the usability and practicality of an approach, especially given that most of the submissions rely or proprietary (paid) models, it was excluded from our analysis for practical reasons.
Token prices have generally decreased over time, making cost comparisons across submissions potentially misleading without proper normalization. Performing such normalization would require detailed, often unavailable, information from the submissions (e.g., price per token at the time of experimentation, total tokens consumed).
Due to these limitations in cost reporting and data availability, we chose to exclude monetary cost from our study.

\paragraph{\textbf{Conclusion Validity.}}

Some submissions, especially those without formal publications and described only in blog posts or informal channels, lacked essential details (e.g. related to the solution architecture - single,  multiple or no agent, autonomy) or contained ambiguous information (e.g., the LLMs used in experimentation by the submitters).
To avoid misclassification, we introduced special categories to explicitly denote unknown or missing information.
Additionally, cross-validation of the classification decisions was conducted among the authors to further reduce the risk of bias or error.




\section{Related work}
\label{sec:relatedwork}


\subsection{Empirical studies  of \swebench{} results}



Previous work have studied in details the effectiveness of approaches evaluated on \swebench{} with a deep focus on the generated patches.
For example, 
Meng et al. \cite{meng2024empiricalstudyllmbasedagents} conducted an empirical study of the efficiency of 7 repair systems on SWE-bench Lite.
In particular, they studied the patches from 4 commercial systems (MarsCode Agent~\cite{liu2024marscode}, Honeycomb, Gru, Alibaba Lingma Agent~\cite{ma2025alibabalingmaagent}) and 3 open-source (AutoCodeRover~\cite{Zhang2024AutoCoderRover}, Agentless + RepoGraph \cite{ouyang2024repographenhancingaisoftware} , Agentless~\cite{xia2024agentlessdemystifyingllmbasedsoftware}). 
All of them are published in the ledearboard and analyzed in our study.
They  conducted a comprehensive analysis of the performance differences of each system. 
This research has three main goals: studying repairability and quality, effect on the fault localization capability, and impact of the reproducibility on bugfixing, with a focus on performance difference of each system.
They use \emph{Issue Quality Metrics} initially defined by Xia et al. \cite{xia2024agentlessdemystifyingllmbasedsoftware} to qualify each issue.
The study found that issue quality significantly influences the effectiveness of resolution methods, underscoring the importance of crafting clear and comprehensive issue descriptions to improve resolution rates from the outset.
They authors also found that existing methods have achieved relatively strong performance in file-level fault localization, but there remains room for improvement in line-level localization tasks.
Finally, Reproduction can provide additional information for defect localization when issue information is lacking and help verify the accuracy of generated candidate patches. However, when the issue description is already clear and precise, reproduction may mislead the LLM's judgment, reducing its focus on the issue description.

Aleithan~\cite{Aleithan2025Revisingswebench} presents a manual analysis of 251 patches generated by the \approach{SWE-Agent + GPT-4} submission. The authors found that in 32.67\% of the cases, the issue description included the complete solution, effectively leaking the correct patch. Additionally, 12.75\% of the patches were incorrect yet still passed the test suite, while 14.74\% were only partially correct, missing critical elements necessary for a full fix. 
When the suspicious instances were removed from the evaluation, the performance of the approach dropped  from 12.47\% to 3.97\%.
This findings motivated Aleithan et al. \cite{aleithan2024swebenchenhanced} to build \emph{SWE-Bench+} a refined version of the benchmark that avoids data leakage in both LLM usage and issue description.

Wang et al. \cite{wang2025solvedSWEBench} presents an empirical study on the correctness of plausible patches generated by three systems evaluated on \sweverif{}: CodeStory, Learn-By-Interact, and OpenHands. The study reveals a flaw in the patch validation process of \swebench{}, where only a subset of the available test cases—specifically those modified by the user in the pull request linked to the issue—is executed. The authors revised this validation approach to run all available test cases and found that, on average, 7.8\% of plausible patches were actually incorrect. 
Motivated by this finding, they introduced PatchDiff, a tool for differential patch testing that automatically detects behavioral discrepancies between the generated patch and the ground truth. Using PatchDiff, they discovered that 29.6\% of plausible patches exhibited different behavior from the ground truth. Upon manual inspection, 28.6\% of these were confirmed to be incorrect. Based on these experiments, the authors estimate that current evaluation practices may overstate resolution rates by an average of 6.2 absolute percentage points.


Chen et al.~\cite{chen2025unveilingpitfallsunderstandingaidriven} go beyond evaluating the final patch generated by agents and instead analyze the full trajectories and testing logs from eight agents submitted to \sweverif{}. Their study focuses on Python execution failures and the most frequent error types encountered during the issue-solving phase, examining how these influence the final outcomes. They find that a higher frequency of execution errors is associated with lower-quality patches and increased corrective reasoning efforts, ultimately complicating the repair process.
Bouzenia et Pradel \cite{bouzenia2025understanding} studied the trayectories of three agents, two  with results submitted to \swebench{} (AutoCodeRover, OpenHands) and one RepairAgent~\cite{bouzenia2024RepairAgent}, not previously evaluated in \cite{bouzenia2024RepairAgent} but in Defects4J~\cite{Just2014Defects4J}.  
They categorize the actions (Explore, Locate, Search, Generate fix, etc.) and then  mined frequent sequences of categorized actions in agent trajectories to identify recurrent decision-making pattern.
They found for instance,  that successful trajectories (that conduct to correct patches) balance exploration, explanation, and validation steps, while unsuccessful ones exhibit repetitive, non-adaptive action cycles.

Ceka et al.\cite{ceka2025understandingsoftwareengineeringagents} conducted an empirical study on the decision-making trajectories of automated repair systems and analyzed the resulting patches submitted to \sweverif{}. Their focus is on two types of agent-based repair systems: workflow-based agents (Agentless\cite{xia2024agentlessdemystifyingllmbasedsoftware}, AutoCoderRover~\cite{Zhang2024AutoCoderRover}, and MASAI~\cite{arora2024masaimodulararchitecturesSE}) and open-process agents (SWE-Agent~\cite{yang2024sweagentagent}, OpenHands~\cite{wang2024openhandsopenplatformai}), all of which are included in our current study.
By examining the agent trajectories during issue resolution, they construct a taxonomy in the form of a graph, where nodes represent discrete actions (e.g., code retrieval, patch synthesis, test invocation), and edges capture transitions between phases. Each transition is annotated with the corresponding agent that executed it. They further incorporate issue difficulty annotations provided by OpenAI~\cite{chowdhury2024swebenchverified}.
Their findings show that agents perform best on easier issues and that different agents excel at different problem types, with no clear overall winner. Compared to human-written fixes, agent-generated patches are typically small and localized—especially for simple issues—while more complex problems reveal greater divergence from developer behavior, as agents rarely apply broad, multi-location changes.

Liang et al.~\cite{liang2025swebenchillusionstateoftheartllms} investigate whether the performance of LLM-based approaches on \swebench{} is driven by genuine coding capabilities or by memorization. To this end, they designed experiments in which models attempt to repair issues using only the issue descriptions, with all repository structure and code context withheld. 
Their findings show that, under this constrained setting, models such as {OpenAI o3-mini}  are still able to repair the majority of issues from \sweverif{}.  However, performance drops when evaluating issues outside of that benchmark, particularly those written in other programming languages (e.g., C\#).



\subsection{Evaluations on \swebench{} not included in the leaderboards}



Further issue repair approaches have been evaluated on \swebench{}, although their results were either not submitted to or not published on the leaderboards at the time of writing.

For example, Antoniades et al.~\cite{antoniades2024Moatless} introduced \approach{SWE-Search}, a multi-agent framework that combines Monte Carlo Tree Search (MCTS) with a self-improvement mechanism to enhance the performance of software agents on repository-level tasks. SWE-Search emphasizes three key principles: adaptability, feedback-driven refinement, and collaborative decision-making. \approach{SWE-Search} is built on top of the Moatless tool, which is also included in our analysis.
In their evaluation on \swebench{} Lite, SWE-Search achieved precision scores ranging from 17\% (using LLaMA-3.1-70B-Instruct and GPT-4o-mini) to 31\% with GPT-4o.

Heshkov et al.~\cite{cheshkov2024conversationaltest} evaluated a \emph{Conversational Patch Generation (CPG)}, which frames the repair process as a question–answer dialogue between a developer and an LLM-based repair approach. 
They tested the approach on 192 single-function bugs from \swelite{} using LLaMA 3.1 and GPT-4o-mini, achieving issue resolution rates of 47\% and 46\%, respectively.


Ma et al.\cite{ma2025thinkinglongerlargerenhancing} explore the use of open-source LLMs —specifically those with 32 billion parameters— as an alternative to the proprietary (and likely larger) models from OpenAI or Anthropic, which are commonly used in \swebench{} submissions. To support this, they propose a unified Test-Time Compute (TTC) scaling framework that leverages increased inference-time computation. Their hybrid approach combines two components: internal TTC, which enhances reasoning depth via extended chain-of-thought (CoT) training\cite{wei2023chainofthought}, and external TTC, which applies reward-guided search and verification to identify optimal solutions.
In their evaluation on \sweverif{}, their fine-tuned version of the \llm{Qwen2.5 Coder 32B} model, named \llm{SWE-Reasoner}, achieves a precision of 46\%.

\subsection{Agent and LLM-based repair approaches evaluated on other benchmarks}

A preliminary effort in agent-based program repair is \approach{RepairAgent}~\cite{bouzenia2024RepairAgent}, which augments an LLM—specifically OpenAI's GPT-3.5—with a set of bug repair-specific tools that the model can invoke to interact with the codebase. \approach{RepairAgent} introduces three key features: (1) a state machine that mimics the cognitive steps a human developer would take when fixing a bug, (2) dynamic prompts that incorporate the current world state, the target goal, and the set of next possible actions, and (3) a long-term memory mechanism to retain information across interaction cycles.
\approach{RepairAgent} has been evaluated on Defects4J, one of the most widely used bug benchmarks in software engineering research (e.g.,\cite{martinez2017automatic,Durieux2019:empirical,Liu2020:ontheefficiency}). Results show that it outperforms several state-of-the-art non-agent-based APR systems, including ChatRepair\cite{Xia2024CharRepair} and SelfAPR~\cite{Ye2022SelfApr}, by repairing a greater number of bugs.

He et al.~\cite{ye2025adversarial} present \approach{AdverIntent-Agent}, a multi-agent system designed to infer developer intent, localize defects, and generate test cases that validate both the inferred intentions and the correctness of proposed patches. The approach supplements developer-provided tests and iteratively refines patches through conversational interactions to produce high-quality repairs.
In their evaluation on Defects4J v2.0, \approach{AdverIntent-Agent} achieves state-of-the-art performance, successfully repairing 77 out of 835 bugs.

Rondon et al.\cite{rondon2025AgentGoogle} investigate the feasibility of applying agent-based automated program repair (APR) in an enterprise setting. To this end, they constructed an evaluation dataset of 178 bugs from Google's internal issue tracking system (GIST), comprising 78 human-reported and 100 machine-reported bugs. They introduce Passerine, a minimalist agent-based APR system inspired by SWE-Agent\cite{yang2024sweagentagent}.
For human-reported bugs, Passerine produced plausible patches (i.e., passing all test cases) in 25.6\% of cases, with 17.9\% being semantically equivalent to the ground-truth fixes. For machine-reported bugs, the system achieved 73\% plausible patches and 43\% semantically equivalent patches.

\section{Conclusion}
\label{sec:conclusion}

In this paper, we presented the first comprehensive study of the submissions to the SWE-Bench Lite and SWE-Bench Verified leaderboards. 
We analyzed \totalentries{} entries corresponding to \totalapproaches{} unique approaches from \totalsubmitters{}, examining their origins, product characteristics, LLM usage, and architectural design of the approaches. 
Our results showed that the majority of high-performing submissions relied on proprietary LLMs, particularly Claude 4, and that both agentic and non-agentic solutions were capable of achieving competitive performance. 
We also found that most submissions came from industry, including both large tech companies and small startups, as well as independent developers. 
By systematically analyzing the submissions to the \swebench{} leaderboards, we uncovered the range of architectural choices and organizational profiles behind them, contributing to a deeper understanding of the current state of the field.


\bibliographystyle{plain}
\bibliography{references}

\end{document}